\documentclass[11pt,twoside]{report}
 \usepackage{a4,hangcaption,amsmath,amssymb,ctagsplt,
   latexsym,fancyheadings,epsfig}
\usepackage[active]{srcltx}
\def\beas{\begin{eqnarray*}}
\def\eeas{\end{eqnarray*}}
\def\bea{\begin{eqnarray}}
\def\eea{\end{eqnarray}}
\newcommand{\be}{\begin{equation}}
\newcommand{\ee}{\end{equation}}
\newcommand{\bes}{\begin{equation*}}
\newcommand{\ees}{\end{equation*}}

\newcommand{\dd}{\displaystyle}
\newcommand{\de}{\partial}
\def\slash#1{#1\!\!\!/}
\begin{document}
\def\Black{}
 \def\AliasBlue{}
 \def\Blue{}
 \def\Brown{}
  \Brown\hfill{BARI-TH 431/02} \vskip1.5cm
\begin{center} \AliasBlue {\Huge EFFECTIVE DESCRIPTION \\OF QCD \\AT VERY HIGH DENSITIES}\\
\vskip1cm\Black{\bf Giuseppe Nardulli}\\
\vskip0.5cm{Physics Department, University of Bari, Italy\\
 I.N.F.N., Sezione di Bari, Italy}\end{center}
\vskip.8cm\begin{center}\Black{\bf Abstract} \end{center} We
describe the effective lagrangian approach to the color
superconductivity. Superconductivity in color interactions is due
to vacuum instability at $T=0$ and high densities, which arises
because the formation of quark-quark pairs is energetically
favored. The true vacuum is therefore populated by a condensate of
Cooper pairs breaking color and baryonic number. The effective
langrangian follows from the Wilson's renormalization group
approach and is based on the idea that at $T=0$ and $\mu\to\infty$
the only relevant degrees of freedom are those near the Fermi
surface. The effective description that arises if one considers
only the leading terms in the $1/\mu$ expansion is particularly
simple. It is based on a lagrangian whose effective fermion fields
are velocity-dependent; moreover strong interactions do not change
quark velocity ({\it Fermi velocity superselection rule}) and the
effective lagrangian does not contain spin matrices. All these
features render the effective theory similar to the Heavy Quark
Effective Theory, which is the limit of Quantum ChromoDynamics for
$m_Q\to\infty$. For this reason one can refer to the effective
lagrangian at high density as the High Density Effective Theory
(HDET). In some cases HDET results in analytical, though
approximate, relations that are particularly simple to handle.

After a pedagogical introduction, several topics are considered.
They include the treatment of the Color-Flavor-Locking and the 2SC
model, with evaluation of the gap parameters by the Nambu-Gorkov
equations, approximate dispersion laws for the gluons and
calculations of the Nambu-Goldstone Bosons properties. We also
discuss the effective lagrangian for the crystalline color
superconductive (LOFF) phase and we give a description of the
phonon field related to the breaking of the rotational  and
translational invariance. Finally a few astrophysical applications
of color superconductivity are discussed.
\Black\thispagestyle{empty}
  \newpage $$$$
  \vfill\thispagestyle{empty}
  \newpage
\setcounter{page}{1}
\tableofcontents
% ----------------------------------------------------------------
%\input{capitolo1.tex}
\chapter{A pedagogical introduction
\label{ch1}}
\section{Overview} The aim of the present review
 is to describe various approaches  to the color
 superconductivity (CSC) \cite{others} that are
  based on the method of the effective lagrangians. CSC is one of the most fascinating
 advances in Quantum Chromo Dynamics (QCD) in recent years
 \cite{alford,wilczekcfl,wilczekcfl1,wilczekcfl2,wilczekcfl3,wilczekcfl4,Schafer:1998na,generali,shuryak,alford1,pisarski,altri} (for reviews see
 \cite{rassegne,rassegne2,hsu}); it
 offers a clue to the behavior of strong interactions at very
 high baryonic densities, an issue of paramount relevance both for
 the understanding of heavy ion collisions and the physics of
 compact stars.

 The origin of the superconductivity in color interactions
 is related to the instability of the naive vacuum at $T=0$; at
 low densities this instability  is supposed to arise
because the formation of a quark-antiquark
 pair is energetically favored; therefore
 Cooper pairs formed by a quark and an antiquark
   (in a color singlet representation) populate the true vacuum. This phase
   is indeed characterized by the order parameter
\be \langle\bar \psi\psi\rangle\neq 0,\label{uno}\ee i.e.  by a
condensate of $\bar q q$ pairs, similarly to the phenomenon of the
Cooper pairs in ordinary superconducting materials at low
temperatures, described by the theory of Bardeen, Cooper and
Schrieffer (BCS)\cite{bardeen}. The condensate (\ref{uno})
violates chiral invariance, which is a symmetry of the QCD
Lagrangian if one neglects the current quark masses.  Therefore a
non vanishing condensate (\ref{uno}) is tantamount to
 the spontaneous breaking of the
chiral symmetry and the existence, because of the Goldstone's
theorem,  of a certain number of massless particles, the
Nambu-Goldstone-Bosons (NGBs).

At high baryonic densities a different phenomenon sets in.
 As a
matter of fact, for sufficiently high baryon chemical potential
$\mu$,
 the color interaction favors the formation of a
quark-quark condensate in the color antisymmetric channel.
 This condensate
\be\langle\psi^T_{i\alpha }C\psi_{j\beta} \rangle
 ,\label{condensatedue}\ee ($\alpha,\,
\beta=1,2,3$ color indices; $i,\,j=1,2,3$ flavor indices) acts as
the order parameter of a new phase where both the $SU(3)_c$ color
symmetry and the $U(1)_B$ baryonic number are spontaneously
broken. As to chiral symmetry, the color interaction favors in
general pairing of quarks of total spin 0
 and opposite momenta; therefore
the quarks in (\ref{condensatedue}) have the same helicities and
the condensate (\ref{condensatedue}) does not break chiral
invariance explicitly\footnote{The presence of an external
chromagnetic field may affect CSC: for an analysis see
\cite{klimenko}.}.

Different phenomena take place depending on the value of the order
parameter (\ref{condensatedue}). One could have indeed:
\be\epsilon^{\alpha\beta\gamma}\epsilon_{ij}<\psi^T_{i\alpha
}C\psi_{j\beta}>=\Delta\delta^{\gamma 3}\label{lab2}\ee where the
sum over the flavor indices run from 1 to 2 and $\psi$ represents
a left handed 2-component Weyl spinor (the right handed field
satisfies the same relation with $\Delta\to - \Delta$); moreover a
sum over spinor indices is understood and $C=i\sigma_2$. This case
correspond to the decoupling of the strange quark
($m_s\to\infty;\, m_u=m_d=0$) and is called in the literature 2SC
model. From dynamical analyses \cite{alford}, \cite{rassegne} one
knows that, for $\mu$ sufficiently large, the condensate
(\ref{lab2}) is non vanishing. Therefore it breaks the original
symmetry group $SU(3)_c\otimes SU(2)_L\otimes SU(2)_R\otimes
U(1)_B$ down to \be SU(2)_c\otimes SU(2)_L\otimes SU(2)_R\otimes
Z_2\ .\ee The chiral group remains unbroken, while the original
color symmetry group is broken to $SU(2)_c$, with generators $T^A$
corresponding to the generators $T^1,T^2,T^3$ of $SU(3)_c$. As a
consequence, three gluons remain massless whereas the remaining
five acquire a mass. The $Z_2$ group means that the quark fields
can still be multiplied by -1.  Even though the original $U(1)_B$
is broken there is an unbroken global symmetry that plays the role
of $U(1)_B$. As for $U(1)_A$, this axial symmetry is broken by
anomalies, so that in principle there  is no Goldstone boson
associated to its breaking by the condensate; however at high
densities explicit axial symmetry breaking is weak and therefore
there is a light would be Goldstone boson associated to the
breaking of the axial color. One can construct an effective theory
to describe the emergence of the unbroken subgroup $SU(2)_c$ and
the low energy excitations, much in the same way as one builds up
chiral effective lagrangian with effective fields at zero density.
For the two flavor case this development can be found in
 \cite{casalbuonisannino}.

For the three flavor case ($m_u=m_d=m_s=0$)  the following
interesting case\footnote{The more general case is\bes
\langle\psi_{i\alpha}^L\psi_{j\beta}^L\rangle=
-\langle\psi_{i\alpha}^R\psi_{j\beta}^R\rangle=\gamma_1\,\delta_{ai}\delta_{bj}+
\gamma_2\,\delta_{aj} \delta_{bi}~,\ees but the solution with
$\gamma_1=-\gamma_2$ corresponding to (\ref{condensates}) is
energetically favored.} has been widely discussed
\cite{wilczekcfl}: \be
\langle\psi_{i\alpha}^L\psi_{j\beta}^L\rangle=
-\langle\psi_{i\alpha}^R\psi_{j\beta}^R\rangle=\Delta
\sum_{K=1}^3\epsilon_{\alpha\beta K}\epsilon_{ijK}~
.\label{condensates}\ee The condensate (\ref{condensates}) breaks
the original symmetry group $SU(3)_c\otimes SU(3)_L\otimes
SU(3)_R\otimes U(1)_B$ down to \be SU(3)_{c+L+R}\otimes Z_2\ .\ee
Both the chiral group and the  color symmetry are broken but a
diagonal $SU(3)$ subgroup remains unbroken in a way that locks
together color and flavor (Color-Flavor-Locking=CFL model). We
have 17 broken generators; since there is a broken gauge group,
 8
of these generators correspond to 8 longitudinal degrees of the
gluons, because the gauge bosons acquire a mass; one has 9 NGBs
organized in an octet associated to the breaking of the flavor
group and in a singlet associated
 to the breaking of the baryonic number.
 The effective theory describing the NGB for the CFL model
 has been studied in \cite{casalbuonigatto}.

The 2SC model might be applicable for intermediate chemical
potential $\mu$, while for very high $\mu$ the CFL phase should
set in; therefore the QCD phase diagram should display different
phases as schematically depicted in Fig. 1.1.
\begin{figure}[htb]
\epsfxsize=9.5truecm \centerline{\epsffile{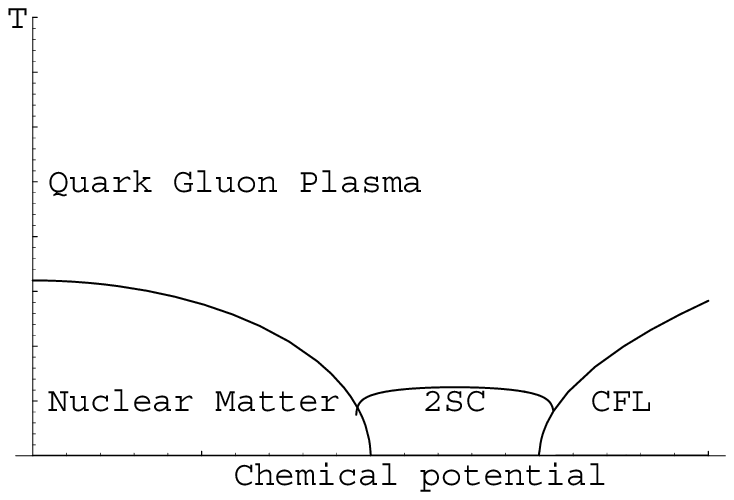}}
\begin{center} {Fig. 1.1}  {Schematic view of the QCD phase diagram}
\end{center}\end{figure}

Similarly to the chiral symmetry breaking condensate (\ref{uno}),
the phase transition and the non vanishing of the condensates
(\ref{lab2}) or (\ref{condensates}) result from a mechanism
analogous to the formation of an electron Cooper pair in a BCS
superconductor. At $T=0$ the only QCD interactions are those
involving fermions near the Fermi surface. Quarks inside the Fermi
sphere cannot interact because of the Pauli principle, unless the
interactions involve large momentum exchanges. In this way the
quarks can escape the Fermi surface, but these processes are
disfavored, as large momentum transfers imply small couplings due
to the asymptotic freedom property of QCD. Even though
interactions of fermions near the Fermi surface involve momenta of
the order of $ \mu$, their effects are not necessarily negligible.
As a matter of fact, even a small attractive interaction between
fermions near the Fermi surface and carrying opposite momenta can
create an instability and give rise to coherent effects. This is
what really happens \cite{alford}, \cite{wilczekcfl} and the
result is the formation of a diquark condensate, as expressed by
(\ref{lab2}) or (\ref{condensates}). We stress again that the only
relevant fermion degrees of freedom are therefore those near the
Fermi surface.

This property and the effective theory for the fields at the Fermi
surface is discussed for electromagnetism and ordinary BCS
superconductivity in
\cite{Polchinski:1992ed,Shankar:1993pf,benfatto} where the Wilson
Renormalization Group approach \cite{wilson} is used to prove the
Landau Theory of the Fermi liquid \cite{landau} and to prove the
marginal character of the BCS interaction.  First works towards
the implementation of these ideas in the context of color
superconductivity can be found in \cite{Evans:1998ek,hong,beane}.
In \cite{cflgatto} this approach was used to build up an effective
theory for the CFL model based on the approximation of the neglect
of the negative energy states. This results in a rather terse
formalism resembling the heavy quark effective theory \cite{HQET}
and displaying as a characteristic note the existence of a Fermi
velocity superselection rule and effective velocity-dependent
fermion fields. Due to these features, we refer to this effective
lagrangian as the High Density Effective Theory (HDET). In
\cite{2fla} the 2SC model has been studied by the same formalism,
while in \cite{gattoloff,nardulliloff,gattoloff2} this effective
theory has been applied to a different color superconducting
phase, the crystalline color superconducting phase
\cite{LOFF,LOFFbis,LOFF3,LOFF4,earlier1}, also called the LOFF
phase from the authors of the original model of the crystalline
electromagnetic superconductivity \cite{originalloff}.

The aim of this paper is to review the developments in the
description of Color Super Conductivity that are based on the
effective lagrangian approach, in particular the HDET approach.
While there are good reviews of the whole subject, the approach
based on effective lagrangian has not been extensively discussed
yet and this paper aims to do that. In the remaining part of this
first section we will present an elementary introduction to some
basic aspects of color superconductivity, most notably the
Wilsonian version of the Landau theory of Fermi liquids and the
derivation of the gap equation. The effective lagrangian approach
for the two basic models (CFL and 2SC) is discussed extensively in
the subsequent Section 2. In Section 3 we review some theoretical
developments that in a way or in another are related to the main
theme of this survey. In Section 4 we present a detailed
discussion of the LOFF state of QCD. Finally in Section 5 we
discuss astrophysical implications of color superconductivity in
the context of the compact stars phenomenology.

\section{Color superconductivity}
To start with, we consider a  non interacting fermion gas; its
energy distribution  ($\beta=1/kT$)
 \bes f(E)=\frac{1}{e^{\beta(E-\mu)}+1}\ ,\ees
 becomes, when  the gas is degenerate ($T\to 0$), the step function, see fig. 1.2:
\be f(E)\to\theta(\mu-E)\ .\label{1.1}\ee The condition $E\leq\mu$
defines a domain in the momentum space whose boundary is the
surface $E(p)\equiv \epsilon_F=\mu$,  called Fermi surface;
$\epsilon_F$ is called Fermi energy and the momentum $p_F$
satisfying $E(p_F)=\mu$ is the Fermi momentum. Any quantum state
inside the Fermi surface, i.e. $E(p)\leq\epsilon_F$ is occupied by
a fermion, and the other ones are empty. For $m=0$, $E=p$ and the
Fermi surface is spherical $p_x^2+p_y^2+p_z^2=p_F^2=\mu^2$. This
results applies  also for non relativistic non interacting fermion
(with $p_F^2=2m\mu$).

In presence of interactions the situation is more complicated,
because, in general $f=f(\vec r,\vec p)$. At very large chemical
potentials one  can however take advantage of the asymptotic
(large $p$) properties of QCD, which implies that at $T=0$ the
only QCD interactions are those involving fermions near the Fermi
surface, as we remarked already.  We shall discuss in the sequel
the role of the surviving interactions for the quarks at the Fermi
surface that, even if small, can be not negligible. For the time
being we observe that the chemical potential is related to the
fermion density (number of fermions per unit of volume $n$ by \bes
n=g\int \frac{d^3p}{(2\pi\hbar)^3}\,f\ ,\ees where $g$ is the
degeneracy of the state. If $f(\vec r,\vec, p)$ does not depend on
$\vec r$ this results in a constant density $n$. Clearly the
density increases with $\mu$; in particular, for degenerate free
fermions
 one gets
 \be  n =\frac{4\pi \,g}{(2\pi\hbar)^3}\int_0^{p_F}p^2dp=
\frac{4\pi g p_F^3}{3h^3}\ .\label{density} \ee
 In a similar vein the
energy density and the pressure are given by: \be \dd
\epsilon=g\int \frac{d^3p}{(2\pi\hbar)^3}\,E\,f\
,\label{energydensity} \ee
 and
 \be
 \dd P=g\int
\frac{d^3p}{(2\pi\hbar)^3}\frac{pv}{3}\,f \label{pressure} \ee
respectively.
\begin{figure}[htb]
\epsfxsize=7truecm \centerline{\epsffile{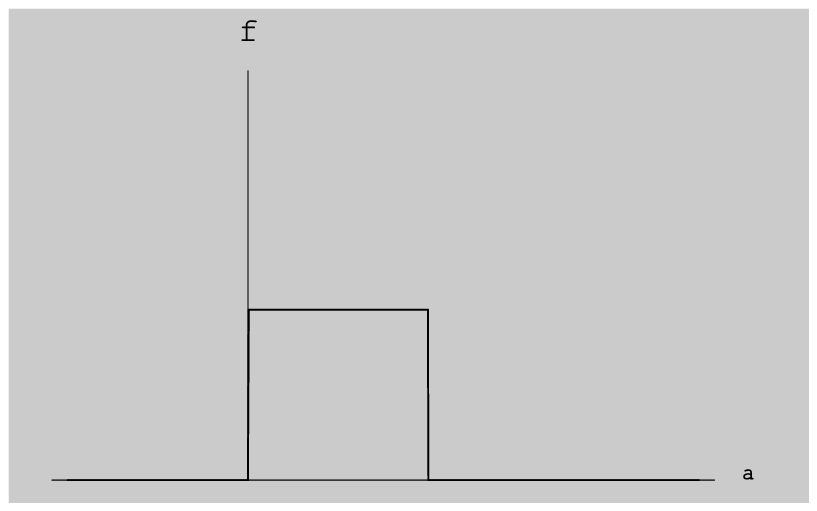}} \noindent
%\begin{center}
Fig. 1.2  {Fermi distribution $f(E)$ at $T=0$. $f(E)=0$ for
$E>\mu=\epsilon_F$, and $f(E)=1$ for $0\le E\le\mu=\epsilon_F$.}
%\end{center}
\end{figure}
For non interacting fermions the state we have described is the
ground-state, i.e. the vacuum. Let us now suppose that the
fermions are quarks and let us discuss the origin of the color
superconductivity; we show by a simple argument that the naive
vacuum is unstable provided an attractive interaction exists. Let
us consider a pair of non interacting fermions near the Fermi
surface; their energy is $\approx 2\mu$. Let us substitute them
with a bound state, i.e. a pair (a boson) with binding energy
$\epsilon_{int}<0$. The fermion energy decreases by the amount
$-2\mu$; on the other hand the boson energy increases by
$2\mu+\epsilon_{int}$, so that the total energy is decreased by
the amount $-|\epsilon_{int}|$. Therefore the formation of a bound
state is favored. This might appear a small thing, but it is not,
as one can immediately realize since there is an infinite number
of ways to do that. The Cooper pair has preferably total momentum
and total spin zero, but there is an infinite number of possible
pairs, corresponding to the different directions of the Fermi
velocity, see fig. 1.3 \be \vec v_F=\frac{\vec p_F}{\mu}\ .\ee
\begin{figure}[htb]
\epsfxsize=1.8truecm \centerline{\epsffile{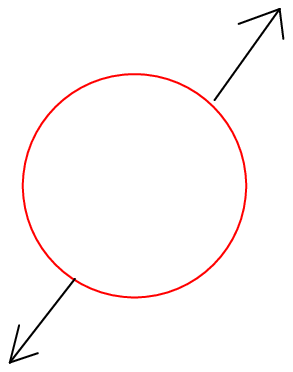}} \noindent
%\begin{center}
{Fig. 1.3}  {Cooper pair of two fermions living at the Fermi
surface with total momentum and total spin zero.}
%\end{center}
\end{figure}

For the phenomenon of the superconductivity to occur one therefore
needs an attractive interaction between the fermions. In the case
of electronic superconductivity in the metals the attraction is
the effect of the interaction of the electrons with the phonon:
the resulting interaction between the fermions is indeed
attractive and provides the required binding energy of the
electronic Cooper pair. In the case of quarks the situation is
more favorable because the QCD interaction in the color
antitriplet ${\bf {\bar 3}}$ (antisymmetric color channel) is
attractive; therefore color superconductivity can occur even in
absence of a lattice. The result of the Cooper interaction is the
formation of a colored diquark condensate, with total baryonic
number 2.

The possibility of color superconductivity was studied long ago
\cite{others}, but only recently \cite{alford} it has been
realized that the effect offers several physical possibilities and
is indeed sufficiently robust to survive the thermal fluctuations.
The physical mechanism at its basis arises from interactions
between two fermions at the Fermi surface. They will be discussed
in the next paragraph.
\section{Effective action \`a la Wilson \label{wilson}}
The previous results can be expressed in a more formal way by
making use of the Landau theory of the Fermi liquid \cite{landau};
we will describe it in the  Renormalization Group language \`a la
Wilson
  \cite{Polchinski:1992ed,Shankar:1993pf,benfatto,wilson}
  \footnote{Our presentation follows
\cite{Polchinski:1992ed}; a similar discussion is in
\cite{Evans:1998ek}.}. The idea of this approach is that the
effective interactions of the fermions at $T=0$ can be described
substituting the particles with their excitations, the {\it
quasiparticles} that are in  one-to-one correspondence
 with the  particles. If one is
interested only in special values of the momenta (in our case
momenta near the Fermi surface), one can ignore the details of the
interaction, which considerably simplifies
 the description
without loss of physical insight. The simplification consists in
the fact that  the quasiparticles can be treated as free. However
the form of the energy dependence on the momentum, i.e. the
dispersion law $\epsilon=\epsilon(\vec p)$, differs from the free
particle case.

To derive in the framework of the Landau theory
  the results already discussed in previous
sections, let us start with a gas of fermions at $T\simeq 0$ and
the free lagrangian \be {\cal L}=\bar \psi \left( i
\slash\partial+\mu\gamma_0\right)\psi\ , \ee from which the free
action can be constructed \be S=\int dt \,d\vec p\,\psi^\dag(\vec
p)\left( i\partial_t+\mu-\epsilon(\vec p)\right)\psi(\vec p) \
.\label{action}\ee For free particles one obtains (\ref{action})
using  the Dirac equation to write $\vec \alpha\cdot\vec
p\,\psi=\gamma_0\vec \gamma\cdot\vec p\,\psi=\epsilon(\vec
p)\,\psi$ (see below, Eq. (\ref{2.1})); therefore, for free
particles, $\epsilon(\vec p)=\sqrt{\vec p^{\,2}+m^2}$. However the
interactions can modify this simple dispersion law and therefore
we do not use this result. Because of this possible modifications,
as we have already stressed, the particles are substituted, in the
Landau theory of the normal Fermi liquid, by quasi-particles, that
we can imagine as particles dressed by their interactions.

Let us now consider what happens when we scale  energies and
momenta down towards the Fermi surface. Let us divide the fermion
momentum as follows \be\vec p=\vec k+\vec \ell\ ,\ee with $\vec k$
on the Fermi surface and $\epsilon(\vec k)=\mu$ (i.e. $\vec k$ is
a Fermi momentum),
 while $\vec\ell=\ell\, \vec n$, with $\vec n$ a unit vector
orthogonal to the Fermi surface. We scale down the momenta by a
factor $s\to 0$ so that they approach the Fermi surface; clearly
\bea E&\to&s E\ ,\cr \vec k\, &\to&\, \vec k\ ,\cr \vec\ell\,
&\to\, &s \vec\ell\ .\label{scale} \eea We note that these
conditions can  be also expressed as follows: \be E,\,\,\,|\vec
\ell|<\delta\ .\ee with $\delta\le\mu$. Moreover \be \epsilon(\vec
p)-\mu\to\epsilon(\vec k )-\mu+(\vec p-\vec
k)\cdot\frac{\partial\epsilon}{\partial\vec p}\Big|_{\vec p=\vec
k} =\vec\ell\cdot\frac{\partial\epsilon}{\partial\vec
p}\Big|_{\vec p=\vec k}=\ell v_F(k)\ , \ee where \be \vec v_F(\vec
k)=\frac{\partial\epsilon}{\partial\vec p}\Big|_{\vec p=\vec k}
\ee is a vector orthogonal to the Fermi surface. Therefore \be
S=\int dt \,d\vec k\,d\ell\,\psi^\dag(\vec p)\left(
i\partial_t-\ell v_F(\vec k)\right)\psi(\vec p) \
.\label{actionbis}\ee

 Under the scale
transformation (\ref{scale}) we also have \bea dt&\to&s^{-1} dt\
,\hskip1cm d^3\vec p=d^2\vec k\,d\ell\to sd^2\vec kd\ell\ ,\cr
\partial_t&\to& s\partial_t\ ,\hskip 2cm\ell\to s\ell\  ,\label{scale2}\eea
which means that the fields $\psi$ in the action $S$ of Eq.
(\ref{action}) scale as $s^{-1/2}$, while $S$ scales as $s^0$. Let
us note explicitly that for spherical Fermi surfaces the
integration is 2-dimensional and the measure is \be d\vec
k=p_F^2d\Omega\ .\ee

Let us now consider the terms in the action that might result from
the
 interactions;
they are constrained by the symmetries of the problem. We classify
them according to the number of fermion fields:
\begin{enumerate}
  \item A term quadratic in the fermion fields, such as
  \be
\int dt\,d^2\vec k\, d\ell\, m(\vec k)\,\psi^\dag(\vec p)\psi(\vec
p)\ ,
  \label{quadratic}
  \ee scales as $s^{-1}$. Potentially this is a {\it relevant} operator
  \footnote{In the language of the Renormalization Group,
   an operator that becomes less and less important
  when the momenta scale is called {\it irrelevant}; an
  operator that becomes more and more important
  is called {\it relevant}; an operator equally important
  at all scales is called {\it marginal}.}. However
  $m(\vec p)$ can be absorbed \footnote{Since we can absorb $m(\vec p)$ into
  $\epsilon(\vec p)$ we cannot assume for the dispersion law the free particle expression;
  the specific form of the dispersion law depends on the
  nature of the interaction.} in the definition of
  $\epsilon(\vec p)$, so that (\ref{quadratic}) is not a new operator at all. As to other quadratic
  terms such as, for example,
  those containing time derivatives and/or factor of
  $\ell$  they can be either absorbed in the corresponding terms
  in (\ref{action}) or are irrelevant.
  \item A quartic term such as
  \bea
  \int dt&\prod_{j=1}^4 &\left(d^2\vec k_j d\ell_j\right)
  \left(\psi^\dag(\vec p_1)\psi(\vec p_3)\right)
  \left(\psi^\dag(\vec p_2)\psi(\vec p_4)\right)\cr&&\cr&\times&V(\{\vec k_j\})
  \delta(\vec p_1+\vec p_2-\vec p_3-\vec p_4)\label{quartic}
  \eea
scales as $s=s^{-1}s^{-4/2+4}$ multiplied by the scale factor of
the Dirac delta; it is therefore irrelevant if one can assume that
\bea \delta(\vec p_1+\vec p_2-\vec p_3-\vec p_4)&=&
  \delta(\vec k_1+\vec \ell_1+\vec k_2+
  \vec \ell_2-\vec k_3-\vec \ell_3-\vec k_4-\vec \ell_4)
  \cr&&\cr&\to& \delta(\vec k_1+\vec k_2-\vec k_3-\vec k_4) \ .\label{1.21}
  \eea
   \item Operators containing $2n$ ($n>2$) fermion fields
   scale as $s^{n-1}$ and are therefore irrelevant.
\end{enumerate}
This is the Landau theory of the normal Fermi liquid derived by
the renormalization group approach. Let us now explain how this
theory is modified for superconducting materials. Let us consider
an interaction term with  a quartic coupling (\ref{quartic}) such
that the two incoming momenta $\vec p_1$ and
 $\vec p_2$ satisfy  $\vec p_1+\vec p_2=0$ (see Fig. 1.3). If the Fermi surface is parity
 symmetric, which we assume,  then
 $\vec k_1+\vec k_2=0$ and also $\vec k_3+\vec k_4=0$. In other
  terms,
 for the special case of scattering with zero momentum, instead of
 (\ref{1.21}) we have
\be \delta(\vec p_1+\vec p_2-\vec p_3-\vec p_4)=
  \delta(\vec \ell_1+\vec \ell_2-\vec \ell_3-\vec \ell_4)
 \ ,\label{1.21bis}
  \ee which scales as $s^{-1}$.

 It follows that the
  corresponding operator becomes marginal and, since there are
  no relevant
  interactions, it may dominate. This is the mechanism at the basis of
  the phenomenon of  Cooper pairing in the BCS \footnote{The BCS theory,
  valid in the limit of weak electron-phonon coupling, is
  generalized to strong  coupling by the
  Eliashberg theory \cite{Eliashberg}.} superconductors.
 The Cooper pairing dominates over the electronic
 electrostatic
 repulsion, in spite of the fact that  the Cooper pair
 has a size of $10^4$ \AA\,
  and the binding
 energy between the electrons
is of order
 $10^{-3}$ eV, while typical electronic interactions are of the order
 of the electron volt.
\begin{figure}[htb]
\epsfxsize=11truecm \centerline{\epsffile{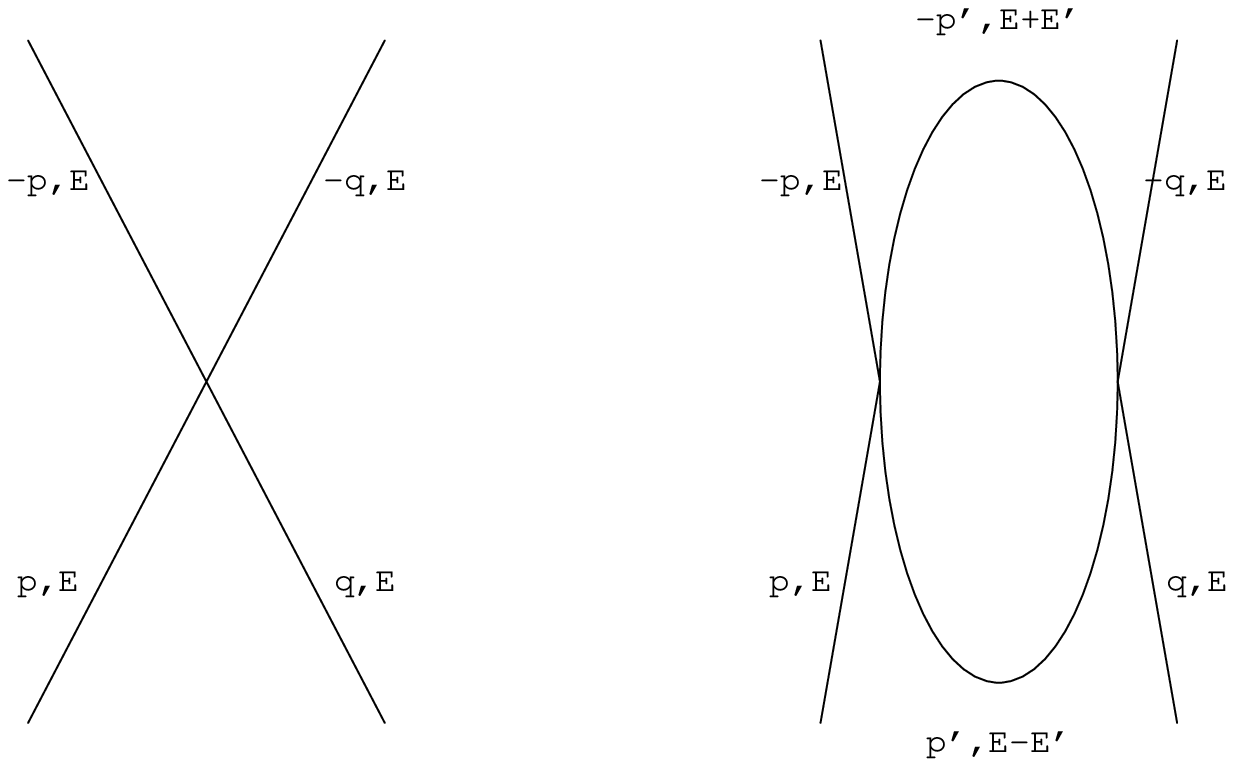}} \noindent
{Fig.1.4}  Four-fermion vertex with the marginal coupling (sum of
incoming momenta $\vec p+(-\vec p)=0$), together with the loop
correction; $+\vec q$ and $-\vec q$ are the final momenta.
\end{figure}

 The origin of the attractive interaction lies in the
 electron-phonon interactions
 in the metal. In order to study them we substitute them with an effective
 four-fermion interaction that, as we know, for special values of
 the momenta is marginal.

 The importance of the
 marginal interactions depends on the quantum corrections because they can
 produce the effect of reducing or increasing the coupling. Let us
 therefore show that, if one considers the loop diagram of Fig 1.4, the attractive
 interaction becomes stronger.

 To compute Fig 1.4 we use the coupling (\ref{quartic}) assuming, for
 simplicity a constant interacting potential $V$. The
 fermion propagator can be easily derived from the (\ref{actionbis}). The energy and the longitudinal
 momentum integration are cut-off, so that one is actually considering only momenta near the Fermi surface.

  The sum of the two diagrams is
 given by

 \bea &&V(E)\ =\
V\ -\ V^2\times\cr&&\cr&\times& \int\frac{dE'd^2\vec k d\ell}{
 (2\pi)^4[(E+E^\prime)(1+i\epsilon)-v_F(\vec k)\ell][(E-E^\prime)(1+i\epsilon)-
 v_F(\vec k)\ell]}\cr&&\cr&=&
 V-V^2N\ln\frac {\delta}{E}\ ,\label{efficace}
 \eea
where \be\ N=\int\frac{d^2\vec k}{(2\pi)^3v_F(\vec k)} \ee and
$\delta$ is the upper cutoff in the energy integration. From
(\ref{efficace}) we get, at the same order of approximation, \be
V(E)\approx \frac{V}{1+NV\ln\frac{\delta}{E}} \ee which shows
that, when $E\to 0$, if $V>0$ (repulsive interaction),
 $V(E)$ becomes weaker and weaker
while, if  $V<0$ (attractive interaction), the coupling becomes
more negative and stronger. Therefore one expects that, if $V<0$,
at zero temperature the naive vacuum is unstable and the formation
of fermion pairs is favored.

For metals at low temperatures the situation is as follows. If at
some intermediate scale $E_1=\left(\frac m M\right)^{1/2}\delta$
($m=$ electron mass, $M=$ nucleus mass) the attractive interaction
is stronger than the Coulomb interaction, then at smaller energies
the ordinary vacuum, filled with gapless fermions, will become
unstable and  the electrons will condense; if, one other hand,
this condition is not satisfied, the metal is an ordinary
conductor. In a superconductor there is a non-vanishing vacuum
expectation value of the electron bilinear: \be
\langle\psi_e\psi_e\rangle\neq 0 \ . \label{ordbcs} \ee The
condensate (\ref{ordbcs}) acts as an
 order parameter and its non-vanishing value
 signals the phenomenon of
 spontaneous symmetry breaking. As the
 broken symmetry is a gauge symmetry, $U(1)_{e.m.}$,
  there are no Goldstone bosons associated to it, but the photon
  acquires a mass, which produces, for example, the Meissner effect.
   This is
 what happens for BCS electromagnetic superconductors. For QCD with light quarks at zero density and zero temperature
 this mechanism is supposed to favor the formation of a $\bar q q$ condensate, with breaking
 of the chiral symmetry and associated Nambu-Goldstone bosons (the pions). For QCD at high density
an attractive coupling ($V<0$) between a quark pair
 is provided by the color antitriplet channel; therefore
it should ultimately dominate for $T\to 0$ and $\mu\to\infty$.

\section{The true vacuum} Once we have been convinced that the
naive vacuum is unstable we should try to find  the true
superconducting vacuum state.  To do that we have to show that the
formation of a quark-quark condensate occurs because of the color
interaction.
 The interaction term is fundamentally provided by gluon exchanges and to compute it and derive the gap equation
 one can follow different approaches, such as, for example, the one of Bailin and Love in \cite{others} based on
 the resummation of bubble diagrams. For this method and similar approaches we refer to the reviews
\cite{rassegne}.

For pedagogical purposes it is however sufficient to consider a
Nambu-Jona Lasinio \cite{NJL} (NJL) interaction  (4 fermion
interaction) that mimics the QCD interaction. In this
approximation the hamiltonian is: \be H=\int d^3x\left(\bar
\psi(i\,\slash\partial+\mu\gamma_0)\psi+F\bar \psi\gamma^\mu
T^A\psi\bar \psi\gamma_\mu T^A\psi\right)\ . \label{true}\ee Here
$F$ is a form factor that is introduced to take into account the
asymptotic freedom property of QCD; its Fourier transform behaves
as follows: $F (p)\to 0$  for $p\to\infty$, and this property also
ensures the good behavior of the integrals for $p\to\infty$.

In order to obtain the gap equation we make a  mean field ansatz:
\be \psi^T_{\alpha i} C\psi_{\beta j}~\to \langle\psi^T_{\alpha i}
C\psi_{\beta j}\rangle . \label{condensatetrue0}\ee We repeat that
Greek indices $\alpha,\,\beta$ are color indices, Latin indices
$i,\,j$ are flavor indices. Let us now show that one expects \be
\langle\psi^T_{\alpha i} C\psi_{\beta j}\rangle=\frac{\Delta}2
\epsilon_{\alpha\beta I}\epsilon_{ijK}\Omega_{IK}\
.\label{condensatetrue}\ee It is useful to note that, by
evaluating (\ref{true}) in mean field approximation, i.e.
substituting to a pair of fermion fields   in (\ref{true}) their
mean value given by (\ref{condensatetrue}) and
(\ref{condensatetrue0}), the interaction term takes the form
\be\sim\Delta\psi\psi\ +\ h.c.\ ,\ee i.e. the fermions acquire a
Majorana mass.

To prove (\ref{condensatetrue}) we note that,
 as we have already discussed, QCD interaction favors antisymmetry in color; the pair has  zero
angular momentum (in this way the entire Fermi surface is
available and the effect becomes macroscopic); if it has also spin
zero the fermion pair must be in an antisymmetric state of flavor
to produce an antisymmetric wavefunction \footnote{With one flavor
this possibility does not exist and one remains with the sole
possibility of a condensate with angular momentum equal to 1; in
this case, however the superconducting effect is more modest, with
gaps of the order of 1 MeV
 \cite{oneflavor1}.}.

Clearly the form  of the matrix
 $\Omega_{IK}$ corresponding to the true vacuum state depends on dynamical
 effects. In the literature two cases have been widely
 discussed
\begin{enumerate}
  \item Color-Flavor-Locking (CFL): $\Omega_{IK}
 =\delta_{IK}$, see (\ref{condensates});
  \item  2 flavor SuperConductivity (2SC): $\Omega_{IK}
 =\delta_{I3}\delta_{K3}$, see (\ref{lab2}),
\end{enumerate}
and we will concentrate our attention on them (other cases have
been also considered; we refer to the literature, as reviewed for
example in \cite{rassegne})

The CFL model takes its name from the well known property\be
\sum_{K=1}^3\,\epsilon_{\alpha\beta
K}\epsilon_{ijK}\,=\,\delta_{\alpha i}\delta_{\beta
j}-\delta_{\alpha j}\delta_{\beta i}\ ,\label{locking}\ee by which
the color and flavor indices become mixed and indeed locked
together.

 The
2SC model corresponds to the presence in the condensate of only
the $u$ and $d$ flavors: this means that this ansatz may be
relevant in the situation characterized by an intermediate
chemical potential $m_u,\,m_d\ll\,\mu\,\ll\, m_s$. Therefore one
can imagine of having taken the limit $m_s\to\infty$ and the
theory is effectively a two-flavor theory.

Let us now introduce $a,a^\dag,b,...$ annihilation/creation
operators of particles and holes:
  \be \psi_{\alpha
 i}=\sum_k u_k\left(a_{k}e^{ikx}+
 b^\dag_{k}e^{-ikx}\right)_{\alpha
i }\ .\ee

 The existence of the  condensate gives rise
 to the following hamiltonian
\beas H&=&\sum_{\vec k}\left(
|k-\mu|a^\dag_ka_k+(k+\mu)b^\dag_kb_k\right)\cr &+&\sum_{\vec
k}\frac{\Delta F^2}2 e^{-i\Phi}a_k a_{-k}+\frac{\Delta F^2}2
e^{+i\Phi}b^\dag_k b^\dag_{-k}+hc\eeas where we have omitted for
simplicity any color-flavor indices as we wish to stress the
mechanism that produces the condensate. Let us now perform  a
unitary Bogolubov-Valatin \cite{bogolubov}  transformation. To do
that  we introduce annihilation and creation operators for
quasiparticle and quasiholes \bea y_k&=&\cos\theta
a_k-e^{i\Phi}\sin\theta a^\dag_{-k}\ ,\cr z_k&=&\cos\varphi
b_k-e^{i\Phi}\sin\varphi b^\dag_{-k}\ .\eea By an appropriate
choice of the parameters of the transformation we can transform
the original hamiltonian in a new hamiltonian describing a gas of
non interacting quasiparticles \be  H=\sum\left[ \epsilon_y(k)
y^\dag_k y_k\,+\, \epsilon_z(k) z^\dag_k z_k
\right]\label{free}\ee  with \bea
\epsilon_y(k)&=&\sqrt{(k-\mu)^2+\Delta^2F^4}\ ,\cr&&\cr
 \epsilon_z(k)&=&\sqrt{(k-\mu)^2+\Delta^2F^4} \ .\eea
These equations show two effects:
\begin{enumerate}
\item The quasiparticles and quasiholes are free, there is no
interaction term;
  \item in the dispersion laws of the quasiparticles a mass term
  proportional to $\Delta$, called {\it gap parameter}, appears.
\end{enumerate}
 To obtain (\ref{free}), i.e. a free hamiltonian, the parameters have to be chosen as follows:
 \bea
\cos2\theta&=&\frac{|k-\mu|}{\sqrt{|k-\mu|^2+\Delta^2F^4}} \ ,\cr
\cos2\varphi&=& \frac{k+\mu}{\sqrt{|(k+\mu)^2+\Delta^2F^4}} \eea
 We note that, differently from the original $a_k,\,b_k$ operators
 that annihilate the false vacuum, the quasiparticles annihilation operators destroy the true vacuum
 \be y_k|0>=0,~~~~z_k|0>=0\ .\ee

   We have still to
prove that $\Delta\neq 0$. We do that and get an equation for
$\Delta $  by substituting  for $y_k, z_k$ in eq.
(\ref{condensatetrue}). We get in this way an integral equation
that has the following schematic form
 ({\it gap equation}):
 \be
  \Delta = C \int dk F^2(k)
  \Big[\frac{\Delta}{\sqrt{(k-\mu)^2+\Delta^2F^4}}+
    \frac{\Delta}{\sqrt{(k+\mu)^2+\Delta^2F^4}}\Big]\ .
\label{gap}\ee We see immediately that the origin of the
instability of the false vacuum lies in the first of the two terms
in the r.h.s. of (\ref{gap}): If $\Delta=0$ there is no
compensation for the divergence at $k=\mu$.

To be more quantitative we have to consider the two models in more
detail. We differ this quantitative analysis to Section \ref{ch4},
while giving here only the results. \vskip.5cm {\bf  CFL model}
($m_s= m_u= m_d=0$). \vskip.5cm In this case all the $3\times 3=9$
quarks acquire a Majorana mass. The CFL condition gives two
different set of eigenvalues. The first one comprises 8 degenerate
masses $\Delta_1=\Delta_2=...=\Delta_8$ and the second set the non
degenerate mass $\Delta_9$. The actual values of the gaps depends
on the model and the approximations involved; typical values are
\bea\Delta_1&=&...=\Delta_8=\Delta\approx 80 MeV\ ,\cr&&\cr
\Delta_9&\approx&-2\Delta \ ,\label{gapcfl}\eea  for $\mu\approx
400$ MeV (a numerical analysis for this case is contained in
Section \ref{2.7}). The interaction couples quarks of different
flavor and color. Moreover, since we start with a massless theory,
the pairing occurs for left-handed and right handed quarks
separately. Therefore these condensates  break $SU(3)_c\otimes
SU(3)_L\otimes SU(3)_R$; finally, since the pair has baryonic
number 2/3, also $U(1)_B$ is broken. It should be observed,
however, that a diagonal $SU(3)_{c+L+R}$ symmetry remains
unbroken. \vskip.5cm {\bf  2SC model} ($m_s\gg\mu\gg m_u, m_d$).
\vskip.5cm In this case there are 4  massive quarks (the up and
down quarks with colors 1 and 2), while the strange quark with any
color
 and the $u$ and $d$ quarks with color 3 remain ungapped.
Numerical results for this case are qualitatively in agreement
with the results (\ref{gapcfl}) of the CFL model (see for example
Section \ref{2.7}).

One can show, using the Schwinger-Dyson equation (also called in
this context the Nambu-Gorkov equation), that, for arbitrary large
$\mu$, the CFL model is favored \cite{wilczekcfl3}; however, for
intermediate $\mu$ the 2SC state can be more stable and the phase
diagram should resemble that depicted in Fig. 1.1.

The spontaneous breaking of global symmetries implies the
existence of Nambu Goldstone bosons; for internal symmetries there
are as many NGBs as there are broken generators $G_a$. The NGBs
are
 massless scalar particles that interpolate among different degenerate vacua
 \be [H,\,G^a]\,=\,0\hskip1cm
  G^a|0>\,\neq \,0\ .\ee
  As they are massless, they are the lowest energy quasiparticles
of the effective theory. On the other hand, for broken local gauge
symmetries, by the Higgs-Anderson mechanism the gauge bosons
acquire masses and there are no NGBs. In CFL model the diquark
condensate (\ref{condensates}) breaks  $SU(3)_L\times
SU(3)_R\times SU(3)_c\times U(1)_B$  to  $SU(3)_{c+L+R}\times
Z_2$. All the 9 quarks are massive and they belong to a
$SU(3)_{c+L+R}$ singlet and a $SU(3)_{c+L+R}$ octet. All the 8
gluons are massive and are degenerate. There are 8+1
Nambu-Goldstone bosons. In the 2SC model the condensate
(\ref{lab2})  breaks  $SU(3)_c\otimes SU(2)_L\otimes SU(2)_R\times
U(1)_B$  down to $ SU(2)_c\times SU(2)_L\times SU(2)_R$. While the
chiral group is unbroken, $SU(3)_c$ is broken to $SU(2)_c$.
Therefore 3 gluons remain massless and 5 acquire a mass.  As to
the other quasiparticles, there is one would-be NGB associated to
the breaking of the axial color; moreover of the 6 quarks (2
flavors in 3 colors)
 4 are massive and 2 are massless.
\section{Effective lagrangian for the NGB in the CFL model\label{cflngbeff}}
We wish now derive an effective lagrangian for the low energy
excitations (NGBs) of the CFL model \cite{casalbuonigatto}. In
order to do that  let us introduce two coset fields $X$ and $Y$
transforming under the symmetry group as the left handed and right
handed quark fields \bea X&\to& g_cXU_L^T\ ,\cr&&\cr Y&\to& g_c Y
U_R^T\ . \eea $X$ and $Y$ contain both 8 scalar fields \bea
X&=&\exp\left\{i\frac{\tilde\Pi_X^AT^A}F\right\}\cr&&\cr
Y&=&\exp\left\{i\frac{\tilde\Pi_Y^AT^A}F\right\}\eea corresponding
to the 16 NGBs induced by the spontaneous symmetry breaking
$SU(3)_c\times SU(3)_L\times SU(3)_R\to SU(3)_{c+L+R}$ (we neglect
for the moment the $U(1)_B$ factor). We introduce the gauge
covariant currents \bea J^\mu_X&=&X\partial^\mu X^\dag +g^\mu\
,\cr&&\cr J^\mu_Y&=&Y\partial^\mu Y^\dag +g^\mu\ , \eea where \be
g_\mu=i g T^A A_\mu^A\ee is the gluon field. Under the gauge group
these currents transform as follows: \bea J^\mu_X&\to&
g_cJ^\mu_Xg_c^\dag\cr&&\cr J^\mu_Y&\to& g_c J^\mu_Y g_c^\dag \ .
\eea The most general gauge invariant (but not Lorentz invariant)
lagrangian containing $X$, $Y$ and $g_\mu$ fields is as follows
\bea {\cal L}_{eff} &=& -\frac{F^2}4 \tilde g_{\mu\nu}\Big\{
Tr\left[(J^\mu_X -J^\mu_Y)(J^\nu_X-J^\nu_Y)\right]\cr&+& \alpha
Tr\left[(J^\mu_X +J^\mu_Y)(J^\nu_X+J^\nu_Y)\right]\Big\} \cr
&+&{\rm (kinetic\,term\,for\,}g)\ ,\label{leff}\eea where $\tilde
g_{\mu\nu}=diag(1,\,-v,\,-v,\,-v)$ and $F,\,\alpha$ are
parameters. We will prove in par. \ref{ngbcfl} that an effective
lagrangian of this type for the scalar fields
 is generated by loop corrections.
We rewrite this equation as follows \bea {\cal L}_{eff} &=&
-\frac{F^2}4 \tilde g_{\mu\nu}\Big\{ Tr\left[(X\partial_\mu
X^\dag- Y\partial_\mu Y^\dag)(X\partial_\nu X^\dag- Y\partial_\nu
Y^\dag)\right]\cr &+&\alpha Tr\Big[(X\partial_\mu X^\dag+
Y\partial_\mu Y^\dag+2g_\mu)(X\partial_\nu X^\dag+ Y\partial_\nu
Y^\dag+2g_\nu)\Big]\Big\}\cr &+&{\rm (kinetic\,term\,for\,}g)\
\label{su3ter}.\eea Note that in this equation one has both a
Debye and a Meissner mass for the gluon, $m^2_D=\alpha g^2F^2$,
$m^2_M=v^2\alpha m^2_D$ corresponding to a mass term for $A_0$ and
$\vec A$, respectively; they
 can be read from (\ref{su3ter}). We will show in par. \ref{ngbcfl}
 that one has to put $\alpha=1$ in order to have the correct normalization for the gluon masses.
Since $F={\cal O}(\mu)$, in the limit of small momenta, one can
neglect the kinetic gluon term and, because of the equations of
motion, the gluon field becomes an auxiliary field given by\be
g_\mu=-\frac 1 2 (X\partial_\mu X^\dag- Y\partial_\mu Y^\dag) \
.\label{af}\ee
 Therefore the lagrangian (\ref{su3ter}) reduces to
 \be {\cal L}_{eff} = -\,\frac{F^2}4 \tilde g_{\mu\nu}
Tr\left[(X\partial_\mu X^\dag- Y\partial_\mu Y^\dag)(X\partial_\nu
X^\dag- Y\partial_\nu Y^\dag)\right]\ . \label{su3bis}\ee

If we put \be \Sigma=Y^\dag X=\exp\left\{\frac{2\,i\Pi^A
T^A}{F}\right\}\ , \label{gauge}\ee $T^A=\lambda^A/2$,  from
 (\ref{su3bis}) we finally obtain
\be {\cal L}_{eff} = \frac{F^2}4
Tr\left(\partial_t\Sigma\partial_t\Sigma^\dag\,-\,v^2\,
\vec\nabla\Sigma \cdot\vec\nabla\Sigma^\dag \right)\ .
\label{su3}\ee
  We note that the matrix $\Sigma$ transforms under
the chiral group as \be\Sigma\to U_L\Sigma U_R^\dag \ .\ee The
values of the parameters in (\ref{su3bis}) have been derived by
several authors (the derivation in the framework of the effective
theory will be presented in Section \ref{ch4}). One finds
\cite{sonstephanov,zarembo,rho1,rho2,honglee,tytgat} : \be
F^2=\frac {\mu^2(21-8\ln 2)}{36\pi^2}\ , \hskip1cm v^2\,=\,\frac 1
3\ .\ee

A similar analysis can be done for the superfluid mode
\cite{casalbuonigatto},\cite{sonstephanov}; we can therefore write
down the low energy effective lagrangian including both the
$SU(3)$ and the $U(1)$ NGBs as follows:
 \bea {\cal L}_{NGB} &=& \frac{F^2}4
Tr\left(\partial_t\Sigma\partial_t\Sigma^\dag\,-\,\frac 1 3\,
\vec\nabla\Sigma \cdot\vec\nabla\Sigma^\dag \right)\cr
&+&\frac{f^2_\sigma}2 \left(\partial_t U\partial_tU^\dag\,-\,\frac
1 3\, \vec\nabla U \cdot\vec\nabla U^\dag \right)\
,\label{su3u1}\eea where \be
U=\exp\left\{\frac{i\sigma}{f_\sigma}\right\}\ ,\hskip1cm
f^2_\sigma=9\mu^2/\pi^2\ ,\ee
 see below, eq. (\ref{fsigma}).

 The lagrangian (\ref{leff}) is completely analogous to the hidden gauge symmetry version of the chiral
lagrangian of QCD \cite{bando}, but for the breaking of the
Lorentz invariance: $g_{\mu\nu}\neq \tilde g_{\mu\nu}$, due to the
high density.

We finally note that there is an extra $U(1)_A$ symmetry in the
lagrangian, which is broken by instantons. Even though the
contribution of the instantons is parametrically small at
asymptotic densities, it is sufficient to forbid the presence of
an extra massless scalar \footnote{Nevertheless, because of this
suppression, this extra pseudoscalar is expected to be rather
light. It can be observed that the instanton interaction is
suppressed not only by the asymptotic energies, but also by the
fact that, being a 6-quark coupling, it becomes irrelevant at the
Fermi surface.}. Notice that one has no NGB associated to the
breaking of the chiral symmetry, because, as the non-vanishing
condensates only couple quarks with the same helicities, chiral
symmetry is broken only indirectly, by the fact that the mechanism
of Color Flavor Locking locks left handed
 quarks to color as well as right handed quarks to color.

\chapter{High Density Effective Theory
\label{ch4}}
\section{The 2-D effective theory}
We wish now to derive an effective description of QCD at very high
baryonic chemical potential $\mu$ and very small temperature
($T\to 0$).

To begin with we consider the equations of motion for a massless
quark at finite density \beas  \left(i\slash D
+\mu\gamma_0\right)\psi \ =\ 0\ ,\eeas  where  $\mu=$ is the quark
number chemical potential. The $\mu\gamma_0$ term corresponds to a
$\mu N=\mu\psi^\dagger\psi$ term in the hamiltonian. We note
explicitly the breaking of Lorentz invariance of this equation,
which, however, is still  symmetric under rotations. At $T=0$ the
vacuum is characterized by the fact that the fermions fill in all
the low energy states up to a maximum energy, the Fermi energy.
Due to the Pauli principle, the interaction of low energy quarks
implies the exchange of high
 momenta, which is disfavored by the Asymptotic Freedom property of QCD.
 Therefore
only the fermions that are near the Fermi surface can interact.
Since they have high momentum ($p\simeq \mu$) they are almost free
and we can write: \bes D^\mu\rightarrow
\partial^\mu=-ip^\mu\ ,\ees i.e.
in momentum space: \bes \left(\slash p +\mu\gamma_0\right)\psi=0 \
,\ees or ($\vec\alpha=\gamma_0\vec\gamma$): \be p_0\psi\ =\
\left(\vec \alpha\cdot\vec p-\mu\right)\psi\ .\label{2.1}\ee

Let us now consider the eigenvalue equation for the Dirac
hamiltonian:\be H_D=p_0=\vec\alpha\cdot\vec p-\mu\ ,\ee whose
eigenvalues can be found solving the equation $(p_0+\mu)^2=\vec
p^{\,2}$. There are positive and negative energies, given by \be
p_0=E_\pm \equiv-\mu\pm|\vec p\,|~\label{eq:3.3}\ee with
 eigenvectors \be \dd{\psi_\pm\ =\ P_\pm\psi\ \equiv\
\frac{1}2 \left(1\pm\frac{\vec\alpha\cdot\vec p}{|\vec p|}\right)\
\psi}\label{2.4}\ . \ee   Therefore at high-density
 the energy spectrum of a
massless fermion is described by states $|\pm\rangle$ with
energies given by (\ref{eq:3.3}). As we have already stressed, for
energies much lower than the Fermi energy $\mu$, among the states
$|+\rangle$ only those close to the Fermi surface. i.e. with
$|\vec p\,|\approx\mu$, can be excited. As to the states
$|-\rangle$, they  have $E_-\approx -2\mu$ and therefore decouple;
this means that in the physical spectrum only the positive energy
states $|+\rangle$ and the gluons remain.

After this quantum mechanical introduction, let us consider the
field theoretical version of the previous argument. The main idea
of the effective theory is the observation that the quarks
participating in the dynamics have large ($\sim \mu$) momenta.
Wherefore one can introduce velocity dependent fields by
extracting the large part $\mu\vec v$ of this momentum. One starts
with the Fourier decomposition of the quark field $\psi(x)$: \be
\psi(x)=\int\frac{d^4p}{(2\pi)^4}e^{-i\,p\cdot x}\psi(p)\
,\label{4.0.dec}\ee and introduces the quark velocity by
 \be p^\mu=\mu
v^\mu+\ell^\mu\ , \label{4.1.dec} \ee where \bes v^\mu=(0,\vec v )
\ees with $|\vec v|=1$. Let us put \bes \ell^\mu=(\ell^0,\vec\ell)
\ees and \bes \vec\ell=\vec v \ell_\parallel+\vec\ell_\perp \ees
with \bes \vec\ell_\perp= \vec\ell-(\vec\ell\cdot\vec v )\vec v\
.\ees We can always choose the velocity parallel to $\vec p$, so
that $\vec\ell_\perp=0$ and \be \int d^4 p=\mu^2\int d\Omega\int
d\ell_\parallel\int_{-\infty}^{+\infty}d\ell_0=
4\pi\mu^2\int\frac{d\vec v}{4\pi} \int  d\ell_\parallel
\int_{-\infty}^{+\infty} d\ell_0 \ .\label{4.0.8}\ee
 In general we will be interested only in the degrees of freedom
 at the
 Fermi surface\footnote
 {There are some important cases where this limitation
 does not apply, see below.}, therefore we can limit the
 integration limits as follows:
 \bes
 -\delta\leq\ell_\parallel\leq+\delta\ees with $\delta\le\mu$ and
 $\delta$ much larger than any other energy scale, in particular
the energy gap masses, if any. By this hypothesis
 \be \int d^3p=4\pi\mu^2\sum_{\vec v}{}^\prime\int_{-\delta}^{+\delta}
 d\ell_\parallel\ ,\label{4.0.8bis}\ee with\bes\sum_{\vec
 v}{}^\prime\equiv\int\frac{d\vec v}{4\pi}\ ;\ees eventually we
 take the limit $\delta\to+\infty$. In this way
   the Fourier decomposition (\ref{4.0.dec}) takes the form
\bes \psi(x) = \sum_{\vec v}{}^\prime e^{-i\mu v\cdot
x}\int\frac{d^4\ell}{(2\pi)^4}e^{-i\ell\cdot x}\,\psi_{\vec v
}(\ell)\ ,\ees where\be
\int\frac{d^4\ell}{(2\pi)^4}=\frac{4\pi\mu^2}{(2\pi)^4} \int
d^2\ell\ ,\ee while  $\psi_{\vec v }(\ell)$ are velocity-dependent
fields. The projection operators  in (\ref{2.4})
 can be approximated as follows:
\be P_\pm=\frac{1\pm\vec\alpha\cdot\vec v}2\ee  and one can
write\be\psi(x)=\sum_{\vec v}{}^\prime e^{-i\mu v\cdot x}
 \left[\psi_+(x)+\psi_-(x)\right]\ ,\ee where
\be  \psi_\pm(x)=\ P_\pm \psi_{\vec v}(x)=P_\pm\int
\frac{d^4\ell}{(2\pi)^4}e^{-i\ell\cdot x} \psi_{\vec v}(\ell)\ \
.\ee It is clear that $\psi_\pm$ are velocity dependent fields
corresponding to positive and energy solutions of the Dirac
equation.

Let us now define \bea V^\mu&=&(1,\,\vec v)\
,~~~~~~~~~~~~~~~\tilde V^\mu=(1,\,-\vec v)\, ,\cr&&\cr
\gamma^\mu_\parallel&=&(\gamma^0,\,(\vec v\cdot\vec \gamma)\,\vec
v)\ ,~~~~~~ \gamma_\perp^\mu=\gamma^\mu-\gamma_\parallel^\mu\
.\eea Moreover we use \bea
\bar\psi_+\gamma^\mu\psi_+&=&V^\mu\bar\psi_+\gamma^0\psi_+\ ,\cr
\bar\psi_-\gamma^\mu\psi_-&=&V^\mu\bar\psi_-\gamma^0\psi_-\ ,\cr
\bar\psi_+\gamma^\mu\psi_-&=&V^\mu\bar\psi_+\gamma^\mu_\perp\psi_-\
,\cr
\bar\psi_-\gamma^\mu\psi_+&=&V^\mu\bar\psi_-\gamma^\mu_\perp\psi_+
\ . \eea Substituting into the Dirac part of the QCD lagrangian we
obtain, in the hypothesis that the residual momenta $\ell$ are
small, \bea {\cal L}_D&=&\sum_{\vec v}{}^\prime
\Big[\psi_+^\dagger iV\cdot D\psi_++\psi_-^\dagger(2\mu+ i\tilde
V\cdot D)\psi_-\,+\cr &+& (\bar\psi_+i\slash D_\perp\psi_- + {\rm
h.c.} )\Big]\ ;\eea $\slash D_\perp=D_\mu\gamma^\mu_\perp$ and
$D_\mu$ is the covariant derivative: $D^\mu=\partial^\mu+ig
A^\mu$. We note that here quark fields
 are   evaluated at the same Fermi velocity; off-diagonal terms
are subleading due to the Riemann-Lebesgue lemma, as they
 are cancelled by the rapid oscillations of the
exponential factor in the $\mu\to\infty$ limit. One may call this
phenomenon {\it  Fermi velocity superselection rule}, in analogy
with the behaviour of QCD in the $m_Q\to\infty$ limit, where the
corresponding effective theory, the Heavy Quark Effective Theory
exhibits a similar phenomenon \cite{HQET}. By the same analogy we
may refer to the present effective theory as High Density
Effective Theory (HDET),

We can get rid of the negative energy solutions by integrating out
the $\psi_-$ fields in the generating functional; at the tree
level this corresponds to solve the equations of motion, which
gives \be iV\cdot D\, \psi_+=0\ee and\be
\psi_-=-\frac{i}{2\mu+i\tilde V\cdot D}\gamma_0 \, \slash
D_\perp\, \psi_+\ .\ee This equation shows the decoupling of
$\psi_-$ in the $\mu\to\infty$ limit. Therefore, in this limit,
the $\psi_-$ field plays no role. We also note that in the
equation for $\psi_+$ only the energy and the momentum parallel to
the Fermi velocity are relevant and the effective theory is
two-dimensional.

We can write the effective lagrangian
 at the next to leading order in
$\mu$ as follows: \bes {\cal L}=-\frac{1}{4}F_{\mu\nu}^a
F^{a\mu\nu}+{\cal L}_D\ ,\ees where\be {\cal L}_D=\sum_{\vec
v}{}^\prime \left[\psi^\dagger iV\cdot D\psi -
\psi^\dagger\frac{1}{2\mu+i\tilde V\cdot D}\slash
D_\perp^2\psi\right]\ .\label{4.2} \ee \vskip.5cm \noindent Using
the identity \bes\psi^\dagger_+
\gamma_\perp^\mu\gamma_\perp^\nu\psi_+=\psi^\dagger_+P^{\mu\nu}\psi_+
\ees where\vskip0.1cm \be P^{\mu\nu}=g^{\mu\nu}-\frac 1
2\left[V^\mu\tilde V^\nu+V^\nu\tilde V^\mu\right]\ ,\label{219}\ee
we can write (\ref{4.2}) as follows

 \be{\cal L}_D=\sum_{\vec v}{}^\prime \left[\psi^\dagger
iV\cdot D\psi  - P^{\mu\nu}\psi^\dagger\frac{1}{2\mu+i\tilde
V\cdot D}D_\mu D_\nu\psi\right]\ .\label{4.2nuovo} \ee

We can make this equation more symmetric by introducing positive
energy fields with opposite velocity: \be
\psi_\pm\equiv\psi_{+,\pm\vec v}\ .\label{NUOVA}\ee Even though we
use the same notation $\psi_\pm$ here and for the
positive/negative energy fields, we stress that $\psi_\pm$ in
(\ref{NUOVA}) and in the sequel are positive energy fields, with
opposite Fermi velocities: $\pm\vec v$. Eq. (\ref{4.2nuovo}) can
be therefore written as follows: \bea {\cal L}_D& =& \sum_{\vec v}
\Big[\psi_+^\dagger iV\cdot D\psi_+\ +\ \psi_-^\dagger i\tilde V
\cdot D\psi_-\cr& -& \psi_+^\dagger\frac{P^{\mu\nu}}{2\mu+i\tilde
V\cdot D}D_\mu D_\nu \psi_+ -
\psi_-^\dagger\frac{P^{\mu\nu}}{2\mu+iV\cdot D}D_\mu D_\nu
  \psi_-\Big]\ .\label{220bis}
  \eea
  To this lagrangian involving only the left
  handed fields, one should add a similar one containing the right
  handed fermionic fields. We observe that a factor $\dd \frac 1 2$
  has been introduced in the
definition of the sum over the Fermi velocities that now reads:
\be \sum_{\vec v}=\int\frac{d\vec v}{8\pi}\ . \label{sumvel} \ee
This extra factor is needed because, after the introduction of the
fields with opposite $\vec v$, $\psi_-=\psi_{+,-\vec v}$, we have
doubled the degrees of freedom and therefore
 we must integrate only over half solid
angle. Apart from this factor there should be another \footnote{In
\cite{cflgatto} this factor explicitly appears in the lagrangian;
here we have simplified the notation.} factor $\dd \frac 1 2$ in
(\ref{220bis}), but we got rid of it by a redefinition of the
fermion field, \bes \psi\to\frac{1}{\sqrt 2}\psi\ .\ees

We close this section by
 a definition that will be useful in the sequel:
 \be {\cal L}_D={\cal L}_{0}+{\cal L}_1+{\cal L}_2\,,
\ee with \bea {\cal L}_{0}& =& \sum_{\vec v} \Big[\psi_+^\dagger
iV\cdot \partial\psi_+\ +\
\psi_-^\dagger i\tilde V \cdot \partial\psi_-\Big]\ ,\\
{\cal L}_1& =&i\,g\, \sum_{\vec v} \Big[\psi_+^\dagger iV\cdot
A\psi_+\ +\ \psi_-^\dagger i\tilde V \cdot A\psi_-\Big]\
,\label{l1}\eea and\be {\cal L}_2=-\sum_{\vec v}
\,P^{\mu\nu}\,\Big[ \psi_+^\dagger\frac{1}{2\mu+i\tilde V\cdot
D}D_\mu D_\nu \psi_+ + \psi_-^\dagger\frac{1}{2\mu+iV\cdot D}D_\mu
D_\nu
  \psi_-\Big]\ .\label{l2}
  \ee
${\cal L}_{0}$ is the free quark lagrangian (without gap mass
terms); ${\cal L}_{1}$ represents the coupling of quarks to one
gluon; ${\cal L}_{2}$ is a non local lagrangian containing
couplings
 of two quarks to any number of gluons. Its effect will be discussed in section
 \ref{evaluating}.

\section{HDET for superconducting phases}

 The construction described above  is  valid for any theory
describing massless fermions at high density provided one excludes
 degrees of freedom very far from the Fermi surface. As discussed
 in Section \ref{ch1}, however, for small temperature and high  density
  the fermions are likely to be gapped due to the phenomenon of the color superconductivity, We shall examine the
  modification in the formalism
 in two models, i.e. the Color Flavor Locking, CFL, model (3 massless
quarks: $m_u,m_d,m_s\ll\mu$) and the Two-Flavor, 2SC, model
($m_s\to\infty$)

\subsection{HDET in the CFL phase}

In the Color Flavor Locking phase the symmetry breaking is induced
by the condensates
 \be
\langle\psi_{\alpha i}^{L\,T} C\psi_{\beta  j}^L\rangle=
-\langle\psi_{\alpha i}^{R\,T}C\psi_{\beta
j}^R\rangle=\frac{\Delta}2\, \epsilon_{\alpha\beta
I}\epsilon_{ijI}~,\label{3.2.1}\ee where $\psi^{L,\,R}$ are Weyl
fermions and $C=i\sigma_2$. Eq. (\ref{3.2.1})  corresponds to the
invariant coupling ($\psi\equiv \psi_L$): \bes -\frac{\Delta} 2
\sum_{I=1,3}\psi^T C\epsilon_I\psi\epsilon_I~-(L\to R)\ +
h.c.,\ees and \be \left(\epsilon_I\right)_{ab}=\epsilon_{Iab}\
.\ee For the Dirac fermions $\psi_\pm$ we introduce the compact
notation
\be \chi=\begin{pmatrix}\psi_+\\
C\psi^*_-\end{pmatrix}\label{chi}\ee and we use a different basis
for quark fields:\be\psi_\pm=\frac{1}{\sqrt{2}}\sum_{A=1}^9
\lambda_A\psi^A_\pm~.\ee The CFL fermionic lagrangian has
therefore the form:  \bea {\cal L}_D&=& {\cal L}_{0}\,+\, {\cal
L}_1\,+\, {\cal L}_\Delta\,=\cr &=&\sum_{\vec v}\sum_{A,B=1}^9
\chi^{A\dagger}\begin{pmatrix}iTr[T_A\,V\cdot D\,T_B] &
\Delta_{AB}\\
\Delta_{AB} &i Tr[T_A\,\tilde V\cdot
D^*\,T_B]\end{pmatrix}\chi^B\cr&&\cr &+& (L\to R)
\label{cflcomplete0} \eea where \be \Delta_{AB}
=\,\Delta\,Tr[\epsilon_IT_A^T\epsilon_IT_B]\label{2.31} \ee and
\be T_A=\frac{\lambda_A}{\sqrt 2}\ . \ee Here
$\lambda_9=\lambda_0=\sqrt{\frac 2 3 }\times\bf 1$. We use the
identity ($g$ any 3$\times$3 matrix): \be
\epsilon_Ig^T\epsilon_I\,=\,g\,-\,Tr\,g \ ; \label{identity}\ee we
obtain \be \Delta_{AB}\,=\,\Delta_A\delta_{AB} \label{2.38}\ee
where \be \Delta_1=\cdots=\Delta_8=\Delta \label{2.39}\ee and \be
\Delta_9=-2\Delta\ . \label{2.40}\ee The CFL free fermionic
lagrangian assumes therefore the form: \be {\cal L}_0+\, {\cal
L}_\Delta\,=\sum_{\vec v}~\sum_{A=1}^9
\chi^{A\dagger}\begin{pmatrix}iV\cdot \partial & \Delta_A\\
\Delta_A &i\tilde V\cdot \de\end{pmatrix}\chi^A\ +\ (L\to R)\
.\label{cflcomplete}\ee Here the average over the velocities is
given by (\ref{sumvel}). From this equation one can immediately
obtain the free fermion propagator that  in momentum space is
given by:\be S_{AB}(p)=\frac{\delta_{AB}}{V\cdot p\,\tilde V\cdot
p-\Delta_A^2+i\epsilon}\begin{pmatrix}\tilde V\cdot p &
-\Delta_A\cr-\Delta_A & V\cdot
p\end{pmatrix}\label{propagatorcfl}\ee We note that this
propagator is analogous to the Nambu-Gorkov propagator of the full
theory.

\subsection{
HDET in the 2SC phase} For the two flavour case, which encompasses
both the 2SC model and the existing calculation in the LOFF phase
we follow a similar approach.

The symmetry breaking is induced by the condensates
 \bes
\langle\psi_{\alpha i}^{L\,T}C\psi_{\beta  j}^L\rangle=
-\langle\psi_{\alpha i}^{R\,T}C\psi_{\beta
j}^R\rangle=\frac{\Delta}2\, \epsilon_{\alpha\beta
3}\epsilon_{ij3}~,\ees
  corresponding to the
invariant coupling ($\psi\equiv \psi^L$): \bes {\cal L
}_\Delta=-\frac{\Delta}2\, \psi^T C\epsilon\psi\epsilon\ -(L\to
R)+{\rm h.c.}~,\ees where \be\epsilon=i\sigma_2\ .\ee As in the
previous section we use a different basis for the fermion fields
 by writing \bea \psi_{+,\alpha i}&=&
\sum_{A=0}^3\frac{(\sigma_A)_{\alpha i}}{\sqrt 2}\psi_{+}^A
~~~~~~~~(i,\,\alpha=1,\,2)~\cr \psi_{+,3 1}&=&\psi_{+}^4\cr
\psi_{+,3 2}&=&\psi_{+}^5\ ,\eea where $\sigma_A$ are the Pauli
matrices for $A=1,2,3$ and $\sigma_0=1$. As usual $\psi_{+}^A$ are
positive energy, velocity dependent fields and we also introduce
$\psi_{-}^A$ according to the formula: \be
\psi_{\pm}^A\equiv\psi_{+,\pm\vec v}^A\ .\ee

A different, but also useful notation
 for the fields $\psi_{+,\,\alpha i}$ uses
 a combination of $\lambda$ matrices, as follows
\be \psi_{+,\alpha i}= \sum_{A=0}^5\frac{(\tilde\lambda_A)_{\alpha
i}}{\sqrt 2}\psi_{+}^A ~.\ee The $\tilde\lambda_A$ matrices are
defined in terms of the usual $\lambda$ matrices as follows: $\dd
\tilde\lambda_0=\frac 1{\sqrt 3}\lambda_8\,+\, {\sqrt\frac 2
3}\lambda_0$; $\dd \tilde\lambda_A=\lambda_A\,(A=1,2,3)$; $\dd
\tilde\lambda_4=\frac{\lambda_{4-i5}}{\sqrt 2}$; $\dd
\tilde\lambda_5=\frac{\lambda_{6-i7}}{\sqrt 2}$.

 After the introduction, analogously to (\ref{chi}),
of the fields $\chi^A$, the 2SC fermionic lagrangian assumes the
form:  \bea {\cal L}_D&=& {\cal L}_{0}\,+\, {\cal L}_1\,+\,\,
{\cal L}_\Delta\,= \cr&=&\sum_{\vec v}\sum_{A,B=0}^5
\chi^{A\dagger}\begin{pmatrix}iTr[\tilde T_A\,V\cdot D\,\tilde T_B
]&
\Delta_{AB}\\
\Delta_{AB} &iTr[\tilde T_A\,\tilde V\cdot D^*\,\tilde T_B
]\end{pmatrix}\chi^B\cr&&\cr &+& (L\to R)\ . \label{2sccomplete0}
\eea Here \bea \Delta_{AB} &=&\,\frac{\Delta}{2}\,Tr[\epsilon
\sigma_A^T\epsilon \sigma_B] \hskip1cm(A,B=0,...3)\cr  \Delta_{AB}
&=&\,0\hskip3.3cm(A,B=4,5)\ . \eea and \be \tilde
T_A=\frac{\tilde\lambda_A}{\sqrt 2}\hskip1cm(A=0,...,5)\ .\ee

 Analogously to
(\ref{identity}) we use the identity ($g$ any $2\times 2$ matrix):
\be \epsilon g^T\epsilon\,=\,g\,-\,Tr\,g \ ; \ee we obtain \be
\Delta_{AB}\,=\,\Delta_A\delta_{AB} \ee where
  $\Delta_{A}$ is defined as follows:
\be \Delta_A=\left(
-\,\Delta,\,+\Delta,\,+\Delta,\,+\Delta,\,0,\,0\right)\
.\label{delta2sc}\ee Therefore the effective lagrangian for free
quarks in the 2SC model
 can be written as follows
 \be {\cal L}_0 \,+\,{\cal L}_\Delta=\sum_{\vec v}~\sum_{A=0}^5
\chi^{A\dagger}\begin{pmatrix}iV\cdot \partial & \Delta_A\\
\Delta_A &i\tilde V\cdot \de\end{pmatrix}\chi^A\ +\ (L\to R)\
.\label{2sccomplete}\ee From this equation one can immediately
obtain the free fermion propagator that  in momentum space is
still given by (\ref{propagatorcfl}),
 with the $\Delta_A$ given by (\ref{delta2sc}).
\section{NGB and their parameters}
In the CFL model the diquark condensate breaks $SU(3)_L\otimes
SU(3)_R\otimes SU(3)_c\otimes U(1)_B$ to $SU(3)_{c+L+R}\otimes
Z_2$.
 By the Goldstone theorem we have 8 Nambu-Goldstone Bosons +1
  related to the breaking of $U(1)_B$. We have described the
  effective theory for this case in par. \ref{cflngbeff}.
  On the other hand in the 2SC model there is only
  one would-be NGB associated to the axial color
   because the chiral group is unbroken and there is an unbroken
   subgroup that plays the role of $U(1)_B$.

 How to describe the NGB couplings to the fermions?
In order to derive an effective low energy lagrangian for the NGBs
and obtain these couplings
 we use the gradient expansion (for a clear exposition see e.g.
 \cite{eguchi}). First we
introduce as many  external fields as the NGBs, with the same
quantum numbers of the goldstones; subsequently we perform a
derivative expansion of the generating functional. This gives rise
to the effective action for the NGBs.

\subsection{Gradient expansion for the $U(1)$ NGB in the CFL model and in the 2SC
model\label{gradient}} To describe the procedure of the gradient
expansion we consider the effective lagrangian for the NGB
associated to the spontaneous breaking of the  $U(1)$ symmetry.
For pedagogical purposes  this calculation will be described in
detail.

To start with we consider the  CFL model. The method of gradient
or derivative expansion  proceeds as follows. Let us introduce the
external field $\sigma$ by the substitution
\be \chi=\begin{pmatrix}\psi_+\\
C\psi^*_-\end{pmatrix}\ \to \ \begin{pmatrix}U^\dag\psi_+\\
U C\psi^*_-\end{pmatrix}\ ,\label{chiexternal}\ee where \be
U=e^{i\sigma/f_\sigma}\ .\label{u1sigma}\ee The quark lagrangian
(\ref{cflcomplete}) becomes
  \be {\cal L}_0 \,+\,{\cal L}_\Delta\,  +\,{\cal L}_\sigma=
  \sum_{\vec v}~\sum_{A=1}^9
\chi^{A\dagger}\begin{pmatrix}iV\cdot \partial & \Delta_A\,U^2\\
\Delta_A \,U^{\dag\,2} &i\tilde V\cdot\partial\end{pmatrix}\chi^A\
+\ (L\to R)\ .\label{cflcompleteexternal0}\ee At the lowest order
in the field $\sigma$ we have \be {\cal L}_\sigma\approx
\sum_{\vec v}~\sum_{A=1}^9 \chi^{A\dagger}
\Delta_A\begin{pmatrix}0 &\dd \frac{2i\sigma}{f_\sigma}\,-\,
\frac{2\sigma^2}{f^2_\sigma}\\ \dd -\frac{2i\sigma}{f_\sigma}\,-\,
\frac{2\sigma^2}{f^2_\sigma} &\end{pmatrix}\chi^A\ +\ (L\to R)\
,\label{cflcompleteexternal}\ee which contains the couplings
$\sigma\chi\chi$ and  $\sigma\sigma\chi\chi$. Since $\sigma$ is an
external field there is no associated kinetic term. However such
term is generated by quantum corrections. To show this we consider
the generating functional depending on the external fermion
 sources $\eta$ and $\eta^\dag$ (we consider only left-handed fields for simplicity):
 \be
 W[\eta,\,\eta^\dag]\,=\,\int[d\chi][d\chi^\dag][d\sigma]
 \exp\left\{i\sum_{\vec  v}\int\chi^\dag
 A\chi+\eta^\dag\chi+\chi^\dag\eta\right\}
\ee where we have introduced \be
A=S^{-1}+\frac{2i\sigma\Delta}{f_\sigma}\,\Gamma_0\,-\,
\frac{2\sigma^2\Delta}{f^2_\sigma}\,{\Gamma_1} \ee and \be
\Gamma_0=\begin{pmatrix}0 &+1\\ -1 &0\end{pmatrix} \ ,\hskip1cm
\Gamma_1=\begin{pmatrix}0 &1\\ 1 &0\end{pmatrix} \ .\ee Clearly we
have \bea
&&W[\eta,\,\eta^\dag]=\int[d\chi][d\chi^\dag][d\sigma]\times\cr&&\cr&\times&
\exp\left\{i\sum_{\vec  v}\int\left(\chi^\dag+\eta^\dag
A^{-1}\right) A
 \left(\chi+ A^{-1}\eta\right)-\eta^\dag A^{-1}\eta
 \right\}\ .
\eea After a change of variables in the functional integration we
obtain \bea
 &&W[\eta,\,\eta^\dag]\,=\,\int[d\chi][d\chi^\dag][d\sigma]
 \exp\left\{i\sum_{\vec  v}\int\chi^\dag A\chi-\eta^\dag A^{-1}\eta
 \right\}\cr&&\cr&=&
\int[d\sigma] \exp\left\{-i\sum_{\vec  v}\int\eta^\dag
A^{-1}\eta\right\} \,
\int[d\chi][d\chi^\dag]\,\exp\left\{i\sum_{\vec  v}\int \chi^\dag
A\chi
 \right\}\cr&=&\,\int[d\sigma]
  \,\exp\left\{-i\sum_{\vec  v}\int\eta^\dag A^{-1}\eta\right\}
 \exp\left\{Tr\ln A\right\}\cr&&\cr&=&
\int[d\sigma]
  \,\exp\left\{i\sum_{\vec  v}\int\left[-iTr\ln A- \eta^\dag A^{-1}\eta\right]\right\}
\ .\eea Now we have \bea &&-\,i\,Tr\ln A\,=\,-i\,Tr\ln
S^{-1}\left(1\,+\, S\,\frac{2i\sigma\Delta}{f_\sigma}\Gamma_0
\,+\,S\,\frac{(-2)\sigma^2\Delta}{f^2_\sigma}{\Gamma_1}\right)\cr&&=
-i\,Tr\ln S^{-1}+i\sum_{n=1}^\infty\frac{(-1)^{2n}}{n}
\left(iS\,\frac{2i\sigma\Delta}{f_\sigma}i\Gamma_0
\,+\,iS\frac{(-2)\,\sigma^2\Delta}{f^2_\sigma}{i\Gamma_1}
\right)^n\ .\nonumber \eea This is a loop expansion. At the lowest
order  it produces the effective action \bea {\cal
S}_{eff}&=&+\frac i 2 \,Tr\int dx dy\ \left[
\frac{i\,S(y,x)2i\sigma(x)\Delta}{f_\sigma}\,i\,\Gamma_0\,
\frac{i\,S(x,y)2i\sigma(y)\Delta}{f_\sigma}\,i\,\Gamma_0\right]\cr&&\cr
&+&\,i\,Tr\,\int dx \,\left[
\frac{i\,S(x,x)\,(-2)\,\Delta\,\sigma^2(x)}{f^2_\sigma}\,i\,\Gamma_1\right]\
.\eea
\begin{figure}[ht]
\epsfxsize=8truecm \centerline{\epsffile{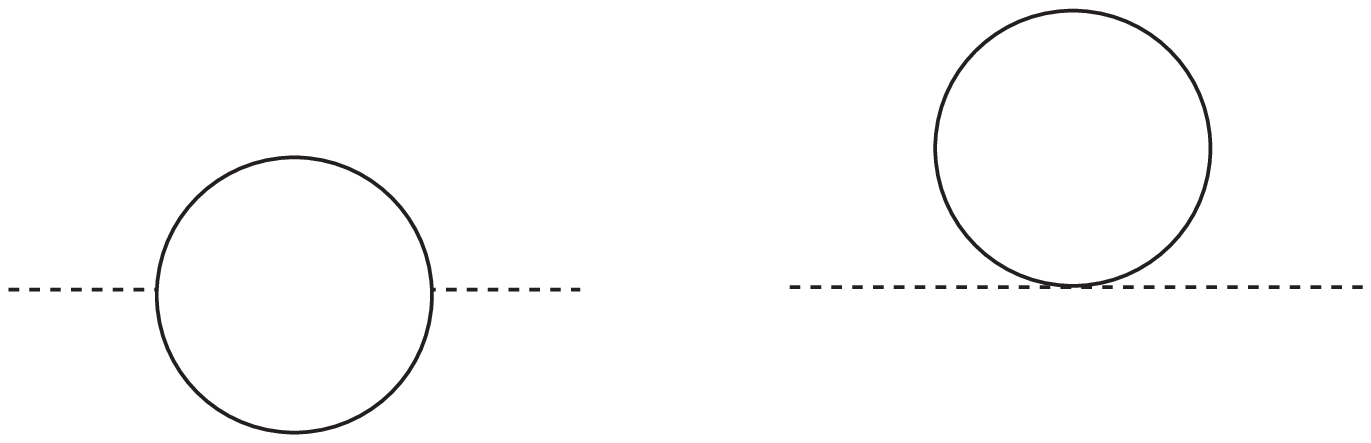}} \noindent
Fig. 2.1  One-loop diagrams. External lines represent the currents
$J_\mu^a,~J_\nu^b$. Full lines are fermion propagators.
\end{figure}
The two terms correspond to the diagrams in Fig. 2.1, i.e. the
self-energy, Fig. 2.1a, and the tadpole, Fig. 2.1b. They can be
computed by a set of Feynman rules to provide \be \nonumber i
{\cal L}_{eff}\ .\nonumber\ee In momentum space the Feynman rules
are as follows:
\begin{enumerate}
  \item For each fermionic internal line with momentum $p$ the propagator
  \be iS_{AB}(p)\,=\,i\delta_{AB}S(p)\,=
  \,\frac{i\,\delta_{AB}}{V\cdot p\,\tilde V\cdot
p-\Delta_A^2+i\epsilon}\begin{pmatrix}\tilde V\cdot p &
-\Delta_A\cr-\Delta_A & V\cdot p\end{pmatrix}\label{propagator}\
;\ee
  \item For each vertex  a term $i{\cal L}_{int}$ that
  can be derived from the effective lagrangian;
for example the $\sigma$ couplings to quarks can be derived from
(\ref{cflcompleteexternal});
\item For each internal momentum not constrained by the momentum conservation the factor
 \be \frac{4\pi\mu^2}{(2\pi)^4}\int d^2\ell\,=\,\frac{\mu^2}{4\pi^3}\int_{-\delta}^{+\delta}
 d\ell_\parallel\,
 \int_{-\infty}^{+\infty}d\ell_0\ ;\ee
  \item A factor of 2 for each fermion loop to take into account the spin ($L+R$);
  \item A statistical factor arising from the Wick theorem if needed.
\end{enumerate}The result of the calculation
of the effective lagrangian in momentum space is as follows: \bea
i\,{\cal L}_I&=&2\frac 1 2\frac{\mu^2}{4\pi^3}\sum_{\vec v,\, A,B}
\int d^2\ell\,\cr&\times& Tr \left[iS_{AB}(\ell+p)
\frac{2i\Delta_B\sigma}{f_\sigma} i\Gamma_0 iS_{BA}(\ell)
\frac{2i\Delta_A\sigma}{f_\sigma}i\Gamma_0\right]\ ,\cr&&\cr
i\,{\cal L}_{II}&=&2\,\frac{\mu^2}{4\pi^3}\sum_{\vec v,\, A,B}
\int d^2\ell\, Tr \left[iS_{AB}(\ell)
\frac{-2\,\Delta_B\sigma^2}{f^2_\sigma}\,i\,\Gamma_1 \right] \eea
 corresponding  to the two diagrams of
Figs. 2.1a and 2.1b respectively.  After some computation one has
\bea &&i\,{\cal L}_{eff}(p)\,=\,i\,{\cal L}_{I}(p)\,+ \,i\,{\cal
L}_{II}(p)\,=\,\sum_{\vec v,\,
A}\frac{\mu^2\Delta^2_A}{\pi^3f^2_\sigma} \cr&\times& \int
d^2\ell\,\Big[\frac{\tilde V\cdot(\ell+p)\sigma
V\cdot\ell\,\sigma+ V\cdot(\ell+p)\sigma \tilde
V\cdot\ell\sigma-2\Delta^2_A\sigma^2}{D_A(\ell+p)D_A(\ell)}
\cr&-&\,\frac{2\sigma^2}{D_A(\ell)} \Big] \eea where we have put
\be D_A(p)\,=\,V\cdot p\,\tilde V\cdot p-\Delta_A^2+i\epsilon\ .
\ee One can immediately notice that \be \,{\cal
L}_I(p=0)\,+\,{\cal L}_{II}(p=0)\,=\,0\ . \ee This result implies
that the scalar $\sigma$ particle has no mass, in agreement with
Goldstone's theorem.
 To get the effective lagrangian in the CFL model at the lowest order in the $\sigma$ momentum
 we expand the function in $p$ ($|p|\ll|\Delta$) to get, in momentum space \be
i\,{\cal L}_{eff}(p)\,=\,\sum_{\vec v,\,
A}\,\frac{2\mu^2\Delta^4_A}{\pi^3f^2_\sigma}(V\cdot
p)\sigma(\tilde V\cdot p )\sigma\ I_2\ ,\ee where we have defined
\be I_2=\int\,\frac{ d^2\ell}{D_A^3(\ell)} \ee (for other
integrals of this form see the appendix at the end of this
section). Since we have \be I_2=\,\frac{-i\,\pi}{2\Delta_A^4}\
,\ee one gets, in  the configuration space, \be {\cal
L}_{eff}(x)\,=\,\frac{9\mu^2}{\pi^2f^2_\sigma}\sum_{\vec
v}(V\cdot\partial \sigma)(\tilde V\cdot\partial\sigma)\ .\ee Since
\begin{equation*}
\sum_{\vec v}V^\mu \tilde V^\nu=\frac 1 2 \int\frac{d\vec v}{4\pi}
V^\mu \tilde V^\nu=\frac 1 2
  \begin{pmatrix}
    1 & 0&0&0 \\ 0 & -\frac 1 3 &0&0 \\ 0 & 0&-\frac 1 3 &0
    \\ 0 & 0&0&-\frac 1 3   \end{pmatrix}_{\mu\nu}\ ,
\end{equation*}we obtain
 \be
{\cal L}_{eff}(x)\,=\,\frac{9\mu^2}{\pi^2f^2_\sigma}\frac 1
2\left((\partial_0\sigma)^2 \,-\,v^2_\sigma(\vec\nabla\sigma)^2
\right) \ ,\ee with \be v^2_\sigma\,=\,\frac 1 3 \
.\label{velocity}\ee This kinetic lagrangian has the canonical
 normalization factor provided
 \be f_\sigma^2=\frac{9\mu^2}{\pi^2}\hskip 1cm(\rm CFL)\ ,\label{fsigma}\ee
 which fixes the
 $\sigma$ constant.
Therefore the effective lagrangian for the NGB $\sigma$ particle
is:
 \be
{\cal L}_{eff}=\frac{1} 2 \left((\partial_0\sigma)^2 \,-\,\frac 1
3(\vec\nabla\sigma)^2 \right) \ ,\label{efflagsigma}\ee
 We note that
the value of the velocity (\ref{velocity}) is a consequence of the
average over the Fermi velocities and reflects the number of the
space dimensions, i.e. 3. Therefore it is universal and we expect
  the same value in all the calculations of this type.

In the case of the 2SC model the formalism applies with obvious
changes. In particular again we introduce the external field
$\sigma$ as in (\ref{u1sigma}); the effective quark-NGB
lagrangian is given by the analogue of (\ref{cflcomplete}) where
now the sum over $A$ runs from $A=0$ to $A=5$ and $\Delta_A$ is
given by (\ref{delta2sc}). The final result differs from the CFL
case only in the coefficient in front of (\ref{fsigma}) which
reflects the number of color-flavor gapped degrees of freedom,
 9 in the CFL case and 4 in the 2SC case; therefore one has
 \be f_\sigma^2=\frac{4\mu^2}{\pi^2}\hskip 1cm(\rm 2SC)\ ,\label{fsigma2sc}\ee whereas
 the result (\ref{velocity}) being universal holds also in this case. The NGB effective lagrangian
 is still given by
 (\ref{efflagsigma}). The NGB  boson is in this case only a
 would-be NGB because the axial color is explicitly broken, though
 this breaking is expected to be small at high densities.
    \subsection{Gradient expansion for the $SU(3)$ NGBs\label{ngbcfl}}
The gradient expansion for the $SU(3)$ NGBs is  more complicated
 because of the color-flavor
structure of these fields. However, the procedure is conceptually
identical to the case of $U(1)$: the NGBs are introduced as
external fields and acquire a
 kinetic term, thus becoming dynamical fields, by integrating
 out the fermion fields.
Let us  outline this calculation in the present section.

From the invariant coupling \be -\frac{\Delta} 2
\sum_{I=1,3}\psi^T C\epsilon_I\psi\epsilon_I~\ee
 the coupling of the left-handed Weyl spinors
$\psi$'s to the octet of external scalar fields is:
 \bes
-\frac{\Delta}2\sum_{I=1,3}Tr[(\psi X^\dagger)^T C\epsilon_I(\psi
X^\dagger)\epsilon_I]\ ,\ees with an analogous relation (with
matrix $Y$) for the right handed fields.
 Both $X$ and $Y$ have
 v.e.v. given by   \bes\langle
X\rangle=\langle Y\rangle=1\ees and shall use the gauge \be
X=Y^\dag\ . \ee  The lagrangian giving the the coupling of the
quarks to the external fields can be obtained from
(\ref{cflcomplete0}) and is given by \bea {\cal L}_D&=&\sum_{\vec
v}\sum_{A,B=1}^9 \chi^{A\dagger}\begin{pmatrix}i\,V\cdot
\partial\delta_{AB} & \Xi_{BA}^*
\\
\Xi_{AB} &i \,\tilde V\cdot
\partial\delta_{AB}\end{pmatrix}\chi^B\cr&&\cr &+& (L\to R)
\label{cflcomplete00}
 \ ,\eea
 where
\bea \Xi_{AB}
&=&\,\Delta\,Tr\left[\epsilon_I\left(T_AX^\dag\right)^T\epsilon_I
\left(T_BX^\dag\right)\right]\,=\cr&=&\,\Delta\left(\,Tr[T_AX^\dag
T_BX^\dag]\,-\,Tr [T_AX^\dag]\,Tr[T_BX^\dag]\right)\, . \eea One
can now expand the NGB fields \bes  X=
 \exp{\displaystyle{ i\left(
\frac{\lambda_a\Pi^a}{2F}\right)}},~~~~~~a=1,\cdots,8~,\ees and
obtain  the 3-point $\chi\chi\Pi$, and the 4-point
$\chi\chi\Pi\Pi$  couplings. The result is, for the 3-point
coupling  \bea {\cal L}_{\chi\chi\Pi}& = &\sum_{\vec v}
\frac{-\,i\,\Delta}{F}
 \Big\{\sum_{a=1}^8\frac{\Pi^a}{\sqrt 6}
\Big[\chi^{9\dagger}\Gamma_0\,\chi^a + \chi^{a\dagger}
\Gamma_0\,\chi^9 \Big] - \cr && - \sum_{a,b,c=1}^8
d_{abc}\chi^{a\dagger} \Gamma_0\chi^b\Pi^c\Big\}
~.\label{trilin}\eea On the the other hand the 4-point coupling (2
Goldstones +
 2 fermions) \bea
{\cal L}_{\chi\chi\Pi\Pi}&=&\sum_{\vec v}
 \Big\{~\frac{4}{3} \sum_{a=1}^8 \frac{\Delta}{8F^2}
\chi^{9\dagger}\Gamma_1\, \chi^9~\Pi^a\Pi^a~ \nonumber
\\&& + 3~\sqrt{\frac{2}{3}}
\sum_{a,b,c=1}^8\left(\frac{\Delta}{8F^2} d_{abc}
\chi^{c\dagger}\Gamma_1\,\chi^9~\Pi^a\Pi^b + {\rm
h.c.}\right)~\nonumber
\\ &&+~ \sum_{a,b,c,d=1}^8
\frac{\Delta}{8F^2}h_{abcd}\chi^{c\dagger} \Gamma_0\,
\chi^d\Pi^a\Pi^b
 \Big\}~,\label{quadrilin}\eea
 \bes h_{abcd}=2 \sum_{p=1}^8
\left(g_{cap}g_{dbp}+d_{cdp}d_{abp}\right) \, -\frac 8 3
\delta_{ac}\delta_{db} +\frac 4 3 \delta_{cd}\delta_{ab}~.\ees

Using the Feynman rules given above and the interaction
lagrangians (\ref{trilin}) and (\ref{quadrilin} we get the
effective lagrangian as follows:
 \be  {\cal
L}_{eff}^{kin}=  \frac {\mu^2(21-8\ln 2)}{36\pi^2F^2}\,\frac
1{2}\sum_{a=1}^8 \left(\dot\Pi^a\dot\Pi^a-\frac 1
3|\vec\nabla\Pi_a|^2\right) \ .\ee The canonical normalization for
the kinetic term is obtained provided \be F^2=\frac {\mu^2(21-8\ln
2)}{36\pi^2}\ . \ee The NGB  velocity is again $1/\sqrt{3}$
because of the
 universality noted above.

\section{Gluon masses and dispersion laws}

\subsection{Evaluating the bare gluon mass\label{evaluating}}
 Let us go back to Eq. (\ref{l2}).  ${\cal
L}_{2}$ is a non local lagrangian containing an infinite number of
local terms;
 let us discuss in the formalism of the two component fermion field
 $\chi$ introduced in eq. (\ref{chi}):
\bes \chi=\begin{pmatrix}\psi_+\\
C\psi^*_-\end{pmatrix}\label{chi2}\ .\ees Let us work
 in the quark basis $\chi\equiv \chi_{\alpha j}$,
($\alpha=$color, $j=$flavor); ${\cal L}_{2}$  can be written as
follows:
 \be {\cal L}_{2}= - \sum_{\vec v}~
\chi^{\dagger}\begin{pmatrix} \{2\mu+i\tilde V\cdot D \}^{-1}&
0\cr 0 &\{2\mu+iV\cdot D\}^{-1} \end{pmatrix}P^{\mu\nu} D_\mu
D_\nu\chi\ .\label{4.4}\ee Let us consider this coupling in the
momentum space where \be \{2\mu+i\tilde V\cdot D \}^{-1}P^{\mu\nu}
D_\mu D_\nu =\frac{(-i\ell_\mu+g
A_\mu^c\lambda^c/2)(-i\ell_\nu+gA_\nu^d\lambda^d/2)}{2\mu+i\tilde
V\cdot(-i\ell+g A^e\lambda^e/2)} P^{\mu\nu} \ .\ee
 Since the effective
theory is bidimensional we can put $\ell^\mu_\perp=\ell_\rho
P^{\rho\mu}=0$. The simplest coupling generated by this
interaction term is a quadrilinear coupling (two quarks, two
gluons); it
 is obtained by
putting to zero the coupling to the gluon in the denominator and
one gets: \be {\cal L}_{\chi\chi AA}= - g^2\sum_{\vec v}~
\chi^{\dagger}A_\mu^c
A_\nu^d\frac{\lambda^c}{2}\frac{\lambda^d}{2}
\begin{pmatrix}\frac 1
{2\mu+\tilde V\cdot \ell }& 0\cr 0 &\frac 1 {2\mu+V\cdot \ell}
\end{pmatrix} P^{\mu\nu}\chi\ .\label{4.5} \ee

Let us now consider the CFL model. Introducing the quark  basis
$\chi^A\ (A=1,...,9)$, we get the coupling in the following
form:\be {\cal L}_{\chi\chi AA}= - g^2\sum_{\vec v}~
\chi_A^{\dagger}A_\mu^c A_\nu^d  \begin{pmatrix}\frac 1
{2\mu+\tilde V\cdot \ell }& 0\cr 0 &\frac 1 {2\mu+V\cdot \ell}
\end{pmatrix}P^{\mu\nu}\chi_B\xi^{AB}_{cd}\ ,\label{4.6} \ee where
the Clebsch Gordan coefficient is given by \be \xi^{AB}_{cd}=\frac
1 8 tr(\lambda^A\lambda^c\lambda^d\lambda^D)\ . \ee ${\cal
L}_{\chi\chi AA}$ contributes to the gluon self-energy through a
tadpole graph analogous to that in Fig. 2.1b (the external lines
should be interpreted now as gluon lines). It is important to
stress that, due to the denominator in (\ref{4.6}), the loop
integral takes relevant contribution from the region where
$\ell_0\pm\ell_\parallel\approx -2\mu$. This means that  one
cannot integrate only near the Fermi surface, but the longitudinal
integration must be extended up to $2\mu$ or, in terms of
$\ell_\parallel=p-\mu$,
 from $-\mu$ to $+\mu$.

The tadpole contribution to the gluon self energy can be computed
by using the set of Feynman rules discussed above and by
(\ref{4.6}). One gets \beas
&&\Pi^{\mu\nu}_{cd}=-i~4\,g^2\sum_{\vec v}
P^{\mu\nu}\int_{-\mu}^{+\mu}\frac{d\ell_\parallel}{2\pi}(\mu+\ell_\parallel)^2
\int_{-\infty}^{+\infty}\frac{d\ell_0}{2\pi}\times\cr&\times&\left(
\sum_{A=1}^8\xi^{AA}_{cd}\frac{1}{8D(\ell)}
+\frac{\delta_{cd}}{6D^\prime(\ell)}\right)\cdot\left(\frac{\tilde
V\cdot\ell }{2\mu+\tilde V\cdot\ell+i\epsilon}+\tilde V\to
V\right) \eeas where
$D(\ell)=\ell_0^2-\ell^2_\parallel-\Delta^2+i\epsilon$ and
$D^\prime(\ell)=\ell_0^2-\ell^2_\parallel-\Delta^{\prime\,2}+i\epsilon$.
The factor 4=$2\times 2$ originates from a statistical factor and
a spin factor. Since \be\sum_{A=1}^8\xi^{AA}_{cd}=\frac 4 3
\delta_{cd}\ee and \be \sum_{\vec v}P^{\mu\nu}=-\frac 1 3
\delta^{\mu\nu} \delta^{\mu i} \delta^{\nu j} \ee we obtain \bea
\Pi^{00}_{cd}&=&0\ ,\hskip1cm \Pi^{0j}_{cd}\ =\ 0\ ,\label{debs}
\\&&\cr
\Pi^{ij}_{cd}&=&-\omega_p^2\delta^{ij} \delta_{cd}\hskip 2cm {\rm
(CFL)}\eea where $\omega_p=g\mu/{\sqrt 2}\pi$ is the plasma
frequency and we have used the result \be
\int_{-\mu}^{+\mu}\frac{d\ell_\parallel}{2\pi}(\mu+\ell_\parallel)^2
\int_{-\infty}^{+\infty}\frac{d\ell_0}{2\pi}\frac{\tilde
V\cdot\ell }{(2\mu+\tilde
V\cdot\ell+i\epsilon)(\ell_0^2-\ell^2_\parallel-\Delta^2+i\epsilon)}=
\frac{i\mu^2}{8\pi}\ .\ee

For the 2SC case the calculation is similar; one obtains the
following results \footnote{We  note that in this case the tadpole
diagram presents infrared divergences (in $\ell_0$).
 To control these divergences one
considers the system in a heat bath and  substitutes the energy
euclidean integration
 $\dd \ell_4= - i \ell_0$ with a sum over the Matsubara frequencies
$\dd \ell_4 \rightarrow \omega_n = 2 \pi \left(n+\frac 1 2
\right)T$;
 eventually  one
performs the limit $T\  \rightarrow 0$. For details see
 \cite{2fla} and the appendix.}
($c,d=1,\cdots,8$): \bea \Pi^{00}_{cd}&=&0\ ,\hskip1cm
\Pi^{0j}_{cd}\ =\ 0\ ,\label{deb2sc}
\\&&\cr
\Pi^{ij}_{cd}&=&-m^2_g\delta^{ij} \delta_{cd}\hskip 2cm {\rm
(2SC)}\eea where $m_g=g\mu/{\sqrt 3}\pi$. Note that the tadpole
graphs only contributes to the Meissner mass.
\subsection{Debye and Meissner mass and gluon approximate dispersion
laws: CFL case\label{dbcfl}} For both the CFL and the 2SC model
the calculation of Section \ref{evaluating} provides a mass term
to the gluons;
 it is a chromo-magnetic mass term (Meissner mass)
 since it gives mass to $\vec A$; there is however a second
 diagram contributing to the
 Meissner mass, i.e. the graph of Fig. 2.1a,
 with external line interpreted as a gluon.
 As we shall see in this section it
 contributes also to the chromo-electric (Debye) mass and to
 the dispersion law, i.e.
\be E=E(p)\ ;\ee masses and dispersion laws can be derived by the
polarization tensor\be \Pi^{\mu\nu}(p)\ .\ee For example
\bea\Pi^{00}(0)&=& m^2_D\cr \Pi^{ij}(0)&=&-\delta^{ij}m^2_M\
.\label{masses}\eea To compute $\Pi^{\mu\nu}$ from Fig. 2.1a we
need ${\cal L}_1$ i.e. the interaction term between the quarks and
one gluon. The fermionic current could be extracted from eq.
(\ref{l1}). However it is more convenient to work in the quark
basis $\chi^A$. We use eq. (\ref{cflcomplete0}) to write, in the
CFL case \be {\cal L}_1\,=\,i\,g\,A^\mu_aJ_\mu^a\ , \ee where \be
J_\mu^a\,=\, \sum_{\vec v}\sum_{A,B=1}^9
\chi^{A\dagger}\begin{pmatrix}i\,V_\mu h_{AaB}&
0\\
0 &-i\,\tilde V_\mu h^*_{AaB}\end{pmatrix}\chi^B+(L\to R)
\label{cflcomplete0new} \ee where \be h_{AaB} =\frac 1 2
\,Tr[T_A\lambda_a T_B]\ .\label{cflcomplete0newnew} \ee We obtain
 \bea
J_\mu^a&=&\frac i 2 \sqrt{\frac 2 3} \sum_{\vec v}
%\sum_{a=1}^8
\left(\chi^{9\dagger}\begin{pmatrix}V_\mu & 0\cr 0 &-\tilde
V_\mu\end{pmatrix}\chi^a+ {\rm h.c.}\right)\ +\cr&&\cr
&+&\frac{i}{2} \sum_{\vec v} \sum_{b,c=1}^8\chi^{b\dagger}
\begin{pmatrix}V_\mu g_{bac} &
0\\ 0 &-\tilde V_\mu g_{bac}^*\end{pmatrix}\chi^c\eea where \be
g_{abc}=d_{abc}+if_{abc}\ .\label{current}\ee and $d_{abc}$,
$f_{abc}$ are the usual $SU(3)$ symbols. The result of the self
energy diagram (Fig. 2.1a, with external lines representing
fermionic currents) can be written as follows: \bea
&&i\,\Pi^{\mu\nu}_{ab}(p)=-2\sum_{\vec
v}\sum_{A,C,D,E}\left(\frac{-ig}{2}\right)^2
\frac{4\pi\mu^2}{(2\pi)^4}\int d^2\ell\,
Tr\Big[iS_{CD}(\ell+p)\times\cr&&\cr&\times& \begin{pmatrix}V_\nu
h_{DbE}&
0\\
0 &-\tilde V_\nu h^*_{DbE}\end{pmatrix}
iS_{EA}(\ell)\begin{pmatrix}V_\mu h_{AaC}&
0\\
0 &-\tilde V_\mu h^*_{AaC}\end{pmatrix}\Big] \eea where the
propagator is given by eq. (\ref{propagator}). We note the minus
sign on the r.h.s of (\ref{pimunu}), due to the presence of a
fermion loop and the factor 2 due to the spin ($L+R$).  To this
result one should add the contribution arising from the tadpole
diagram of Fig. 2.1b (with, also in this case,  external lines
representing fermionic  currents $J_\mu^a,~J_\nu^b$); this diagram
was computed in \ref{evaluating}:  \be \Pi^{\mu\nu}_{ab}\Big
|_{1b}~=~-~\delta_{ab}\delta^{j k}\delta^{\mu j}\delta^{\nu
k}\omega_p^2\ . \label{1b}\ee
 To derive the dispersion law for the gluons,
we write the equations of motion for the gluon field $A^b_\mu$ in
momentum space and high-density limit:

\be \left[\left(-p^2 g^{\nu\mu}+p^\nu p^\mu
 \right)\delta_{ab}+\Pi^{\nu\mu}_{ab}\right]A^b_\mu~=0\ .\label{105}\ee
  We define the invariant quantities $\dd \Pi_0,
\Pi_1, \Pi_2$ and $\Pi_3$ by means of the following equations, \be
\Pi^{\mu \nu}(p_0,\,\vec p) = \left\{ \begin{array}{ll}
    \Pi^{00}=\Pi_0(p_0,\,\vec p)  \\
    \Pi^{0i}=\Pi^{i0}=\Pi_1(p_0,\,\vec p) \, n^i \\
    \Pi^{ij}=\Pi_2 (p_0,\,\vec p) \,\delta^{ij}\, +\, \Pi_3(p_0,\,\vec p)\,
n^i n^j
    \end{array}
    \right.
\label{dec} \ee with $\dd \vec n = \frac{\vec p}{p}$ . It is clear
that in deriving the dispersion laws we cannot go beyond momenta
at most of the order $\Delta$, as the Fermi velocity
superselection rule excludes gluon exchanges with very high
momentum; it is therefore an approximation, but nevertheless a
useful one, as in most cases hard gluon exchanges are strongly
suppressed by the asymptotic freedom property of QCD.

Besides  the longitudinal and transverse gluon fields defined by
\bea A_{i\, L}^{a} &=& \left( \vec n \cdot \vec A^a \right) \, n_i
~, \cr A_{i\, T}^{a} &=& A_{i}^a-A_{i\, L}^{a}~, \eea it is useful
to consider also the scalar gluon field $A_0^{a}$. By the equation
\be p_{\nu} \Pi^{\nu \mu}_{a b} A^b_{\mu} = 0 \ ,\label{108} \ee
one obtains the relation \be \left( p_0 \,\Pi_0\, -\, |\vec p|\,
\Pi_1 \right) A_0 = \vec n \cdot \vec A
 \left( p_0\, \Pi_1
\,-\,|\vec p|\, (\Pi_2+\Pi_3) \right) \ , \label{7} \ee between
the scalar and the longitudinal component of the gluon fields. The
dispersion laws for the scalar, longitudinal and transverse gluons
are respectively \bea \left(\Pi_2\,+\,\Pi_3 \,+\,p_0^2\right)
\left(|\vec p|^2\,+\,\Pi_0 \right)&=&p_0|\vec p|
\left(2\,\Pi_1\,+p_0|\vec p|\right)\ ,\cr
\left(\Pi_2\,+\,\Pi_3\,+\,p_0^2\right) \left(|\vec
p|\,p_0\,+\,\Pi_0\right)&=&p_0|\vec p|
\left(2\,\Pi_1\,+\,p_0^2\right)+\Pi_1^2 \ ,\cr p_0^2-|\vec
p|^2+\Pi_2&=&0\ .\label{cflpig} \eea The analysis of these
equations is rather complicated \cite{shovkovy}. We give an
approximate evaluation of them
 by developing $\Pi^{\mu\nu}$ in powers of $p$.
By including in the previous formulas terms up to the second order
in momenta we get \cite{cflgatto}:
\begin{equation}
\Pi^{\mu\nu}_{ab}(p)\Big
|_{1a}=\Pi^{\mu\nu}_{ab}(0)+\delta\Pi^{\mu\nu}_{ab}(p)
\label{pimunu}
\end{equation} where \be \Pi^{\mu\nu}_{ab}(0)=\frac{\mu^2
g^2}{2\pi}\delta_{ab} \sum_{\vec
v}\Sigma^{0,\mu\nu}~,\label{3.103}\ee
 \be
\delta\Pi^{\mu\nu}_{ab}(p)=\frac{\mu^2 g^2}{2\pi}\delta_{ab}
\sum_{\vec v} \Sigma^{\mu\nu}~,\ee and
\be\Sigma^{0,\mu\nu}=k_1(V^\mu V^\nu)+k_2(V^\mu\tilde V^\nu)\ +\
(V\to\tilde V)~,\ee \be\Sigma^{\mu\nu}=aV^\mu V^\nu\frac{(\tilde
V\cdot p)^2}{\Delta^2}\,+\,bV^\mu\tilde V^\nu \frac{V\cdot
p\,\tilde V\cdot p}{\Delta^2}\ +\ (V\to\tilde V)\ .\ee The
coefficients are given by \bea a&=&\frac{1}{108\pi}
\left(31-\frac{32}{3}\ln 2 \right)
\\
b&=&\frac{10 }{108\pi} \left(1-\frac{8}{3}\ln 2 \right)
\\
k_1&=&\frac{3 }{2\pi}
\\ k_2&=&-\,  \frac{1}{9\pi}
\left(3+4\ln 2 \right)~.
 \eea

Since one has, including the one-loop ${\cal O}(g^2)$ correction
in the gluonic lagrangian, \be {\cal L}_G\,=\,-\,\frac 1 4\, G^2
\to \,-\,\frac 1 4\, G^2\,+\,\frac 1 2\, A_{a\nu}\Pi^{\mu\nu}_{ab}
A_{b\mu}\ ,\ee one  can compute the one-loop contributions to the
Debye and Meissner mass from $\Pi^{00}_{ab}(0)$ and
$\Pi^{ij}_{ab}(0)$ respectively, according to (\ref{masses}); to
the results of (\ref{3.103}) for the diagram  of Fig. 2.1a, one
must add the contribution (\ref{1b}); the two masses
 are identical for all the eight gluons in the CFL model and  are given by:
\be m^2_D\,=\,\frac{\mu^2 g^2}{36\pi^2} \left(21-8\ln 2
\right)~=\,g^2F^2\ ,\ee and \be m^2_M\,=\,\frac{\mu^2 g^2}{\pi^2}
\left(-\frac{11}{36}-\frac{2}{27} \ln 2+\frac{1}{2} \right)
\,=\,\frac{m^2_D}{3}~,
 \ee
where the first two terms are the result of the diagram of
Fig.2.1a and the last one  is the result of the diagram of Fig.
2.1b (this contribution is called the bare Meissner mass in
\cite{sonstephanov}). These results have been obtained by the
method of the effective lagrangian in \cite{cflgatto} and agree
with the findings of other authors \cite{sonstephanov},
\cite{zarembo}. In particular $m_M^2 = v^2 m_D^2$, where $v$, as
for  the NGB velocity, is equal to $1/\sqrt{3}$.

The Debye and Meissner masses do not exhaust  the analysis of the
dispersion laws for the gluons in the medium. In particular we
wish to show the existence of a light excitation for small momenta
in the CFL phase. To obtain the dispersion laws for the gluons in
the small momenta limit we proceed as follows. From (\ref{7}) one
obtains, for small momenta: \be p^0A_0\approx\frac{1}{3}(\vec p
\cdot\vec A)\ .\label{eqmot2}
 \ee
Substituting this result in ({\ref{108}) taken for $\nu=0$ one
gets the dispersion law for the time-like gluons $A^b_0$: \be
p^0=\pm \,E_{A_0},~~~E_{A_0}=\frac{ 1}{\sqrt 3}\sqrt{|\vec
p\,|^2+\frac{m^2_D}{1+\alpha_1}}~,\ee with \be
\alpha_1=\frac{\mu^2 g_s^2}{6\Delta^2\pi}(a-b)=\frac{\mu^2
g_s^2}{216\Delta^2\pi^2}\left(7+\frac{16}{3}\ln 2\right)~.\ee The
rest mass of $A_0^b$ is therefore  given by \be
m^R_{A_0}=\frac{m_D}{\sqrt{3(1+\alpha_1)}}~.\ee Since for large
values of $\mu$ one has $g^2\mu^2\gg\Delta^2$ it follows that \be
m^R_{A_0}\approx\frac{m_D}{\sqrt{3\alpha_1}}=
\sqrt{6\,\frac{21-8\ln 2}{21+16 \ln2}}\,\Delta\approx
1.70\,\Delta~. \label{127}\ee We also define the so-called
effective mass (to be denoted $m^*$ in the sequel) by \be \vec
v=\frac{\partial E}{\partial\vec p}=\frac{\vec p}{m^*(p)}~\ee in
the limit  $\vec p\to 0$. We get, if $m^*=m^*(0)$, \be m^*_{A_0}=
\frac{\sqrt{3}\, m_D}{\sqrt{1+{\alpha_1}}}=3\,m^R_{A_0}\approx
5.10\, \Delta~.\ee

 Let us now turn to the dispersion law for the longitudinal and transverse
 gluons; they are obtained by considering (\ref{108}) for $\nu=i$
  and use again (\ref{eqmot2}). One gets the dispersion laws for the
results for the longitudinal and transverse cases as follows:
 \bea E_{A_L}&=&\frac{
1}{\sqrt 3}\sqrt{|\vec
p\,|^2\frac{1+\alpha_2}{1+\alpha_1}+\frac{m^2_D}{1+\alpha_1}}~\cr&&\cr
 E_{A_T}&=&\sqrt{|\vec
p\,|^2\frac{1+\alpha_3}{1+\alpha_1}
+\frac{m^2_D}{3(1+\alpha_1)}}~, \eea with
 \bea \alpha_2=\frac{\mu^2
g_s^2}{30 \Delta^2\pi^2}(a-9b)=-~\frac{\mu^2 g_s^2}{3240
\Delta^2\pi^2}\left(59-\frac{688}{3}\ln 2\right)~\cr
\alpha_3=-\frac{\mu^2 g_s^2}{30 \Delta^2\pi^2}(a+b)=-\frac{\mu^2
g_s^2}{3240\Delta^2\pi^2}\left(41-\frac{112}{3}\ln 2\right)~. \eea
We get easily the following results \be
m^R_{A_L}=m^R_{A_T}=m^R_{A_0}=\frac{m_D}{\sqrt{3(1+\alpha_1)}}\equiv
m_R ~.\label{132}\ee  The results (\ref{127}) and (\ref{132}) show
that the rest mass is of of the order of $\Delta$ and one could
therefore wonder if these results are significant, since they have
been obtained in the limit $|p/\Delta|\ll 1$. To estimate the
validity of this approximation we use the exact result, which can
be obtained by equations (\ref{cflpig}). Since $\Pi_3(p_0,0)=0$,
the rest mass of the three species $A_0,\,A_L,\, A_T$ is given by
\be m^2+\Pi_2(m^2,0)=0\ ,\label{equa}\ee with \bea \Pi_2(m^2,0)&
=&\frac{\mu^2g^2}{2\pi^2} \Big[-1+\int_0^{+\infty} dx \,\Big(\,
\frac{1}{e(x)[4e(x)^2-m^2/\Delta^2]} \cr&+&\frac{e(x)+\tilde
e(x)}{9e(x)\tilde e(x)}\frac{2-x^2+e(x)\tilde e(x)}{(e(x)+\tilde
e(x))^2-m^2/\Delta^2}\Big)
  \Big]\ ,
\eea with \bea e(x)&=&\sqrt{1+x^2}\cr \tilde e(x)&=&\sqrt{4+x^2}\
.\eea The numerical result of (\ref{equa}) is \be m\equiv
m_R=\,1.36\, \, \Delta \ . \ee A comparison with (\ref{127}) shows
that the relative error between the two procedures is of the order
of $18\%$ and this is also the estimated difference for the
dispersion law  at small $\vec p$. We notice that the gradient
expansion approximation tends to overestimate the correct result
\cite{shovkovy}.
\subsection{Debye and Meissner mass and gluon approximate dispersion
laws: 2SC case} The analysis of the dispersion laws for the glue
in the case of the 2SC model is complicated by the presence of an
unbroken color subgroup.

The fermionic current appearing in \be{\cal L}_1=i g A_\mu^a
J_a^\mu\ee can be derived from (\ref{2sccomplete0}). One finds \be
J_\mu^a\,=\, \sum_{\vec v}\sum_{A,B=0}^5
\chi^{A\dagger}\begin{pmatrix}i\,V_\mu K_{AaB}&
0\\
0 &-i\,\tilde V_\mu K^*_{AaB}\end{pmatrix}\chi^B+(L\to R)\ .
\label{2sccomplete0new} \ee where the coefficients $K _{AaB}$ are
given by:\be K_{AaB}\,=\,\frac 1
2\,Tr\{\tilde\lambda_A\lambda^a\tilde\lambda_B\}\ .\ee In this way
one can  compute the self energy contribution to
$\Pi^{\mu\nu}_{ab}$. These results are contained in \cite{2fla}
and we briefly summarize them.

We write \be \Pi^{\mu \nu}_{a b}(p)\ =\ \Pi^{\mu \nu}_{a b}(0) +
\delta \Pi^{\mu \nu}_{a b}(p),  \ee with \be \Pi^{\mu \nu}_{a
b}(0)\ =\ \frac{\mu^2 g^2}{ \pi^2} \sum_{\vec v} \Sigma^{0,\mu
\nu}_{ab}, \label{pi1} \ee and \be \delta \Pi^{\mu \nu}_{a b}(p)\
=\ \frac{\mu^2 g^2}{ \pi^2} \sum_{\vec v}
 \Sigma^{\mu \nu}_{ab}(p). \ee It is useful to distinguish gluons of different colours, i.e.
the gluons $A^a_\mu$ ($a=1,2,3$) of the unbroken color subgroup
$SU(2)_c$ and the gluons $A^a_\mu$ with $a=4,...7$ and $a=8$.

\subsubsection{Colors=1,2,3}

In this case one finds from the self energy diagram of Fig. 2.1
($a,b=1,2,3$):
 \bea \Pi^{\mu \nu}_{a b}(p)
&=& i \delta_{a b} \frac{g^2 \mu^2}{2 \pi^3} \sum_{\vec v}
 \int \! d^2 \ell\, \Big( \frac{ V^{\mu}V^{\nu} \tilde V\cdot
 \ell\,
 \tilde V \cdot(\ell+p)
 + (V\leftrightarrow\tilde V)}{D_1(\ell+p) D_1 (\ell)} \,+\cr&&\cr&+& \Delta^2
\frac{V^{\mu} \tilde V^{\nu} + V^{\nu} \tilde V^{\mu}}{D_1(\ell+p)
D_1 (\ell)} \,\Big) \ . \eea

As we noted in the previous sections  working out the CFL case the
extraction of the dispersion law in the general case is rather
difficult. We consider here only the small momentum limit. Working
in this limit one gets for the various components of the tensor
$\Pi^{\mu\nu}$: \be \Pi^{0 0}_{ab}(p)\ =\ \Pi^{0 0}_{ab}(0) +
\delta \Pi^{0 0}_{ab}(p) = \ \delta \Pi^{0 0}_{ab}(p) =\
\delta_{ab}\frac{ \mu^2 g^2}{18 \pi^2 \Delta^2} |\vec p\,|^2\, ,
\label{glue1} \ee \be \Pi^{k l}_{ab}(p)\ =\ \Pi^{k l}_{ab}(0) +
\delta \Pi^{k l}_{ab}(p) = \ \delta_{ab} \delta^{kl} \frac{ \mu^2
g^2}{3 \pi^2} \left( 1 + \frac{p_0^2}{6 \Delta^2} \right)\, ,
\label{glue2} \ee and \be \Pi^{0 k}_{ab}(p)\ =\ \delta \Pi^{0
k}_{ab}(p) =\ \delta_{ab} \frac{ \mu^2 g^2}{18 \pi^2 \Delta^2}
 p^0 p^k \,.
\label{glue3} \ee These results agree with the outcomes of
\cite{rischke1} and \cite{sonrishke}.

We note that there is no contribution to the Debye mass in this
case; moreover the contribution to the Meissner mass from this
term exactly cancels the contribution from the tadpole graph.
 This results reflects the fact that in the
2SC model the $SU(2)_c$ color subgroup generated by the first
three generators $T^c$ ($c=1,2,3)$ remains unbroken.

One can now compute the dispersion laws for the unbroken gluons.
From the previous formulae one gets: \be {\mathcal L}\ =\
-\frac{1}{4} F^{\mu \nu}_{a} F_{\mu \nu}^{a} + \frac{1}{2}
\Pi^{\mu \nu}_{ab} A_{ \mu}^a A_{ \nu}^b\ ~. \ee Introducing the
fields $\dd E_i^a \equiv F_{0i}^a$ and $\dd B_i^a \equiv i
\varepsilon_{ijk} F_{jk}^a$, and using (\ref{glue1}),
(\ref{glue2}) and (\ref{glue3}) these results can be written as
follows (we neglect for the time being the 3 gluons and 4 gluons
vertices to be discussed below): \be {\mathcal L}\ =\  \frac{1}{2}
(E_i^a E_i^a - B_i^a B_i^a) + \frac{k}{2} E_i^a E_i^a ~,
\label{33} \ee with \be k= \frac{g^2 \mu^2}{18 \pi^2 \Delta^2} ~.
\ee These results have been first obtained in \cite{sonrishke}. As
discussed in this paper, these results imply that the medium has a
very high {\it dielectric constant} $\dd \epsilon = k +1$ and a
{\it magnetic permeability} $\dd \lambda = 1$. The gluon speed in
this medium is now \be v=\frac{1}{\sqrt{\epsilon \lambda}} \propto
\frac{ \Delta}{g \mu} \ee and in the high density limit it tends
to zero. The one loop lagrangian (\ref{33}) assumes the gauge
invariant expression \be  {\mathcal L}\ =\ - \frac{1}{4} F^{\mu
\nu}_j F^{j}_{\mu \nu} ~~~~ (j=1,2,3) ~, \label{36} \ee provided
the following rescaling is used \bes A_0^j \rightarrow A_0^{j
\prime} =  k^{3/4} A_0^j ~, \label{A0} \ees \bes A_i^j \rightarrow
A_i^{j \prime} =  k^{1/4} A_i^j ~, \label{Ai} \ees \bes x_0
\rightarrow x_0^{\prime}  =  k^{-1/2} x_0 ~, \label{x0} \ees \bes
g \rightarrow g^{\prime}  =  k^{-1/4} g ~.
 \ees

\subsubsection{Colors=4-7}
For $a,b=4,...,7$ the polarization tensor  is \bea \Pi^{\mu
\nu}_{a b}(p)& = &i \delta_{a b} \frac{g^2 \mu^2}{4 \pi^3}
\sum_{\vec v}
 \int \! d^2 \ell \left( V^{\mu}V^{\nu} \tilde V \cdot\ell\, \tilde V
\cdot(\ell+p)
 + V\leftrightarrow\tilde V
\right)\times\cr&\times&
  \left( \frac{1}{D_1(\ell+p) D_2 (\ell)}
+ \frac{1}{D_2(\ell+p) D_1 (\ell) }\right) \ . \eea
 In the small momentum limit one gets
 in this case
 \be
\Pi^{0 0}_{ab}(p)\ =\ \Pi^{0 0}_{ab}(0) + \delta \Pi^{0
0}_{ab}(p)\ =\ \delta_{ab}\frac{\mu^2 g^2}{2 \pi^2} \left(1 +
\frac{p_0^2+ |\vec p\,|^2/3}{2 \Delta^2} \right), \label{pi004}
\ee
 \be \Pi^{0 i}_{ab}(p)\ =\ \delta \Pi^{0 i}_{ab}(p)\ =\
 \delta_{ab}
\frac{\mu^2 g^2}{6 \pi^2 \Delta^2} p^0 p^i ~, \ee and \be \Pi^{i
j}_{ab}(p)\ =\ \Pi^{i j}_{ab}(0) + \delta \Pi^{i j}_{ab}(p)\ =\
\delta_{ab}\frac{\mu^2 g^2}{6 \pi^2} \left( \delta^{ij}+
\frac{\delta^{ij}p_0^2}{2 \Delta^2}+ \frac{ \delta^{ij} \vec
p^2+2p^ip^j }{10 \Delta^2} \right) . \label{piij4} \ee We can
observe a contribution to both the Debye and Meissner mass in this
case.
\subsubsection{Color=8}
For the gluon 8, one has \bea \Pi^{\mu \nu}_{8 8} &=& i \frac{g^2
\mu^2}{6 \pi^3} \sum_{\vec v}
 \int \! d^2 \ell \Big[\left( V^{\mu}V^{\nu} \tilde V \cdot\ell\, \tilde
  V\cdot (\ell+p)
 + V\leftrightarrow\tilde V
\right)\times\cr&\times& \left( \frac{1}{D_1(\ell+p) D_1 (\ell)} +
\frac{2}{D_2(\ell+p) D_2 (\ell) }\right) \cr&&\cr&-& \Delta^2
\frac{V^{\mu} \tilde V^{\nu} + V^{\nu} \tilde V^{\mu}}{D_1(\ell+p)
D_1 (\ell)} \ \Big] \ , \eea and, in the small momentum limit,
  \be \Pi^{0 0}_{88}(p)\ =\ \Pi^{0
0}_{88}(0) + \delta \Pi^{0 0}_{88}(p)\ =\ \frac{ \mu^2 g^2}{
\pi^2} \left(1 + \frac{p_0^2}{18 \Delta^2} \right), \ee and \be
\Pi^{0 i}_{88}(p)\ =\ \delta\Pi^{0 i}_{88}(p)\ =\ \frac{\mu^2
g^2}{54 \pi^2 \Delta^2} p^0 p^i \,, \ee \be \Pi^{i j}_{88}(p)\ =\
\Pi^{i j}_{88}(0) + \delta \Pi^{i j}_{88}(p)\ =\ \frac{\mu^2
g^2}{18 \pi^2} \left( 4 \delta^{ij}+ \frac{\delta^{ij} \vec
p^2+2p^ip^j}{15 \Delta^2} \right)\, . \ee Also in this case there
are contributions to the Debye and Meissner masses. The results
are summarized in the following table.
\begin{center}
\begin{tabular}{|c|c|c|} \hline
  \hspace{.8in} & \hspace{.8in} & \hspace{.8in}  \\ [-.06in]
  $\dd a$     &  $\dd \Pi^{00}(0)$  & $\dd - \Pi^{ij}(0)$ \\ [.07in] \hline
                    &           &  \\ [-.09in]
$\dd 1 - 3$   &  $\dd  0$  &  $\dd 0 $ \\ [.07in] \hline
                   &           &  \\ [-.09in]
$\dd 4 - 7$ & $ \frac{3}{2} m^2_g $  &  $ \frac{1}{2} m^2_g$
\\[.07in] \hline
                   &           &  \\ [-.09in]
$\dd 8$   &  $ 3 m^2_g $& $ \frac{1}{3} m^2_g $\\[.07in] \hline
\end{tabular}\\
{Table 2.1: Debye and Meissner masses in the 2SC
  phase; $m^2_g=\frac{\mu^2 g^2}{3 \pi^2}$.}
\end{center}These results are in agreement with a calculation
performed by \cite{rischke1} with a different method.

 \subsubsection{Dispersion laws and light plasmons for gluons 4-8}
 To get the dispersion law for the broken color gluons
 we follow the procedure
 used for the CFL case, Eq. (\ref{105}) and its sequel.

Expanding (\ref{7}) at first order in $\dd p$ we find for $\dd b=
\beta = 4,5,6,7$ \be p^0 A_0^{\beta}\ =\ \frac{1}{3} \, \vec p
\cdot \vec A^{\, \beta} ~, \ee while for $\dd b=8$ \be p^0A_0^8\
=\ \frac{1}{9} \, \vec p \cdot \vec A^{\, 8}. \ee
 For colors 4-7 we use eqs. (\ref{pi004}) - (\ref{piij4})
and we get the following dispersion laws for the longitudinal and
transverse modes: \be E_L\ =\ \sqrt{-\, \frac{7}{15}
 |\vec p\,|^2 +2 \Delta^2} ~,\label{5247}
\ee \be E_T\ =\  \sqrt{-\, \frac{1}{5}
 |\vec p\,|^2 +2 \Delta^2} ~, \label{52b47} \ee The rest mass for these
 gluons, in
the gradient expansion approximation, is given by \be m^R_{A_{L}}\
=m^R_{A_{T}}\ =\ {\sqrt 2} \Delta . \label{m1} \ee  To estimate
the validity of this approximation we can use the exact result,
obtained analogously to the method described for the CFL case. The
rest mass of the three species $A_0,\,A_L,\, A_T$ is given by \be
m^2+\Pi_2(m^2,0)=0\ ,\label{equa2}\ee with \be
\Pi_2(m^2,0)=\frac{\mu^2g^2}{3\pi^2} \left[-1+\int_0^{+\infty}  \!
\! dx \, \frac{x + {\sqrt{x^2+1}}}{(x+{\sqrt{x^2+1}})^2-m^2} (1 -
\frac{x}{\sqrt{x^2+1} })
  \right]\ee The numerical result of (\ref{equa2}) is \be m\equiv m_R=0.894
\, \Delta \ . \ee A comparison with (\ref{m1}) shows that the
difference is of the order of $40-50\%$ and this is also the
estimated difference for the dispersion law  at small $|\vec p|$.
We notice that also in this case, as for three flavors,
 the gradient expansion
approximation tends to overestimate the correct result.

In the case of the gluon of color 8 we work as for the colors 4-7.
We obtain the following dispersion laws for the longitudinal and
transverse modes: \be E_L\ =\ \sqrt{ \frac{4}{270}  \frac{m^2_M}{
\Delta^2}
 |\vec p\,|^2 + m_{M}^2} ~,\label{52}
\ee \be E_T\ =\ \sqrt{ - \frac{1}{30}  \frac{m^2_M}{ \Delta^2}
|\vec p\,|^2 + m_{M}^2} ~, \label{52b} \ee where the Meissner
 mass is in table 2.1.
From these equations we see that \be m^R_{A_L}=m^R_{A_T}=m_{M} ~.
\label{54}\ee The large rest mass of the 8th gluon is an exact
result, independent of the small momentum approximation and is due
to the fact that, in this case \be
\Pi_2(m^2,\,0)\,=\,-\,\frac{\mu^2g^2}{9\pi^2}\,=\,-\,m_M^2\ . \ee

\section{One-loop corrections to three- and four-point gluon vertices in the 2SC model}
To complete the effective lagrangian for the eight gluons in the
two-flavor model one can compute the one loop corrections to the
gluon vertices $\dd \Gamma_3$ (3 gluons) and $\dd \Gamma_4$ (4
gluons). We shall not consider in the sequel a possible light
glueball which is discussed in \cite{sannino2}. Therefore we write
the full gluon lagrangian as \be {\cal L}= -\frac{1}{4}
F_{\mu\nu}^a F^{\mu\nu}_a + \frac{1}{2} \Pi^{\mu\nu}_{ab}
A_{\mu}^a A_{\nu}^b + {\cal L}^1_{(3)} + {\cal L}^1_{(4)} ~,
\label{pippo} \ee where ${\cal L}^1_{(3)}$ and  ${\cal L}^1_{(4)}$
are the one loop lagrangian terms for the three and four point
gluon vertices respectively. For the three point gluon vertex only
the diagram in Fig. 2.2 survives
 at leading order in $1/\mu$; for
the four point gluon vertex only the diagram in Fig. 2.3 survives
in the same limit.
\begin{figure}[htb]
\epsfxsize=3truecm \centerline{\epsffile{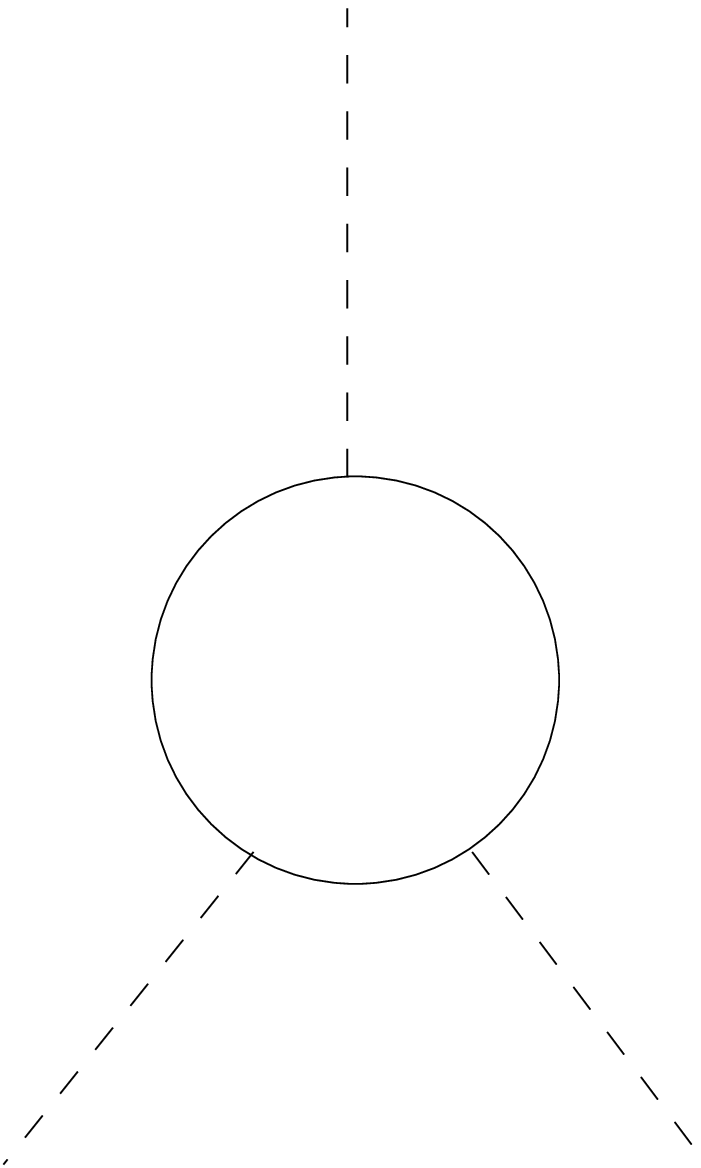}} \noindent
{Fig. 2.2}  {Three gluon vertex. Dotted lines represent gluon
fields; full lines are fermion propagators.}
\end{figure}
\begin{figure}[htb]
\epsfxsize=3truecm \centerline{\epsffile{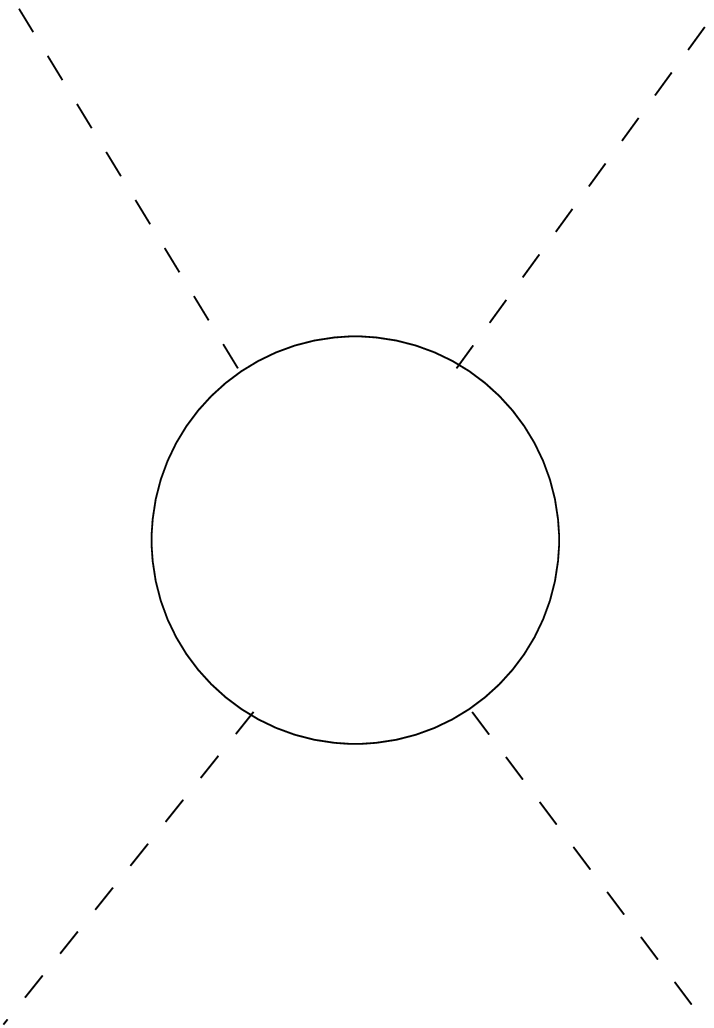}} \noindent {
Fig. 2.3}  {Four gluon vertex. Dotted lines  represent gluon
fields; full lines are fermion propagators.}
\end{figure}
\newline To start with, let us consider the three-point function. At the
tree level (${\cal L} = {\cal L}^0$) the contribution to the
lagrangian can be written in the form: \be {\cal L}_{(3)}^0 = -g
f_{abc} A_{a}^{\mu} A_{b}^{\nu}
\partial_{\mu} A_{c,\nu} ~. \label{L3} \ee At one loop
(${\cal L} = {\cal L}^1$) we have to distinguish between the
contribution of the diagram involving  the gluons in the unbroken
$SU(2)$ gauge group, which we call ${\cal L}_{(3),1}^1$, and those
involving  gluons corresponding to broken generators, which we
call ${\cal L}_{(3),2}^1$. So the one-loop correction at the three
gluon vertex may be written as follows: \be
 {\cal L}^1_{(3)}={\cal L}^1_{(3),1}+{\cal L}^1_{(3),2}.
 \ee
For the $\dd SU(2)$ contribution our result is as follows: \be
{\cal L}_{(3),1}^1 = - g k f_{abc} A_{a}^{\mu} A_{b}^{\nu}
\de^{\mu} A_{c}^{\nu} [\delta_{\mu 0} \delta_{\nu i} + \delta_{\mu
i} \delta_{\nu 0}]~, \label{L31} \ee with $\dd a,b,c \in \{1,2,3
\}$ and $\dd k=\frac{g^2 \mu^2}{18 \pi^2 \Delta^2}$. This term is
consistent with the fact that color superconductivity leaves
invariant the subgroup $SU(2)_c$ generated by the $SU(3)_c$
generators $T^a$,\, a=1,2,3$.$ It can be also obtained in a
simpler manner, by requiring gauge invariance for the $SU(2)$
gluons. We do not write down explicitly ${\cal L}^1_{(3),2}$whose
expression can be found in \cite{2fla}.

The same results are found for the four gluon vertex; again the
structure of the vertex coincides with the one dictated by gauge
invariance for the gluons 1,2 and 3, while this invariance is
broken for the other generators.

\section{Gap equation in HDET
\label{2.7}} Let us conclude this section by deriving the gap
equation in the High Density Effective Theory. Let us begin with
the CFL case; to get the gap equation one can use the effective
action formalism of Cornwall-Jackiw-Tomboulis  \cite{CJT}. In this
context it corresponds to the calculation of the grand potential
\be \Omega=\frac 1 2 \frac{4\pi\mu^2}{(2\pi)^4} \int d^2\ell
\left\{ Tr \ln \left[ S_{free}^{-1}(\ell)S(\ell)\right] \,-\,Tr
S(\ell)\Sigma(\ell)\right\}\ .\label{gp}\ee Here $S^{-1}\equiv
S^{-1}_{AB}$ and $S^{-1}_{free}\equiv S^{-1}_{AB,\,free}$ are
respectively the inverse fermion propagator
 and the free inverse fermion propagator and
 $\Sigma=S^{-1}-S_{free}^{-1}$ is the proper self-energy.
 By extremization of the density energy with respect to the gap one gets
the Schwinger-Dyson  equations; the fermion equation is
 the desired
 gap equation. In our case
the truncated Schwinger-Dyson equation assumes the form \bea
&&S^{-1}_{AB}(p=0)-S^{-1}_{AB,\,free}(p=0)=\cr&&=\,i
\frac{4\pi\mu^2g^2}{(2\pi)^4} \int d^2\ell\
i\,D^{\mu\nu}_{ab}J_\mu^{aAC}iS_{CD}(\ell)J_\mu^{bDB}\ . \eea Here
\be J_\mu^{aAB}\,=\,g\,
\begin{pmatrix}i\,V_\mu h_{AaB}&
0\\
0 &-i\,\tilde V_\mu h^*_{AaB}\end{pmatrix}\ee with $h_{AaB}$ given
in (\ref{cflcomplete0newnew}). We assume a fictitious gluon
propagator \be
i\,D^{\mu\nu}_{ab}=\,i\,\frac{g^{\mu\nu}\delta^{ab}}{\Lambda^2} \
,\ee which corresponds to a local four fermion interactions
similar to
 the Nambu- Jona Lasinio model \cite{alford}.
One obtains two equations: \bea \Delta&=&
-\,\frac{2i\mu^2g^2}{4\Lambda^2\pi^3} \left(\frac{\Delta_9}6
I(\Delta_9)-\frac{\Delta}3 I(\Delta) \right)\cr
 \Delta_9&=& -\,\frac{2i\mu^2 g^2\Delta}{3\Lambda^2\pi^3}
 I(\Delta)\ ,\label{2.186}\eea
 where
\be I(\Delta)\,=\,\int \frac{d^2\ell}{V\cdot\ell\,\tilde
V\cdot\ell-\Delta^2+i\epsilon}= \,-\,i\,2\pi\,{\rm
arcsinh}\left(\frac\delta {|\Delta|}\right)\ . \ee We fix
$\Lambda/g$ by using the same equation and the same model for the
gluon propagator at $\mu=0$, where the effect of the order
parameter, in this case the chiral condensate, is to produce a
constituent mass $M$ for the light quark. This assumption gives
the equation \be
1=\frac{4g^2}{3\Lambda^2\pi^2}\int_0^{K}\frac{p^2}{\sqrt{p^2+M^2}}\
. \ee For $M=400 $ MeV and the cutoff $K=800$ MeV (700 MeV) we get
$\Lambda/g= 181$ MeV ($154 $ MeV). The corresponding values for
the gap parameters are given in tables 2.2 and 2.3.
\begin{center}
\begin{tabular}{|c|c|c|c|}
\hline   $K=\delta+\mu$ & $\Delta$ & $\Delta_9$ &
${\Delta_9}/{\Delta}$ \\
\hline
  800\,MeV & 66\,MeV  & -150\,MeV  & -2.27\\
  700\,MeV & 77\,MeV  & -189\,MeV  & -2.45 \\ \hline
\end{tabular}
\vskip.3cm Table 2.2 $\Delta$, $\Delta_9$ and $\Delta_9/\Delta$ in
the CFL model; $\mu=500$ MeV.
\end{center}\vskip.3cm
\begin{center}
\begin{tabular}{|c|c|c|c|}
\hline   $K=\delta+\mu$ & $\Delta$ & $\Delta_9$ &
${\Delta_9}/{\Delta}$ \\
\hline 800\,MeV & 25\,MeV  & -57\,MeV  & -2.28 \\
  700\,MeV & 51\,MeV  & -114\,MeV  & -2.24 \\ \hline
\end{tabular}\vskip.3cm
Table 2.3 Same as in table 2.2, with $\mu=400$ MeV.
\end{center}
In order to obtain these results we have used the same cutoff $K$
for the two gap equations at $\mu=0$ and $\mu\neq 0$; this fixes
the value of $\delta$, the cutoff on the residual momentum along
the Fermi velocity at $\mu\neq 0$ ($|l_\parallel|\leq \delta$):
\be\delta=K-\mu\ .\ee A similar approach has been used by other
authors, see for example the review \cite{rassegne} and references
therein.
 Their calculation differs from the
present one for two reasons: they include the ${\cal O} (1/\mu)$
corrections to the gap equations and use a smooth cutoff \be
F(p)=\left(\frac{\Lambda^2}{\Lambda^2+p^2}\right)^\nu\ ,\ee with
$\Lambda=800$ MeV and various values of $\nu$.
 They obtain at $\mu=400$ MeV
$\Delta=80\, {\rm MeV}$ and $\Delta_9=-176\, {\rm MeV}$ while for
$\mu=500$ MeV $\Delta=109 \,{\rm MeV}$ and $\Delta_9=-249\, {\rm
MeV}$. Note that in any case the ratio $\Delta_9/\Delta\approx-2$
which means that the CFL ansatz is indeed justified by this
analysis.

The analogous of eq. (\ref{2.186}) for the 2SC case is \be
1=\frac{i\,\mu^2g^2}{3\Lambda^2\pi^3}\,I(\Delta)=
\frac{2\,\mu^2g^2}{3\Lambda^2\pi^2}\,{\rm arcsinh}\left(
\frac{\delta}{|\Delta|}\right) \label{2sc}\ee that can be solved
explicitly: \be |\Delta|=\frac{\delta}{\dd
\sinh\left(\frac{3\Lambda^2\pi^2}{2\mu^2g^2}\right)}\ .
\label{2scbis}\ee With the same values of $\Lambda/g$ ($\Lambda/g=
181$ MeV
  and $154 $ MeV)
we find the values for the gap parameter of the 2SC model reported
in tables 2.4 and 2.5.
\begin{center}
\begin{tabular}{|c|c|}
\hline $K=\delta+\mu$ & $\Delta$ \\
\hline
  800\,MeV & 88 \,MeV \\
  700\,MeV & 105\,MeV \\ \hline
\end{tabular}
\vskip.3cm Table 2.4 Value of the  parameter $\Delta$ in the 2SC
model; $\mu=500$ MeV.
\end{center}\vskip.3cm
\begin{center}
\begin{tabular}{|c|c|}
\hline $K$ & $\Delta$  \\
\hline
  800\,MeV & 39 \,MeV \\
  700\,MeV & 68 \,MeV\\ \hline
\end{tabular}
\vskip.3cm Table 2.5 Same as in Table 2.4 with $\mu=400$ MeV.
\end{center}

\section{Appendix. A few useful integrals\label{appendix}}
We list a few 2-D integrals used to obtain the results of this
Section. Let us define\be I_n=\int
\frac{d^{N}\ell}{(V\cdot\ell\,\tilde V\cdot\ell -
\Delta^2+i\epsilon)^{n+1}}= \frac{i\,(-i)^{n+1}\pi^{\frac N
2}}{n!} \frac{\Gamma(n+1-\frac N 2)}{\Delta^{2n+2-N}}\ ; \ee
therefore, for $N=2-\epsilon$ and with $\gamma$ the
Euler-Mascheroni constant, \bea
I_0&=&-\frac{2i\pi}{\epsilon}+i\pi\ln\pi\Delta^2+i\pi\gamma \ ,\cr
I_1&=&+\frac{i\pi}{\Delta^2}\ ,\ I_2=-\frac{i\pi}{2\Delta^4}\ ,\
I_3=+\frac{i\pi}{3\Delta^6}\ .\eea Moreover we define \bes
I_{n,\,m}=\int \frac{d^{2}\ell}{ (V\cdot\ell\,\tilde V\cdot\ell
-\Delta^2+i\epsilon)^n (V\cdot\ell\,\tilde V\cdot\ell
-\Delta^{\prime 2}+i\epsilon)^m}\ . \ees We get \bea I_{1,\,1}&=&
\frac{i\pi}{\Delta^2-\Delta^{\prime\, 2}}
\ln\frac{\Delta^2}{\Delta^{\prime\, 2}} \ ,\cr I_{2,\,1}&=&i\pi
\left[\frac{1}{\Delta^2(\Delta^2-\Delta^{\prime\, 2})}-
\frac{1}{(\Delta^2-\Delta^{\prime\, 2})^2}
\ln\frac{\Delta^2}{\Delta^{\prime\, 2}} \right]\ ,\cr
I_{3,\,1}&=&\frac{i\pi}2\left[
\frac{-1}{\Delta^4(\Delta^2-\Delta^{\prime\, 2})}-
\frac{2}{\Delta^2(\Delta^2-\Delta^{\prime\, 2})^2}
+\frac{2}{(\Delta^2-\Delta^{\prime\, 2})^3}\ln
\frac{\Delta^2}{\Delta^{\prime\, 2}} \right]\ ,\cr
I_{2,\,2}&=&i\pi\left[
\left(\frac{1}{\Delta^2}+\frac{1}{\Delta^{\prime\,2}}\right)
\frac{1}{(\Delta^2-\Delta^{\prime\, 2})^2}
+\frac{2}{(\Delta^2-\Delta^{\prime\, 2})^3}\ln
\frac{\Delta^2}{\Delta^{\prime\, 2}} \right]. \eea
 The
following integral is used in the context of the 2SC model: \be
\tilde I_1=\int \frac{(V\cdot\ell)^2\
d^{2}\ell}{(V\cdot\ell\,\tilde V\cdot\ell )^{2}}\ . \ee The
divergence is cured going to imaginary frequencies and finite
temperature ($\ell_0=i\omega_n,\ \omega_n=\pi T(2n+1)$; $T\to
0,\,\mu\to \infty$): \be \tilde I_1=2\pi Ti\int_{-\mu}^{+\mu}dx
\sum_{-\infty}^{+\infty}\frac{(i\omega_n-x)^2}{(x^2+\omega_n^2)^2}
=2\pi Ti\left(-\frac{1}T\tanh\frac{\mu}{2T}
\right)\rightarrow-2\pi i \ . \ee Other divergent integrals, such
as \be \tilde I=\int \frac{d^{2}\ell}{(V\cdot\ell\,\tilde
V\cdot\ell )(V\cdot\ell\,\tilde V\cdot\ell -\Delta^2)} \ee are
treated in a similar way.

 Finally, useful
angular integrations are: \bea \int\frac{d\vec v}{4\pi}
v^jv^k&=&\frac{\delta^{jk}}{3}\ ,\cr \int\frac{d\vec
v}{4\pi}v^iv^jv^kv^\ell&=& \frac 1{15}(\delta^{ij}\delta^{k\ell}+
\delta^{ik}\delta^{j\ell}+\delta^{i\ell}\delta^{jk})\ . \eea

\chapter{Theoretical developments
\label{ch6}}
\section{Advances in the CFL model}
In the last three years there has been a tremendous activity in
the study of QCD at high density. To review these results is
beyond the scope of this paper and I refer to some excellent
review papers that have been  already published \cite{rassegne}. I
will survey here only some of the most important developments that
in a way or in another are related to our subject, i.e. the
effective lagrangian approach. To begin with, in this paragraph I
report on some of these advances in the Color-Flavor-Locking
model.
\subsection{Would-be NGBs masses}
In par. \ref{cflngbeff} the effective theory for the NGBs in the
CFL was presented\footnote{In the 2SC model there is one NGB and
three gluons remain massless. The effective theory for the 2SC
model has been developed in \cite{casalbuonisannino}. Its coupling
to electromagnetic field and weak currents is in
\cite{casalbuonisannino2}.}. In the CFL model the Nambu Goldstone
Bosons are the only elementary excitations that are massless in
the limit of zero Dirac quark masses. To
 have a more realistic description one should however include
the effect of Dirac quark masses, which implies that also the
octet of would-be NGBs acquire mass (the superfluid mode remains
massless). The Dirac masses induce a symmetry breaking term in the
lagrangian that becomes: \be {\cal L}_{eff}\ = \frac{F^2}4
Tr\left(\partial_t\Sigma\partial_t\Sigma^\dag\,-\,v^2\,
|\vec\nabla\Sigma|^2\right) \ - \ c\,\left[detM\,Tr
\left(M^{-1}\Sigma\right)+h.c.\right]\label{mass}\ee where
$M=diag(m_u,\,m_d,\,m_s)$. The
 value of
the parameter $c$ is \cite{sonstephanov,rho1,rho2,honglee,tytgat}
: \be  c=\frac{3\Delta^2}{2\pi^2}\ .\ee The masses of the
different bosons in the octet depend linearly on the Dirac quark
masses and have a pattern different from the analogous octet of
pseudo NGBs of the chiral symmetry; for example one finds: \be
m^2_{\pi^\pm}=\frac{2c}{F^2} m_s(m_u+m_d)\,.\hskip1cm
m^2_{K^\pm}=\frac{2c}{F^2} m_d(m_u+m_s)\ .\ee Note that \be
\frac{m^2_{K^\pm}}{m^2_{\pi^\pm}}\simeq\frac{m_d}{m_u+m_d}\ ,\ee
i.e. the {\it kaon} is lighter than the {\it pion}.

Numerically these masses are much smaller than the corresponding
masses at zero densities, due to the suppression factor
$\Delta/\mu$; for example the kaon mass, for $\mu\simeq 500$ MeV
is in the range of $\sim 10$ MeV.
\subsection{Electromagnetic interactions\label{em}}
The diquark condensate breaks $U(1)_{em}$, which is clear as it
couples a quark pair with a non zero total charge. In presence of
spontaneous breaking of electromagnetism photons acquire a
Meissner mass and ordinary conductors become superconductors. In
the case of color superconductors, however, the situation is
different. Here  a new $U(1)$ subgroup appears, arising from a
combination of color and flavor; it remains unbroken and takes the
role of electromagnetism in the high density
 phase, while the associated vector boson remains massless
\cite{wilczekcfl}.
 The situation is similar to the standard model of
 the electroweak interactions, where the photon arises as a combination
 of the two neutral bosons related to
 two diagonal generators of the $SU(2)\times U(1)$
 electroweak group.

 To determine the unbroken  $U(1)$ group,
 let us show that the condensate
 \be Tr [\psi^T\epsilon_I\psi\epsilon_I] \ ,\ee see Eq. (\ref{2.31}),
 is invariant under the $U(1)$ transformation
 \be
\psi_{\alpha i}\to U^{\alpha\beta}_{ij}\psi_{\beta j}\ee with \be
U^{\alpha\beta}_{ij}=
 \left(\exp\left\{\,i\epsilon\left[{\bf 1}\otimes \,Q\,+
 \,{T}\otimes\,{\bf 1}\right]\right\}\right)^{\alpha\beta}_{ij}\ ,\ee
  where $Q=diag(2/3,\,-1/3,\,-1/3)$ is the quark charge matrix in
the flavor space and $T_{\alpha\beta}=1/{3}(-2,\,1,\,1) $ acts on
color indices.
 In infinitesimal
 terms:
 \be \psi \to \psi+\delta\psi\ ,\ee
 where \be\delta\psi_{\alpha i}=i \,\epsilon
 \left[\psi_{\alpha j}\,Q_{ij}\,+
 \,T_{\alpha\beta}\psi_{\beta j}\,\right]\ ,\ee
To prove the invariance under the $U(1)$
 group generated by \be\tilde Q=
 {\bf 1}\otimes Q+{T}\otimes{\bf 1}\ee
it is sufficient to note that \be \delta
Tr[\psi^T\epsilon_I\psi\epsilon_I]=
2\,i\,\epsilon\left\{\left[Tr(\psi T\psi+\psi\psi Q)-Tr(\psi)Tr(
T\psi+\psi Q)\right]\right\}\,=\,0 \,,\ee where we have used
(\ref{identity}). Therefore the unbroken electromagnetism arises
from the combination of the original $U(1)_{em}$ with the $U(1)$
subgroup generated by a color hypercharge.

Let us now consider the vector boson taking the place of the
photon.
 It would appear in the lagrangian through the covariant derivative
\be D_\mu=\partial_\mu
\,-\,i\left(e\,A_\mu\,Q\,+\,g\,G^8_\mu{\sqrt 3} T\right)\ee
($T^8=\sqrt 3\,T$). Writing \bea
A_\mu&=&\cos\theta \tilde A_\mu-\sin\theta\tilde G^8_\mu\ ,\label{a}\\
G^8_\mu&=&+\sin\theta \tilde A_\mu+\cos\theta\tilde G^8_\mu\ ,
\label{atilde}\eea one can see that the field coupled
 with a strength proportional
to $\tilde Q$ is $\tilde A_\mu$, obtained from these equations
with \be \cos\theta=\frac{g}{\sqrt{g^2+\frac{e^2}{3}}}\ .\ee
Therefore the field \be \tilde A_\mu=\cos\theta A_\mu+\sin\theta
G^8_\mu\label{a12} \ee takes the role of the massless photon
field, while \be \tilde G^8_\mu=-\sin\theta A_\mu+\cos\theta
G^8_\mu\label{a12bis}\ee is a massive gluon. We note that, since
$g\gg e$, the angle $\theta$ is small, $\theta\approx 0$ and the
effective photon is mostly made by the ordinary photon. We also
note that the electromagnetic charges $\tilde Q$ of the quarks are
produced by combining two sets of charges, the charges
$q_i=(2/3,\,-1/3,\,-1/3)$ and the charges
$t_\alpha=1/{3}(-2,\,1,\,1)$; the quarks have therefore always
integer electromagnetic charges, as in the old Han-Nambu model.
The unit of the electric charge, corresponding to the absolute
value of the electron charge, is\footnote{The electron charge is
less than $e$ because the electron couples to the photon with a
strength reduced by $\cos\theta$.} \be \tilde e=\frac{e
g}{\sqrt{g^2+\frac{e^2}{3}}}\ .\ee The quarks up with colors {\it
r,\,g,\,b}
 have respectively
charges $ 0,\,+\tilde e,\,+\tilde e$; the  quarks down (or
strange) with colors 1,2,3 have respectively charges $-\tilde
e,\,0,\,0$. It is worth noticing that also the gluons have integer
electric charge $\tilde Q$, exactly as the octet of vector bosons
of low density QCD. The discussion of the electroweak couplings in
the context of the effective theories can be found in
\cite{casalbuonisannino2}.
\subsection{Quark-hadron continuity\label{qhc}}
Let us summarize once again the properties of the low energy
spectrum in the CFL model. Besides the octet+singlet
quasi-particles of spin $\frac 1 2$ with integer charge there is
an octet of massive vector states, again with integer charges and
an octet of pseudoscalar NGBs: they are massless in the limit of
zero quark masses, but acquire masses if the chiral symmetry is
explicitly broken; finally there is a massless NGB associated with
the breaking of $U(1)_B$ that remains massless even in presence of
quark masses. There is also a light pseudoscalar bosons that is
never massless: it is associated to the symmetry $U(1)_A$ which is
broken by instantons. If we now compare these  results with the
symmetries of the 3-flavor hypernuclear matter
 \cite{continuity} we see a considerable similarity. On one side
 we have the integer-charge
 low lying baryonic octet $N,\Sigma,\Lambda$ whose members can pair
 to produce $SU(3)$ condensates, e.g. $\Lambda\Lambda$,
  $\Sigma\Sigma$ etc. breaking the baryonic number;
 on the other side there are the  octet+singlet quasiparticles of spin
  1/2 with integer charges as well.
 On the nuclear side we have the
 low mass pseudoscalar octet of pions, kaons etc. and on the other
 side the octet of NGB of the broken symmetry of the CFL phase. Finally to the
 low mass $J^P=1^-$ octet of vector resonances
 ($\rho,\,K^*$, etc.) would correspond  the massive gluons.
 All that means that probably
 the two phases are contiguous
 and one passes from one to the other without
 phase transition. A further example will be discussed in paragraph
 \ref{large}.

 \subsection{Asymptotic analysis}
We have derived in the paragraph \ref{2.7}
 the gap equation in the
Nambu-Jona Lasinio model where the gluon propagator is substituted
by a local four-fermion coupling.
 Eq. (\ref{2.186}) has the
approximate solution in the CFL model \be \Delta\approx 2\delta \,
\exp\left\{-\,\frac{3\Lambda^2\pi^2}{\mu^2g^2}\right\} \ .
\label{g2} \ee A more refined treatment would involve the use of
the gluon propagator and has been performed in
\cite{son},\cite{rischke} (see also \cite{Barrois} and Barrois and
Frautschi in \cite{others}). The result of this analysis
 is given by the
formula: \be \Delta\approx \frac{\mu}{g^5} \,
\exp\left\{-\,\frac{3\pi^2}{{\sqrt{2}}g}\right\} \ .\label{g5} \ee
One should note the different behavior \be\Delta\sim
\exp(-const/g)\label{diffe}\ee instead of  $\Delta\sim
\exp(-const/g^2)$ between (\ref{g5})
 and (\ref{g2}), which is due to the fact that in the complete
 treatment the gluon propagator at high density has
 a collinear divergence at zero scattering angle.
 We can qualitatively understand the behavior (\ref{diffe})
 as follows \cite{rassegne}. The collinear divergence
 should be regulated by
 the gluon rest mass that, as we know, is of
 the order $\Delta$, see par. \ref{dbcfl}. Therefore
 the gap equation has the form
 \be\Delta\propto g^2\mu^2\int d\ell_0 d\theta
 \frac{\Delta}{\sqrt{\ell_0^2+\Delta^2}}
 \frac{1}{\mu^2\theta+\Delta^2}\label{gap4}\ .\ee
In this way one gets the relation \be 1\propto
g^2\left(\ln\frac{\Delta}{\mu}\right)^2\ee and therefore eq.
(\ref{diffe}).
\section{Other models}
\subsection{One flavor}In the real world $m_u\,\simeq\,m_d\,\simeq\, 0$
might be a good approximation; therefore, if
\be\Delta>m^2_s/(2\mu)\ ,\label{ll0}\ee the color flavor locking
could be a
 good approximation as well. The reason for the condition
  (\ref{ll0})
 can be understood as follows (see \cite{LOFF4} and the discussion on the
 strange quark mass in
 \cite{wilczekcfl1} and  \cite{wilczekcfl4}).
The main effect of the strange quark mass is kinematic and
produces a shifting between the Fermi momenta of the $u,\,d$
quarks, $p_F\simeq \mu$, and that of the strange quark,
$p_F=\sqrt{\mu^2-m_s^2}\simeq\mu-\frac{m_s^2}{2\mu}$. The
difference between the two momenta must be small for the pairing
to occur, in particular it must be smaller than the smearing of
the Fermi surface, of order $\Delta$, which explains (\ref{ll0}).

On the other hand, if $m_s>\sqrt{2\mu\Delta}$, one has to consider
the role played by the strange quark explicitly.
 Quantum Cromo Dynamics at large chemical potential and small $T$
with one flavor (strangeness) has been considered in
\cite{oneflavor1}. Also in this case the theory
 exhibits color superconductivity, however the symmetry breaking
 condensate, being antisymmetric in color, must be in
 an angular momentum 1 state,
 because there is no other way to obtain an antisymmetric wavefunction for the strange quark pair.
In the non relativistic limit the computations are relatively
simple \footnote{The analysis of the ultrarelativistic limit does
not change the conclusions obtained by the non relativistic
approximation.};
 there
are two possibilities for the order parameter $\Delta^a_i$, i.e.
\be\langle\psi^TC\sigma^i\hat\lambda_A^a\psi\rangle\ , \hskip1cm
\langle\psi^TC\hat q^i\hat\lambda_A^a\psi\rangle; \label{order}\ee
here $\hat \lambda_A^a$ ($a=1,\,2,\,3$) are the three Gell-Mann
matrices antisymmetric in color (i.e.
$\lambda_2,\,\lambda_5,\,\lambda_7$); the two terms in
(\ref{order}) are distinct
 because in the non relativistic limit spin and orbital angular
 momentum are separately conserved.
Four possible cases were considered \cite{oneflavor1}: \bea
\Delta^a_i&=&\Delta\delta^a_i\ ,\cr
\Delta^a_i&=&\Delta\delta^{a3}\delta_{i3}\ ,\cr
\Delta^a_i&=&\Delta\delta^{a3}(\delta_{i1}+i\delta_{i2})\ ,\cr
\Delta^a_i&=&\Delta(\delta^{a1}\delta_{i1}+
\delta^{a2}\delta_{i2})\ ,\label{order2} \eea called respectively
{\it Color Spin Locking} (CSL), {\it 2SC}, {\it A} and {\it planar
phase}. For the former condensate in (\ref{order})\footnote{The
latter condensate produces either smaller or anisotropic gaps and
the analysis is more involved, see \cite{oneflavor1}  for further
discussion.}
 the results are as
follows. First one determines the value of the parameter $\Delta$
solving the Schwinger-Dyson  equation for the four cases in
(\ref{order2}); one obtains that the largest gap corresponds to
the $CSL$ case. Next one looks at the grand potential and observes
that the solution minimizing it corresponds again to the Color
Spin Locking phase. One can conclude that the $CSL$ phase is
indeed favored for the case of one-flavor QCD at high density. The
global symmetries of the $CSL$ phase agree with the expectations
based on $N_f$=1 QCD at high density, and therefore one can argue
that also in this case, as for CFL model with $N_f=3$,
 low and high density
phases are continuously connected.

\subsection{QCD at large isospin density\label{large}}
It would be desirable to confront the analytical results obtained
by different approximations for QCD at high density with  the
lattice QCD calculations. However this comparison is at present
impossible because current methods to evaluate the QCD partition
function use path integral with a complex fermion determinant. In
the quenched QCD approximation one ignores the fermionic
determinant, but this procedure leads to inconsistent results in
the case of finite baryonic chemical potential $\mu_B$
\cite{Kogut:eq}, \cite{Kogut:1994wf}. In absence of a consistent
approximation scheme, one may hope to get a clue to the behaviour
of quark matter at finite $\mu_B$ by considering QCD at finite
$\mu_I$, where
 $\mu_I$ is the chemical potential associated to the
 third component of isospin, $I_3$. In \cite{isospin}
 the case of finite $\mu_I$ and $\mu_B=0$ was
 considered, which, though unrealistic, can be of some utility
 because it should be accessible to lattice calculations
\cite{Alford:1998sd}. The results found in \cite{isospin} are as
follows. When  $|\mu_I|<m_\pi$ the vacuum does not differ from the
$\mu_I=0$ case, because to excite a $\pi$ costs $m_\pi-|\mu_I|$.
If  $|\mu_I|$ is small, but larger than $m_\pi$, the excitation of
the pions is favored and the ground state is a pion superfluid
(with one massless pion and the other two massive).
 This state
is characterized by the order parameter $<\pi^->\neq 0.$ For large
$|\mu_I|$ one goes to a region characterized by the order
parameter $<\bar u\gamma_5 d>\neq 0$. Since the two order
parameters
 $<\pi^->\neq 0$
and  $<\bar u\gamma_5 d>\neq 0$ have the same quantum numbers and
break the same symmetries it can be conjectured that there is no
phase transition along the $\mu_I$ axis from the small to the high
chemical potential \cite{isospin} and one can move continuously
from the hadron to the quark phase. This hypothesis is similar to
the conjecture of quark hadron continuity discussed in the CFL
model, see paragraph \ref{qhc}.
\section{Transport theory for quarks in dense matter}
 The transport theory for colored particles in a Yang Mills
external field can be used to derive quasi-particle properties for
QCD at high density. This development not only allows
 a different derivation of known results, but also
  sheds a new light on the dynamics of the
quasi-particles.

 The classical equations describing the motion of a
particle in colored external field have been derived by
Wong\footnote{The original Wong derivation used Dirac equation.
The quadratic form of Dirac equation was used by Heinz
\cite{heinz} to disentangle quark spin.} \cite{wong}. Here we
derive the Wong equations by the Fok Schwinger proper time method
that presents the advantage of being covariant.

To begin with, we write the Dirac equation for the Feynman
propagator \bes (i\slash D_x
-m)S_F(x,x^\prime)=\delta^4(x-x^\prime)\ees  in an operatorial way
\bes (i\slash D -m)S_F\ =\ 1 \ .\ees We have \bes S_F\ =\
\frac{1}{i\slash D -m }\ =\ \frac{i\slash D +m }{-2m}
\frac{-2m}{(i\slash D)^2 -m^2 } =\ \frac{i\slash D +m }{-2m} \,G\
,\ees with \bes G\ =\ \frac{-2m}{(i\slash D)^2 -m^2 }\equiv \
\frac 1 H \ .\ees Let us interpret $H$, given by \be H\ =\
\frac{(i\slash D)^2 -m^2 }{-2m}\ ,\ee
 as a hamiltonian; with $\tau$ proper time of the
quark we have: \be H \ \Psi = i\ \frac{\partial \Psi}{\partial\tau
}\ ,\ee i.e.\bes i\ \frac{\partial U}{\partial\tau }= H\ U \ ,\ees
where
 \bes
U\ =\ U(\tau)\ =\ e^{-i H\tau} \ees is the  propagator. In the
coordinate representation one has \bes U(x,x^\prime,\tau)\ =\
<x|U(\tau)|x^\prime> \ .\ees
 Since one has
 \be G\ =\ \frac 1 H \ =\ -\ i\ \int_{-\infty}^{0}d\tau e^{-i
 H\tau}\ ,\ee
 we also have
\beas i\partial_\tau <x|U(\tau)|x^\prime>& =&<x|H U|x^\prime>\ =
<x|UU^\dag
 HU|x^\prime>=\cr&&\cr
&=&<x(\tau)|H(x(\tau),\pi(\tau), Q(\tau))|x^\prime(0) >    \
,\eeas
 where
\be\pi^\mu=i D^\mu \ =\ p^\mu + g A^\mu\ee
 is the kinetic momentum, and
\beas |x(\tau)>&=& U^\dag (\tau)|x(0)>\ ,\cr |x>& =& |x(0)>\
.\eeas
 Moreover
 \bes U^\dag(\tau)H U(\tau)\ = \ H(x(\tau),\pi(\tau),Q(\tau))\ees
 is the hamiltonian  and
\beas\pi(\tau)& =& U^\dag(\tau)\pi U(\tau)\ ,\cr&&\cr
 x(\tau)& =&U^\dag(\tau)x U(\tau)\ , \cr&&\cr
 Q_a(\tau)& =& U^\dag(\tau)Q_a U(\tau) \eeas
 are operators in the Heisenberg representation.

The Heisenberg equations of motion are as follows: \bea
\frac{dx^\mu}{d\tau} & = & i [H,\ x^\mu ] \ ,\cr&&\cr
\frac{d\pi^\mu}{d\tau}& =& i [H,\ \pi^\mu ]\ ,\cr&&\cr
\frac{dQ_a}{d\tau}& =& i [H,\ Q_a ] \ .\eea
 Since the following expression holds for $H$:
\bea H& =& \frac{-1}{2m}\ (\slash\pi - m)(\slash\pi + m)\, =\cr&=&
\frac{-1}{2m}\left[\left(p^\mu+g\hat A^\mu \right)^2\ + \
\frac{g}{2}\sigma\cdot\hat F\ - \ m^2 \right]\ ,\label{3.38}\eea
 the
Heisenberg equations are transformed as follows. The first
equation in (\ref{3.38}) becomes \be \frac{dx^\mu}{d\tau} \ = \
\frac{\pi^\mu}{m} \, \ee
 as expected.
The analogous of the Lorentz force is as follows:
\bes\frac{d\pi^\mu}{d\tau}\ =\ \frac{g}{2m}\left( \pi_\lambda \hat
F^{\lambda\mu}\ +\ \hat F^{\lambda\mu} \pi_\lambda-
\frac{1}{2}D^\mu(\sigma\cdot \hat F) \right)\ ,\ees while the
dynamics of  the color charge is governed by the equation
 \bes\frac{dQ_a}{d\tau}\ =\
\frac{-g}{2m}f_{abc}\left( \pi_\lambda  A^{\lambda}_b Q^c\ +\
A^{\lambda}_b Q^c \pi_\lambda \ +\ \frac{1}{2} (\sigma\cdot
F^b)Q^c \right)\ .\ees

Since in the classical limit all the operators commute, one has
\bea[\pi, F]&=&0\ ,\cr&&\cr [Q^a,Q^b]&=&\,i\,\hbar\, f_{abc}
Q^c\to 0 \,. \eea The spin terms can also be neglected
\cite{heinz} and one obtains the Wong  equations \cite{wong}
($\hat F^{\lambda \mu}=-Q^aF^{a\lambda \mu}$ and $Q^a$ analogous
of $-\lambda/2$ in the classical limit \cite{heinz}): \bea m \
\frac{dx^\mu}{d\tau} &=&{\pi^\mu}\ ,\cr &&\cr m
\frac{d\pi^\mu}{d\tau}\ &=&\, g\ Q^a \ F^{a\,\mu\lambda}\
\pi_\lambda \ ,\cr&&\cr m\, \frac{dQ_a}{d\tau}&=&-g \
f_{abc}\pi_\lambda\ \ A^{b\,\lambda}\ Q^c .\label{ww}\eea

 Let us
now consider the classical transport equation for the one particle
distribution function $f(x,p,Q)$ which is the probability for
finding the particle in the state $(x,p,Q)$: \be
m\frac{df(x,p,Q)}{d\tau}=C[f](x,p,Q)\ ,\ee where $C[f]$ denotes
the collision integral.  Putting \be D_\mu\ =\ \partial_\mu\
-\,g\, f_{abc}Q^c A_\mu^b
\partial^Q_a\ ,\ee
from equations (\ref{ww}) in the approximation $C[f]=0$ we get
therefore the Boltzmann equation in the collisionless case: \be
\left[D_t\ + \vec v\cdot \vec D\ -\ g\  Q^a\ \left(\vec E^a\ +\
\vec v \wedge\vec B^a\right)\frac{\partial}{\partial \vec
p}\right]\ f\ =\ 0\ .\label{boltzmann}\ee
 Notice that in  (\ref{boltzmann})
the velocity \be\vec v\equiv \frac{\partial \epsilon(\vec
p)}{\partial \vec p}\ee is determined by the dispersion law of the
quasiparticle.

Eq. (\ref{boltzmann}), together with the Yang Mills equations,
 that we do not write here,
represent the complete set of Boltzmann-Vlasov equations for the
non abelian case and have been considered in \cite{lucchesi} to
derive the hard thermal loops of QCD. In  \cite{litim} Litim and
Manuel  apply it to the two-flavor color superconductor. The
method can be summarized as follows. First of all one writes the
colored current associated to the motion of the quasi-particle:
\be J^a_\mu(x)=g\sum\int\frac{d^3p}{(2\pi\hbar)^3} dQ \,v_\mu\,
Q^a f(x,p,Q)\label{cd1}\ ,\ee where the sum is over the helicities
and $v^\mu=(1,\vec v)$. Let us call \bes f^{eq}=\frac 1{1+e^{\beta
\epsilon}} \ees the Fermi equilibrium distribution function
 and let us assume a small, perturbative deviation from equilibrium:
 \bes
 f=f^{eq}+g\delta f\ .\ees
One also defines the color density \be J^a(x,p)=g\int dQQ^a
f(x,p,Q)\label{cd2}\ ,\ee so that (\ref{cd1}) can be written as
follows: \bes J^a_\mu(x)=\int\frac{d^3p}{(2\pi\hbar)^3}v_\mu
J^a(x,p) \ .\ees
 Expanding (\ref{boltzmann}) in $g$ one
obtains the transport equation for the color density:\be
\left[D_t\ + \vec v\cdot \vec D\right] J(x,p)\, =\, g^2N_f\vec
v\cdot\vec E \frac{df^{eq}}{d\epsilon}\ .\label{boltzmann2} \ \ee
The formal solution of (\ref{boltzmann2}) is \be
J^a_\mu=g^2N_f\int\frac{d^3p\,d^4y}{(2\pi)^3}v_\mu G^{ab}\vec
v\cdot \vec E^b(y)\ \frac{df^{eq}}{d\epsilon} \ ,\label{formal}\ee
where $\dd G^{ab}=<y|(v\cdot D)^{-1}|x>_{ab}$ is the Green
function. We observe that, since $D$ depends on the gluon fields
$A$, $J^a_\mu$ is a functional of $A$.
 This treatment is analogous to the derivation
of the Hard Thermal Loop and the Hard Dense Loop approximations
from the transport theory \cite{lucchesi}, \cite{Manuel:1995td},
\cite{Braaten:1989mz}. For the non relativistic and Abelian case
it reduces to the transport theory for the BCS superconductor
\cite{ll}.

 All the relevant information is contained in the
functional $J[A]$ because the effective action $\Gamma$ can be
obtained from $\dd J[A]=-\frac{\delta\Gamma}{\delta A}$ and all
the diagrams can be obtained by deriving $\Gamma$. As a matter of
fact, expanding $J[A]$ in powers of $A_\mu$ one gets \be
J^a_\mu[A]=\Pi_{\mu\nu}^{ab} A_b^\nu+\frac 1 2
\Gamma_{\mu\nu\lambda}^{abc}A^\nu_bA^\lambda_c
 . \ee Therefore from (\ref{formal}) one can obtain the various
 tensors, in particular the polarization tensor
  $\Pi^{ab}_{\mu\nu}$, that in \cite{litim} is
   computed for the 2SC model.

   There are several advantages in computing the
   polarization tensor by this method. First of all
   one obtains at once not only the results at $T=0$, but also the thermal dependence. For
   example the $T-$dependence of the Debye mass
   is given by
   \be
m^2_D=-\frac{g^2N_f}{2\pi^2}\int_0^\infty
dp\,p^2\frac{df^{eq}}{d\epsilon}\equiv
\frac{g^2\mu^2N_f}{2\pi^2}\hat I_0\left(\frac{\Delta}{T},\,
\frac{T}{\mu}\right)
   \,\ee where
   \bes
\hat I_n\left(\frac{\Delta}{T},\,
\frac{T}{\mu}\right)=\,-\frac{1}{\mu^2}\int_0^\infty dp\,p^2
v^n\frac{df^{eq}}{d\epsilon}
   \ .\ees One can check that
   $I_n(\infty,\,0)=0$, which implies that there is no Debye
   screening in the 2SC model at $T=0$, a result we already knew. For small temperatures one gets \be
m^2_D\sim\frac{g^2\mu^2N_f}{2\pi^2}\sqrt{\frac{2\pi\Delta}{T}}e^{-\Delta/T}
\ .\ee From the real part of $\Pi_{\mu\nu}^{ab}$ one can get the
dispersion laws; the resulting expressions are involved, but can
be expanded for small momenta, similarly to the approximations
presented in the par. \ref{dbcfl}. On the other hand the imaginary
part of $\Pi_{\mu\nu}^{ab}$ gives the Landau damping, that,
however, is absent for small energies, i.e. plasmon and transverse
excitations are stable provided $E_{L,T}(k)>k$. In \cite{litim} a
comparison with the results obtained by quantum field theory at
$T\neq 0$ \cite{rischke1} is given and an agreement is found. The
method based on the transport equation is however more powerful
because it allows the calculation of other significant quantities
beyond the leading order, such as, for example, the damping rate,
or transport properties such as thermal and electrical
conductivities or shear viscosity that are dominated by the
excitations of the NGB or the gapless quarks.

\chapter{The LOFF phase
\label{ch5}}
\section{Overview of the
crystalline superconductive (LOFF) state}In nature flavor symmetry
is broken not only explicitly by quark mass terms, but also by
weak interactions. Therefore in the applications of the color
superconductivity one has to take into account this symmetry
breaking; for example in compact stars, considering only two
flavors, isospin is broken by $\delta\mu=\mu_u- \mu_d\neq 0$, due
to the process: \be d\to u e \nu\ . \ee
 When the two quarks in the Cooper pair have different chemical
potentials, the vacuum is characterized, for certain values of
$\delta\mu$, by a non vanishing expectation value of a quark
bilinear breaking translational and rotational invariance. The
appearance of this condensate is a consequence of the fact that
 in a given range of
$\delta\mu$ \cite{LOFF}, the formation of a Cooper pair with a
total momentum\be \vec p_1\,+\,\vec p_2\,=\,2\vec q\neq \vec 0\ee
 is energetically favored in comparison to
the BCS state characterized by $ \vec q=\vec k+(-\vec k )=0$ . For
the two flavor case one finds
  \cite{LOFF}
   \bea \delta\mu& \in&[\delta\mu_1,\delta\mu_2]\ ,\cr
\delta\mu_1&=&0.71\Delta\ ,\hskip.5cm \delta\mu_2=0.74\Delta, \
\delta\mu_j\ll\mu\label{small}\eea where $\Delta$ is the BCS gap.
A similar phenomenon was already observed many years ago in the
context of the BCS theory for superconducting materials in
presence of magnetic impurities by Larkin, Ovchinnikov, Fulde and
Ferrel and the corresponding phase is named LOFF state
\cite{originalloff}.

\begin{figure}[htb]
\epsfxsize=5.5truecm \centerline{\epsffile{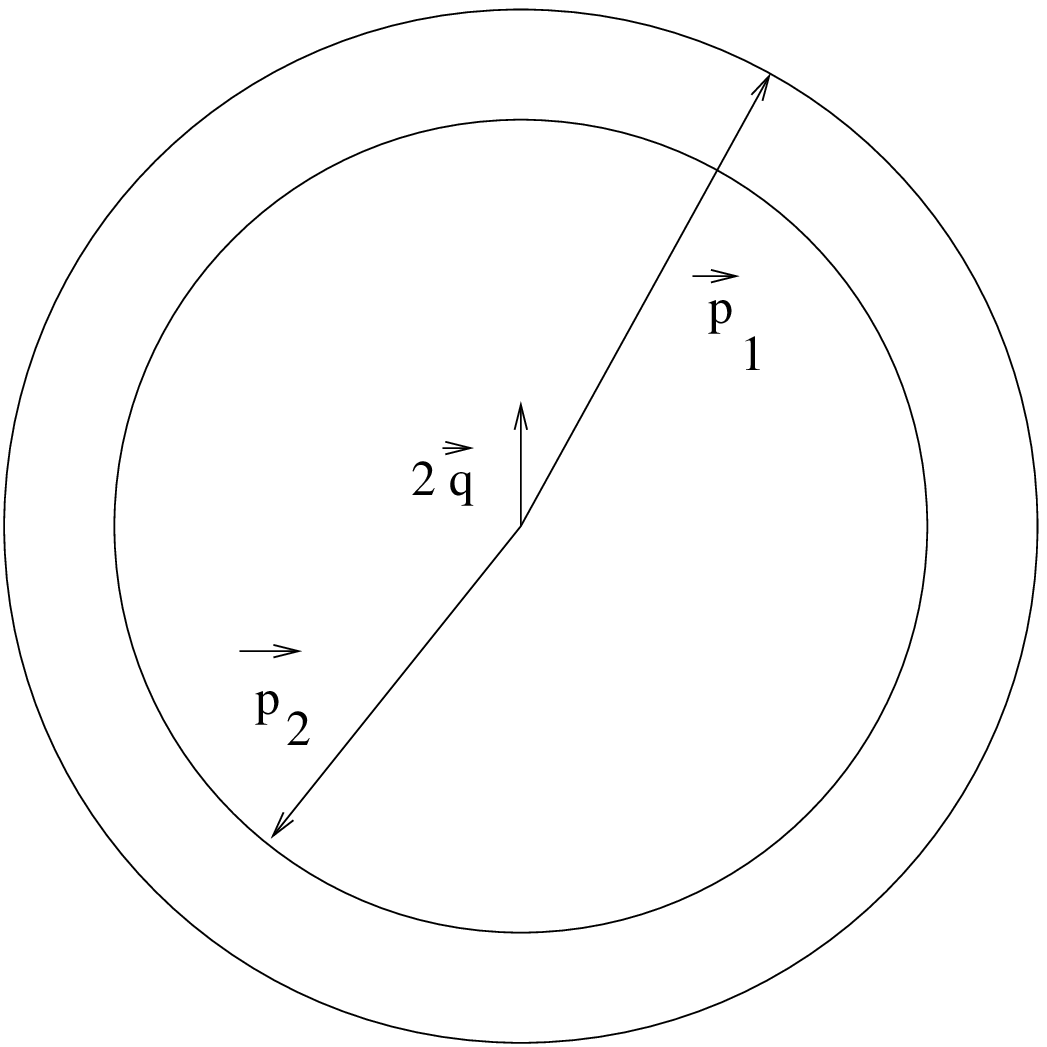}} \noindent
Fig. 4.1  LOFF kinematics. The Cooper pair has a total momentum
$2\vec q\neq 0$.
\end{figure}
 The exact
form of the order parameter (diquark condensate) breaking
space-time symmetries in the crystalline phase is not yet known.
In \cite{LOFF} the following ansatz is made: \be \Delta(\vec
x)=\Delta e^{i2\vec q\cdot\vec x}=\Delta e^{i2q\vec n\cdot\vec x}\
. \label{ansatz}\ee The value of $|\vec q|$ is fixed by the
dynamics, while its direction $\vec n$ is spontaneously chosen. In
the sequel I will assume the ansatz (\ref{ansatz}) as well. The
order parameter (\ref{ansatz}) induces a lattice structure given
by parallel planes perpendicular to $\vec n$:\be \vec n \cdot\vec
x\,=\,\frac{\pi k}{q}\hskip 1cm (k\,=0,\, \pm 1,\,\pm 2,...) \
.\label{planes}\ee We can give the following physical picture of
the lattice structure of the LOFF phase: Due to the interaction
with the medium,  the Majorana masses of the red and green up and
down quarks have a periodic modulation in space, reaching on
subsequent planes maxima and minima.

The Cooper pair can be formed only if quarks are in the
antisymmetric channel, where there is a color attractive
interaction; therefore it must be in antisymmetric flavor state if
it is in antisymmetric spin state ($S=0$). The condensate has
therefore the form \be -<0|\epsilon_{ij}\epsilon_{\alpha\beta 3 }
\psi^{i\alpha}( \vec x)C\psi^{j\beta}(\vec x)|0>= 2\Gamma_A^L
e^{2i\vec q\cdot\vec x}\label{scalar}\ .\ee The analysis of
\cite{LOFF} shows that, besides the condensate (\ref{scalar})
(scalar condensate), another different condensate is possible,
i.e. one characterized by total spin 1 (vector condensate) and by
a symmetric flavor state: \be
i<0|\sigma^1_{ij}\epsilon_{\alpha\beta 3 } \psi^{i\alpha}(\vec
r)C\sigma^{03}\psi^{j\beta}(\vec x)|0>=2\Gamma_B^L e^{2i\vec
q\cdot\vec x}\ .\label{vector}\ee Note that in the BCS state the
quarks forming the Cooper pair have necessarily $S=0$.

It goes without saying that the hypotheses of  \cite{LOFF} are
rather restrictive, as these authors assume only two flavors and
make the ansatz of a plane wave behavior, Eq. (\ref{ansatz}). In
any event their results are as follows. Assuming a point-like
interaction as the origin of the Fermi surface instability, the
LOFF state is energetically favored in the above-mentioned small
range of values of $ \delta\mu$ around $\delta\mu\sim 0.7\Delta$.
The actual value of the window range compatible with the presence
of the LOFF state depends on the calculation by which the
crystalline color state is computed. While the determination of
the interval (\ref{small}) is based on a local interaction,
assuming gluon exchange, as in \cite{LOFFbis}, the window opens up
considerably. The value of the gap parameters will be discussed
below, at the end of par. \ref{effloff}.

The order parameters (\ref{scalar}) and (\ref{vector})
spontaneously break rotational and translational symmetries.
Associated with this breaking there will be Nambu Goldstone Bosons
as in a crystal; these quasi-particles are known as phonons. Let
us discuss them in some detail.

\section{Phonons in the LOFF phase}
For a generic lattice structure it is known that there are three
phonons associated to the breaking of space symmetries. However
one can show \cite{gattoloff} that in order to describe the
spontaneous breaking of space symmetries induced by the
condensates (\ref{scalar}) and (\ref{vector}) one NGB is
sufficient. The argument (see Fig. 4.2) is as follows\footnote{A
similar argument is used in \cite{manohar} in a different
context.}. Rotations and translations are not independent
transformations, because the result of a translation plus a
rotation, at least locally, can be made equivalent to a pure
translation. Let us discuss the consequence of this fact and
consider the goldstone field associated to the spontaneous
breaking of the rotational and translational invariance.
\begin{figure}[htb]
\begin{center}
\epsfig{file=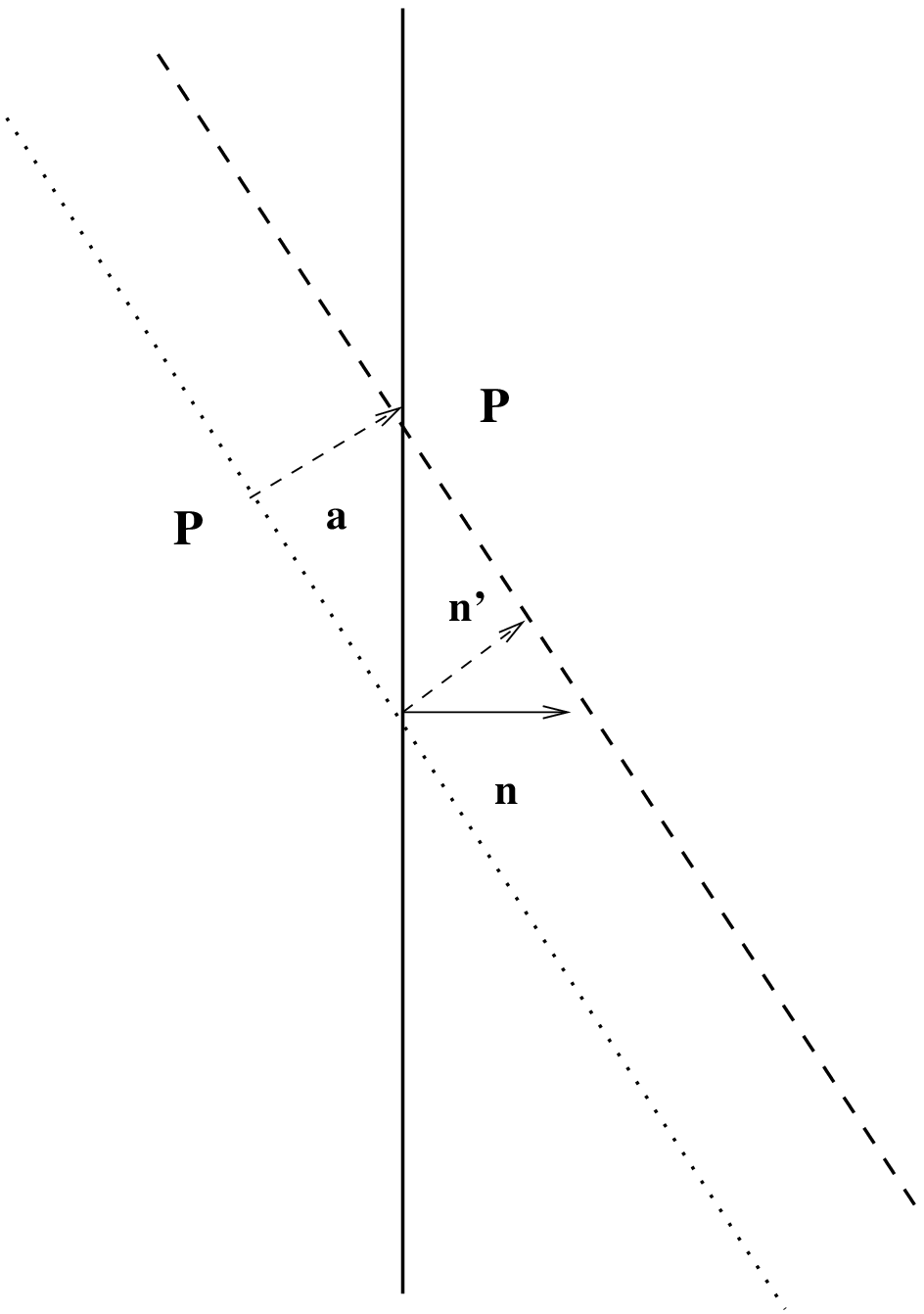,height=2.4in} \vskip.2cm
\end{center}\noindent
Fig. 4.2 {In the point P the effect of the rotation $\vec n\to\vec
n^{\,\prime}$ and the effect of the translation $\vec x\to \vec
x+\vec a$ tend to compensate each other.}

\end{figure} In the present case the NGB is a long wavelength small amplitude variation of the
condensate $\Delta(\vec x)\to \Delta(\vec x)e^{i\phi/f}$, with
 \be
 {\phi}/{f}\,=\,2q(\vec n+\delta\vec n)\cdot(\vec x+
 \delta\vec x)-2q\vec n\cdot\vec x\,=\,2q\vec R\cdot\vec x+ T-2q\vec
  n\cdot\vec x\ ,\label{35}\ee
 where we have introduced  the auxiliary function $T$,
  given by $ T\,=\,2\,q\,\vec R\cdot\delta\vec x$.

Now the lattice fluctuations $\phi/{f}$ must be small; from this
it
  follows that $T$ and $\vec R$ are not independent fields and therefore
  $T$ must depend functionally on $\vec R$, i.e. $T=F[\vec R]$,
 which, using again (\ref{35}), means
  that\be \Phi\equiv2q\vec n\cdot\vec x+{\phi}/{f}
 =2q\vec R\cdot\vec x+{F[\vec R]}\equiv G[\vec R,\,\vec x]\ .\ee
 The solution of this functional relation has the form
 \be \vec R=\vec
 h[\Phi]\ee where $\vec h$ is a vector built out of the scalar function
 $\Phi$. By this function one can only\footnote{In principle
there is a second vector,  $\vec x$, on which $\vec R$ could
depend linearly, but this possibility is excluded because $\vec R$
is a vector field transforming under translations as $\vec R(\vec
x)\to \vec R^{\,\prime}(\vec x^{\,\prime})=\vec R(\vec x)$.} form
the
 vector $\vec\nabla \Phi$ ; therefore  we
 get
 \be
\vec R =\frac{\vec \nabla\Phi}{|\vec \nabla\Phi|}\, ,\label{r3}\ee
which satisfies $|\vec R|=1$ and $<\vec R>_0=\vec n$. In terms of
the phonon field $\phi$ the  field $\vec R$ is given at the first
order in $\phi$ by the expression\be \vec R=\vec
n+\frac{1}{2fq}\left[\vec\nabla\phi-\vec n(\vec n\cdot\vec
\nabla\phi)\right]\,.\label{63}\ee In order to derive the form of
the interaction between the NGB  and the quarks let us now write
the effective lagrangian in the LOFF phase.
\section{HDET  approach to the LOFF state\label{effloff}}

To take into account the two non vanishing vacuum expectation
values we add the term: \bea {\cal L}_\Delta&=&{\cal
L}_\Delta^{(s)}+{\cal L}_\Delta^{(v)}=\cr&&\cr &=&-\frac{
e^{2i\vec q\cdot\vec x} }2\,\epsilon^{\alpha\beta 3}
\psi_{i\alpha}^T(x)C\left(\Delta^{(s)}\epsilon_{ij}
+\vec\alpha\cdot\vec n \Delta^{(v)} \sigma^1_{ij}\right)
\psi_{i\beta}(x)\cr&+&{\rm h.c.} \label{ldeltaeff}\eea to the
lagrangian.

Let us first consider  the scalar condensate: \be {\cal L
}^{(s)}_\Delta=-\frac{\Delta^{(s)}}2\, e^{2i\vec q\cdot\vec x}
\epsilon^{\alpha\beta 3}\epsilon_{ij} \psi_{i\alpha}^T(x)C
\psi_{i\beta}(x)\ -(L\to R)+{\rm h.c.} \label{loff5}~.\ee Here
$\psi(x)$ are positive energy left-handed fermion fields discussed
in the previous section. We neglect the negative energy states,
consistently with the assumptions of HDET.

 In order to introduce velocity dependent positive energy
fields \bes\psi_{+,\,\vec v_i;\,i\alpha}\ees with flavor $i$, we
have to decompose the  fermion momenta: \be \vec p_j=\mu_j\vec
v_j+{\vec \ell}_j \hskip1cm (j=1,2)\ . \label{dec1}\ee Therefore
we have : \bea {\cal L }^{(s)}_\Delta&=&-\frac{\Delta^{(s)}} 2
\,\sum_{\vec v_1,\vec v_2} \exp\{i\vec x\cdot\vec\alpha(\vec
v_1,\,\vec v_2,\vec q
)\}\epsilon_{ij}\cr&\times&\epsilon^{\alpha\beta
3}\psi_{+,\,-\,\vec v_i;\,i\alpha}(x)C \psi_{+,\,-\,\vec
v_j;\,j\beta}(x) -(L\to R)+{\rm h.c.}\label{loff6}\eea where
\be\vec\alpha(\vec v_1,\,\vec v_2,\,\vec q)=2\vec q-\mu_1\vec
v_1-\mu_2\vec v_2\ .\ee We choose
 the $z-$axis along $\vec q$ and the vectors $\vec v_1,\,
 \vec v_2$ in the $x-z$ plane; if $\alpha_1,\,\alpha_2$ are the angles
 formed by the vectors $\vec v_1,\,\vec v_2$ with the $z-$axis we have
\bea \alpha_x&=&\
-\mu_1\sin\alpha_1\cos\phi_1-\mu_2\sin\alpha_2\cos\phi_2\ ,\cr
\alpha_y&=&\
-\mu_1\sin\alpha_1\sin\phi_1-\mu_2\sin\alpha_2\sin\phi_2\ ,\cr
\alpha_z&=&\ 2\,q\, -\,\mu_1\cos\alpha_1\,-\,\mu_2\cos\alpha_2 \
.\label{alfaz}\eea  In the $\mu_1,\,\mu_2\,\to\infty$ limit the
only non vanishing terms in the sum correspond to the condition
\be \alpha_x=\ \alpha_y=0\ .\label{4bis}\ee Let us introduce
\bea\mu&=&\frac{\mu_1+\mu_2}{2}\cr\delta\mu&=&\,-\,\frac{\mu_1-\mu_2}{2}
.\label{dec2}\eea Eq. (\ref{4bis}) implies
\be\alpha_1=\alpha_2+\pi+{\cal O}\left(\frac 1 \mu\right)\ ,\ee
i.e. the two velocities are opposite in this limit.

The limit $\mu\to\infty$ does not imply $\alpha_z=0$; however if
we might take the limit $q\approx \delta\mu\to\infty$ as well, we
would also have $\alpha_z\approx 0$; we will make this
approximation \cite{gattoloff2}, which is justified by the results
of \cite{LOFF}: $q\simeq 1.2 \delta\mu$ and $\delta\mu\sim
0.7\Delta_{BCS}\gg \Delta^{(s,v)}$. This approximation makes the
loop calculations much simpler than the analogous in \cite{LOFF}
and therefore it may be particularly useful in complex
calculations. In any case corrections to this approximation can be
implemented. We have up to terms
${\cal{O}}\left(\frac{1}{\mu^2}\right)$:\bea
 \cos\alpha_1\,&=&\,-\,\frac{\delta\mu}{q}+\frac{q^2-\delta\mu^2}{q\mu}
 \cr
 \cos\alpha_2\,&=&\,+\,\frac{\delta\mu}{q}+\frac{q^2-\delta\mu^2}{q\mu} \ ,
 \eea which, together with (\ref{4bis}), has the solution
\bea \alpha_2&\ \equiv\ &\theta_q\ =\
 \arccos\frac{\delta\mu}q\,-\,\frac\epsilon 2\ ,\label{tetaq}\\
\alpha_1&\ =\ &\alpha_2+\pi-\epsilon\eea with
\be\epsilon\,=\,2\,\frac{\sqrt{q^2-\delta\mu^2}}\mu \ .\ee
Therefore, as anticipated, $\dd\epsilon={\cal O}\left(\frac 1
\mu\right)$ and in the limit $\mu\to\infty$ the two velocities are
almost
 antiparallel:\be\vec v_1\simeq -\vec v_2\ .\ee
Putting
 \be \psi_{+,\,\pm\vec
v_i;\,i\alpha}(x)~\equiv~\psi_{\pm\vec v_i;\,i\alpha}(x)~,
 \ee eq. (\ref{loff6}) becomes: \be{\cal L }^{(s)}_\Delta=-\frac{\Delta^{(s)}}{2} \sum_{\vec
v}\epsilon_{ij}\epsilon^{\alpha\beta 3} \psi_{+\vec
v;\,i\alpha}(x)C \psi_{-\vec v;\,j\beta}(x)-(L\to R)+{\rm
h.c.}\label{eq:20}\ee In a similar way the term corresponding to
the vector condensate in the lagrangian
  can be written as follows: \be {\cal L
}^{(v)}_\Delta=\, - \frac{\Delta^{(v)}}2 \sum_{\vec
v}\sigma^1_{ij}\epsilon^{\alpha\beta 3} \psi_{+\vec
v;\,i\alpha}(x)C(\vec v\cdot\vec n ) \psi_{-\vec v;\,j\beta}(x)
-(L\to R)+{\rm h.c.} \label{eq:23} \ee where $\vec n=\vec q/|\vec
q|$  is the direction corresponding to the total momentum carried
by the Cooper pair and we have used $\psi_-^TC\vec\alpha\cdot \vec
n\psi_+=\vec v\cdot\vec n\psi_-^TC\psi_+$~.

In the usual basis of the HDET the effective lagrangian is
 \bea
  {\mathcal L}&=& {\mathcal L}_0 \ +\ {\mathcal L}^{(s)}_\Delta \
  +\ {\mathcal L}^{(v)}_\Delta\ =\cr&&\cr
&=& \sum_{\vec v}\sum_{A=0}^5 \chi^{A\,\dag}\left(
\begin{array}{cc}
 i\, \delta_{AB}\ V\cdot\partial\ & \,\Delta_{AB}^\dag
\\
\,  \Delta_{AB} & i\,\delta_{AB}\ \tilde V\cdot\partial\
\end{array}\right)\chi^B\ .\eea
Here the matrix $\Delta_{AB}$ is as follows: \be
\Delta_{AB}=0\hskip1cm(A\,{\rm or} \,B=4\,{\rm or} \,5 )\ee while
for $A,B=0,...,3$ we have:\be \Delta_{AB}=
\left(\begin{array}{cccc}
  \Delta_0 & 0 & 0 & -\Delta_1 \\
  0 & -\Delta_0 &- i\Delta_1 & 0 \\
  0 & +i \Delta_1&-\Delta_0& 0 \\
  \Delta_1 & 0 & 0 & -\Delta_0
\end{array}\right)\ ,
\ee with \be\Delta_0\,=\, \Delta^{(s)}\ ,\hskip1cm
\Delta_1\,=\,\vec v\cdot\vec n\, \Delta^{(v)}\ .\ee

Let us now discuss the precise meaning of the average over
velocities. As discussed above we embodied a factor of 1/2
 in the
average over velocities; however in the LOFF case the sum over the
velocities is no longer symmetric and now reads: \be \sum_{\vec
v}\equiv \frac {k_R} 2\int\frac{d\phi}{2\pi}\ , \label{sum3}\ee if
$k_r$ is kinematical factor of the order of 1 induced by the
approximation of the Riemann-Lebesgue lemma \cite{gattoloff2} and
\be\vec v=\left(\sin\theta_q\cos\phi,\,
\sin\theta_q\sin\phi,\,\cos\theta_q\right)\ ,\ee
 and $\theta_q$ given in (\ref{tetaq}). For future reference we observe that
\be \sum_{\vec v}\,v_iv_j\,=\,k_R\left(\frac{\sin^2\theta_q}{4}\,
\left(\delta_{i1}\delta_{j1}+\delta_{i2}\delta_{j2}\right)
\,+\,\frac{\cos^2\theta_q}{2}\delta_{i3}\delta_{j3}\right) \ .\ee

The effective action for the fermi fields in momentum space reads:
\be S=\sum_{\vec v}\sum_{A,B=0}^5 \int \frac{d^4\ell}{(2\pi)^4}
\frac{d^4\ell^\prime}{(2\pi)^4}
\chi^{A\dag}(\ell^\prime)D^{-1}_{AB}(\ell^\prime,\ell)\chi_B(\ell)~,
 \ee
 where $D^{-1}_{AB}(\ell^\prime,\ell)$ is the inverse propagator,
 given by:
\be D^{-1}_{AB}(\ell^\prime,\ell)\,=\,\left(
\begin{array}{cc}
V\cdot\ell  \delta_{AB}& \Delta_{AB}^\dag
\\
\Delta_{AB}
 &  \tilde V\cdot\ell
 \delta_{AB}
\end{array}\nonumber
\right)\,\delta^4(\ell^\prime-\ell)\ .\ee From these equations one
can derive the quark propagator, given by \be \displaystyle
D_{AB}(\ell,\ell^{\prime\prime})=
(2\pi)^4\delta^4(\ell-\ell^{\prime\prime})\times\sum_C\left(
\begin{array}{cc} \displaystyle
 \frac{ \tilde V\cdot\ell\,\delta_{AC}}{\tilde D_{CB}(\ell)}\,
& \displaystyle -\frac{\Delta_{AC}^\dag}{D_{CB}(\ell)}
\\ \\
\displaystyle -\frac{\Delta_{AC}}{\tilde D_{CB}(\ell)}
 &    \displaystyle
 \frac{ V\cdot\ell\,\delta_{AC}}{D_{CB}(\ell)}\,
\end{array}
\right)\label{propagatore}\ee where \bea &D_{CB}(\ell)&=
\left(V\cdot \ell\,\tilde V\cdot
\ell\,-\,\Delta\Delta^\dag\right)_{CB}\cr
 &\tilde D_{CB}(\ell)&=  \left(V\cdot \ell\,\tilde
V\cdot \ell\,-\,\Delta^\dag\Delta\right)_{CB}\ . \eea

 The propagator for  the fields $\chi^{4,5}$
does not contain gap mass terms and is given by \be
D(\ell,\ell^{\prime})=(2\pi)^4\,\delta^4(\ell-\ell^{\prime})\,\left(
\begin{array}{cc}
 (V\cdot\ell)^{-1}&0\\
 0&  (\tilde V\cdot\ell)^{-1}
\end{array}
\right)~.\ee For the other  fields $\chi^A$, $A=0,\cdots,3$, it is
useful to go to a representation where $\Delta\Delta^\dag$ and
$\Delta^\dag\Delta$ are diagonal. It is accomplished by performing
a unitary transformation which transforms the basis $\chi^A$ into
the new basis $\tilde \chi^A$ defined by \be
\tilde\chi^{A}=R_{AB}\chi^{B}\ ,\label{newbasis1}\ee with \be
R_{AB}=\frac 1 {\sqrt 2 } \left(\begin{array}{cccc}
  1 & 0 & 0 & 1 \\
  0 & 1 &- \,i & 0 \\
  0 & +i &-\,1& 0 \\
 1 & 0 & 0 & -\,1
\end{array}\right)\ .\label{newbasis2}
\ee In the new basis we have
 \bea
\left(\Delta\Delta^\dag\right)_{AB}&=&\lambda_A\delta_{AB}\cr
\left(\Delta^\dag\Delta\right)_{AB}&=&\tilde\lambda_A\delta_{AB}\eea
where $\lambda_A,\,\tilde\lambda_A$ can be approximated as\bea
\lambda_A&=&((\Delta_0+\Delta_1)^2, \,(\Delta_0-\Delta_1)^2,
\,(\Delta_0+\Delta_1)^2, \,(\Delta_0-\Delta_1)^2 )~,\cr&&\cr
\tilde\lambda_A&=&\left((\Delta_0-\Delta_1)^2,
\,(\Delta_0-\Delta_1)^2, \,(\Delta_0+\Delta_1)^2,
\,(\Delta_0+\Delta_1)^2\right)\ .
 \eea For further reference we also define
  \be \mu_C=(\Delta_0+\Delta_1, \,\Delta_1-\Delta_0,
\,\Delta_0+\Delta_1, \,\Delta_1-\Delta_0) \label{muc} \ .\ee

As shown in \cite{gattoloff2} the integration measure for the LOFF
case is \be
 \int\frac{d^4\ell}{(2\pi)^4}=\frac{2\pi\mu^2}{(2\pi)^3}
 \int_{-\delta}^{+\delta}d\ell_\parallel\int_{-\infty}^{+\infty}\frac{d\ell_0}{2\pi}
\ . \ee Let us conclude this section by deriving the gap equation
in the LOFF phase. As in the CFL and 2SC cases one writes a
truncated Schwinger-Dyson equation. One obtains two equations:
\bea \Delta_0&=&
\,i\,\frac{\mu^2k_R}{24\Lambda^2\pi^3}\sum_{A=0}^3\,|\mu_A|\,I(\mu_A)
\cr&&\cr
 \Delta_1&=& - \,i\,\frac{\mu^2k_R}{24\Lambda^2\pi^3}\left(\mu_0I(\mu_0)+
 \mu_1 I(\mu_1)\right)\ ,\label{2.186bis}\eea
 where
\be I(\Delta)\,=\,\int \frac{d^2\ell}{V\cdot\ell\,\tilde
V\cdot\ell-\Delta^2+i\epsilon}= \,-\,i\,2\pi\,{\rm
arcsinh}\left(\frac\delta {|\Delta|}\right)\ . \ee The values of
the parameters have been fixed in paragraph \ref{2.7}. The coupled
equations (\ref{2.186}) have a non trivial solution;  one gets
\cite{gattoloff2}: \bea
\Delta_0&=&\Delta^{(s)}=\frac{\delta}{\sinh\left(\frac{3\Lambda^2\pi^2}{\mu^2k_R}\right)}\\
\Delta_1&=&\cos\theta_q\,\Delta^{(v)}=0\ .\label{delta1}\eea These
results are compatible with the findings of \cite{LOFF} where the
scalar condensate is in the range of a few MeV and $\Delta^{(v)}$
is found to be negligible. Since  in HDET one has $\Delta_1=0$ it
can be argued that the result $\Delta_1\neq 0$ found in
\cite{LOFF} comes from a non leading $\dd{\cal O}\left(\frac 1
\mu\right)$ contribution.

\section{Phonon-quark coupling\label{sec5}}
Using the results of the two previous sections one gets the
following 3-point and 4-point phonon-quark-quark couplings:
 \bea
 {\cal L}_{\phi\psi\psi}&=&-\frac{i\phi}{f}
 \sum_{\vec
v}\left[\Delta^{(s)}\epsilon_{ij}+ \vec v\cdot\vec n
\Delta^{(v)}\sigma^1_{ij}\right]\epsilon^{\alpha\beta
3}\psi_{i,\alpha,\vec v}\,C\,\psi_{j,\beta,-\vec
v}\cr&&\cr&&-\frac{1}{2fq} \sum_{\vec v} \vec v\cdot \left[ \vec
\nabla\phi-\vec n(\vec n\cdot\vec\nabla\phi)\right]
\Delta^{(v)}\sigma^1_{ij}\epsilon^{\alpha\beta
3}\psi_{i,\alpha,\vec v}\,C\,\psi_{j,\beta,-\vec
v}\cr&&\cr&&-(L\to R)\ +\ h.c.
 \label{Trilineare}\eea and
 \bea
 {\cal L}_{\phi\phi\psi\psi}&=&\frac{\phi^2}{2f^2}\,
\sum_{\vec v}\left[\Delta^{(s)}\epsilon_{ij}+ \vec v\cdot\vec n
\Delta^{(v)}\sigma^1_{ij}\right]\epsilon^{\alpha\beta
3}\psi_{i,\alpha,\vec v}\,C\,\psi_{j,\beta,-\vec
v}\cr&&\cr&&-\frac{i\phi}{f}\, \sum_{\vec v} \vec v\cdot \left[
\vec \nabla\phi-\vec n(\vec n\cdot\vec\nabla\phi)\right]
\Delta^{(v)}\sigma^1_{ij}\epsilon^{\alpha\beta
3}\psi_{i,\alpha,\vec v}\,C\,\psi_{j,\beta,-\vec v}\cr&&\cr&&
-\frac{ 1}{4f^2q^2}\,\sum_{\vec v} \Big[\frac {\vec v\cdot \vec n}
2 \left(3(\vec
n\cdot\vec\nabla\phi)^2-|\vec\nabla\phi|^2\right)\cr&&\cr&-& (\vec
v\cdot\vec \nabla\phi)(\vec n\cdot\vec\nabla\phi)\Big]\times
\cr&&\cr&&\times \Delta^{(v)}\sigma^1_{ij}\epsilon^{\alpha\beta
3}\psi_{i,\alpha,\vec v}\,C\,\psi_{j,\beta,-\vec v}   -(L\to R)\
+\ h.c.
 \label{quadrilineare}\eea
 In the basis
of the $\chi$ fields one gets
 \be {\mathcal L}_3\,+\,{\mathcal L}_4=
 \sum_{\vec v}\sum_{A=0}^3
\tilde\chi^{A\,\dag}\,   \left(
\begin{array}{cc}
 0 & g_3^\dag\,+\,g_4^\dag
\\
  g_3\,+\,g_4& 0\end{array}\right)
    \,\tilde\chi^B\ .\label{vertex}\ee
Here\bea g_3&=&\,\left[\frac{i\phi\Delta_{AB}}{f}+\sigma_{AB}\hat
O[\phi] \right] \ ,\cr
g_4&=&\,\left[-\frac{\phi^2\Delta_{AB}}{2f^2}+\sigma_{AB}
 \left(\frac{i\phi}{f}\hat O[\phi]
 \,+\,\hat Q[\phi]\right)\right] \ ,\eea with \bea
 \hat O[\phi]&=&\frac{1}{2fq}\vec v\cdot \left[ \vec \nabla\phi-\vec n(\vec
n\cdot\vec\nabla\phi)\right] \Delta^{(v)}\ ,\cr
 \hat Q[\phi]&=&
\frac{\Delta^{(v)}}{4f^2q^2}\left[\frac {\vec v\cdot\vec n}
2\left(3 (\vec n\cdot\vec\nabla\phi)^2-|\vec\nabla\phi|^2\right)-
(\vec v\cdot\vec \nabla\phi)(\vec n\cdot\vec\nabla\phi)\right]\ ,
 \eea
 %and
 %\bea \Delta_{AB}&=& \left(\begin{array}{cccc}
 % 0 & 0 & 0 & \Delta_0+\Delta_1 \\
 % 0 &  \Delta_1-\Delta_0 &0 & 0 \\
 % 0 & 0& -\Delta_0-\Delta_1& 0 \\
 % \Delta_0-\Delta_1& 0 & 0 & 0
%\end{array}\right)\ ,
\bea \sigma_{AB}&=& \left(\begin{array}{cccc}
  0 & 0 & 0 & -1 \\
  0 & 0 & -i & 0 \\
  0 & +i & 0& 0 \\
 +1 & 0 & 0 & 0
\end{array}\right)\ .
\eea Finally the effective action for the field $\phi$, $S[ \phi
]$, is obtained by the lagrangian as follows \be S=\int
dt\,dx\,dy\ \frac{\pi}{q}\sum_{k=-\infty}^{+\infty}{\cal
L}(\phi(t,x,y,k\pi/q)\ .\ee

\section{Dispersion law for the phonon field\label{sec6}}
To introduce formally the NGB in the theory one uses the same
gradient expansion discussed in the paragraph \ref{gradient}. At
the lowest order one has to consider the diagrams in Fig. 2.1,
i.e. the self-energy and the tadpole diagrams. At the second order
in the momentum expansion one gets, after averaging over the Fermi
velocities \be \Pi(p)\,=\,-\,\frac{\mu^2k_R}{4\pi^2 f^2}\Big[
 p_0^2-v_\perp^2(p_x^2+p^2_y)-v_\parallel^2 p^2_z
 \Big]\, .\label{final2bis}\ee
 One obtains canonical normalization for the kinetic term
provided \be
 f^2= \frac {\mu^2k_R}{2\pi^2}
  \ .
\ee On the other hand, at the lowest order in $\Delta^{(v)}$: \bea
v_\perp^2&=&\frac 1
 2\sin^2\theta_q+\left(1-3\cos^2\theta_q\right)\left(1-\log\frac{2\delta}{\Delta_0}\right)
 \left(\frac{\Delta^{(v)}}q\right)^2
 \\ &&\cr
 v_\parallel^2&=&\cos^2\theta_q
  \ . \eea
 In conclusion, the dispersion law for the phonon is \be E(\vec
p)=\sqrt{v_\perp^2(p_x^2+p^2_y)+v_\parallel^2 p^2_z}\ \ee which is
anisotropic. Besides the anisotropy related to
 $v_\perp\neq v_\parallel$, there is another source of
 anisotropy,  due to the fact that
 $p_z$, the component
 of the momentum perpendicular to the planes (\ref{planes}),
 differently from $p_x$ and $p_y$
 is a quasi momentum and not a real momentum.

\chapter{Astrophysical implications
\label{ch7}}
\section{Overview}
Should we look for a laboratory to test color superconductivity,
we would face the problem that in the high energy physics
programmes aiming at new states of matter, such as the Quark Gluon
Plasma, the region of the $T-\mu$ plane under investigation is
that of low density and high temperature. On the contrary we need
physical situations characterized by low temperature and high
densities. These conditions are supposed to occur in the  inner
core of neutron stars, under the hypothesis that, at the center of
these compact stars, nuclear matter has become so dense as to
allow the transition to quark matter. A schematic view of a
neutron star is in Fig. 5.1.
\begin{figure}[htb]
\epsfxsize=5.5truecm \centerline{\epsffile{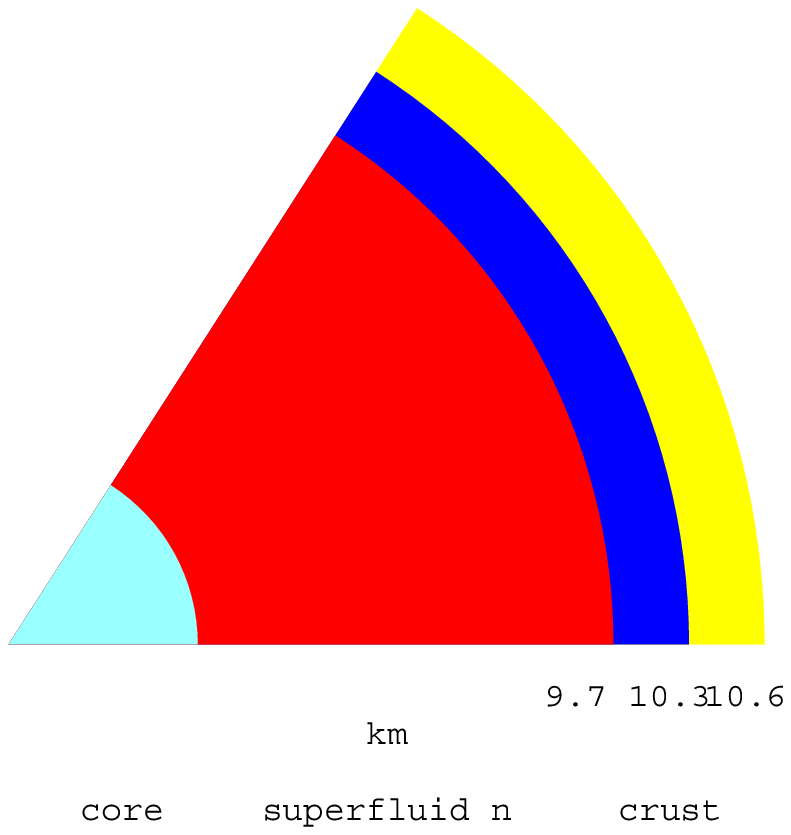}} \noindent
{Fig. 5.1}  {Schematic view of a neutron star as computed by an
equation of state with three nucleon interaction \cite{Shapiro}.}
\end{figure}
In the subsequent sections we shall give a pedagogical
introduction to the physics of compact stars and we shall review
some possible astrophysical implications of the color
superconductivity.
\section{A brief introduction to compact stars}
To begin with, let us show that for a fermion gas high chemical
potential means  high density. To simplify the argument we assume
that the fermions are massless and not interacting, so that
$\mu=\epsilon_F=p_F$. It follows from eq. (\ref{density}) that
\be\rho\propto \mu^3\ ,\ee which means that the chemical potential
increases as $\rho^{1/3}$: This is the reason why we should search
color superconductivity in media with very high baryonic density.

The equations (\ref{density})-(\ref{pressure}) allow to determine
the equation of state, which is  obtained by solving the equations
 \be
P=\Phi(x_F)\ ,~~~\rho\propto x_F^3 \ ,\ee where $x_F=p_F/mc$. For
the simplest case (free massless fermions) we obtain \be P\propto
\mu^4\ ,~~~~\rho\propto\mu^3\ ,\ee and therefore the equation of
state has the form \be P=K\rho^{\frac 4 3}\ .\ee Clearly this
result also holds for massive fermions in the Ultra Relativistic
(UR) case where the mass is negligible. One can prove that in the
Non Relativistic (NR) case one has $ P=K\rho^{\frac 5 3}$. More
generally the equation of state can be approximated by the
expression \be P=K\rho^\gamma \ , \ee and the two cases discussed
above are characterized as follows:
 \beas
&&\rho\ll 10^6 g/cm^3~ {\rm \{electrons\}}\cr
  NR\left( \gamma=\frac 5 3\right)&&\cr && \rho\ll 10^{15}
g/cm^3~{\rm \{neutrons\}} \cr&&\cr&&\cr&& \rho\gg 10^6
g/cm^3,~{\rm \{electrons\}}\cr UR\left(\gamma=\frac 4
3\right)&&\cr&& \rho\gg 10^{15} g/cm^3~{\rm \{neutrons\}}\ .\eeas
Let us note explicitly that, at $T=0$, $P\neq 0$. This is a
quantum-mechanical effect due to the Pauli principle and the Fermi
Dirac statistics (for comparison, for a classical Maxwell
Boltzmann gas $P\to 0$ when $T\to 0$). In absence of other sources
of outward pressure it is the  pressure of the degenerate fermion
gas  that balances the gravity and avoids the stellar collapse.

One can see that the densities that can be reached in the compact
stars are very different depending on the nature of the fermions.
The two cases correspond to two classes of compact stars, the
white dwarfs and the neutron stars. White dwarfs (w.d.s) are stars
that have exhausted nuclear fuel;
 well known examples are Sirius B,
or 40 Eri B. In the Hertzsprung-Russel diagram w.d.s fill in a
narrow corner below the main sequence. In a w.d. stellar
equilibrium is reached through a compensation between  the inward
pressure generated by gravity and the outward the pressure of
degenerate electrons. Typical values of the central density, mass
and radius for a w.d. are $\rho= 10^6 g/cm^3$, $M\sim
 M_\odot$,  $R\sim 5,000 km$.

Suppose now that in the star higher values of $\rho$ are reached.
  If $\rho$ increases, inverse beta decay becomes important: \be
e^-p\to n\nu\ .\ee This process fixes the chemical composition at
equilibrium \be \mu_e+\mu_p=\mu_n\ . \ee On the other hand one has
to enforce  neutrality:\be n_e=n_p\ .\ee One proves that these two
conditions  fix the ratio of the number of proton $n_p$ to the
number of neutrons $n_n$ for ultrarelativistic particles: \be
\frac{n_p}{n_n}=\frac 1 8\, .\ee This number should be seen as a
benchmark value, as it is derived under simplifying hypotheses,
most notably the absence of interactions and the neglect of
masses. In any event it suggests that, for higher densities, the
star tends to have a relatively larger fraction of neutrons and
therefore it is named a {\it neutron star.} It must be stressed
that one of the relevant facts about neutron stars is that the
general relativity effects cannot be ignored and the relevant
equilibrium equations to be used are the Oppenheimer-Volkov
equations of hydrostatic equilibrium.

The following simple argument, due to Landau (1932) can be used to
evaluate the relevant parameters of white dwarfs and neutron stars
(see the textbook \cite{Shapiro}; more recent reviews of compact
stars are in \cite{generaliNS}). Let us consider $N$ fermions in a
sphere of radius $R$ at $T=0$; the number of fermion per volume
unit scales as $\dd n\sim{N}/{R^3}$; the volume per fermion is
therefore $\dd\sim \frac 1 {n}$ and the uncertainty on the
position is of the order of $n^{-1/3}$; the Fermi momentum is of
the order of the uncertainty on the fermion momentum and therefore
\bes p_F\sim n^{1/3}\hbar~,\ees a result we obtained already under
more stringent hypotheses (Fermi-Dirac distribution) and derived
again here using only the uncertainty relations. The Fermi energy
of the baryons is therefore
 \bes \epsilon_F\sim
\frac{\hbar cN^{1/3}}{R}\ ,\ees if $N$ is the total number of
baryons. Note that this applies both to neutron stars and to
electron stars, because also in stars where the pressure mainly
come from electrons there will be a considerable amount of protons
and neutrons and the largest part of the energy comes from the
baryons, not from the electrons.
 On the other hand the gravitational energy per baryon is
\bes E_G\sim -\frac{GNm_B^2}{R}\ ,\ees and the total
 energy  can be estimated as\be
E=E_G+E_F\sim \frac{\hbar cN^{1/3}}{R}-\frac{GNm_B^2}{R}\ . \ee
Now equilibrium can exist only if $E\ge 0$. As a matter of fact if
$E<0$ ($N$ large) then $\lim_{R\to 0} E=-\infty$, which means that
the energy is unbounded from below and the system is unstable.
Therefore, $E\ge 0$ gives the maximum number of baryons as
follows: \be N\le N_{max}=\left(\frac{\hbar
c}{Gm_B^2}\right)^{3/2}\sim 2\times 10^{57}\ .\ee As a consequence
the maximum mass is \be M_{max}=N_{max}m_B=1.5 M_\odot\
.\label{chandra}\ee This mass can be estimated better and its
better determination ($\sim 1.4 M_\odot$) is known as the
Chandrasekhar limit; for our purposes the estimate (\ref{chandra})
is however sufficient. Notice that the Chandrasekhar limit is
similar for compact stars where the degeneracy pressure is mainly
supplied by electrons and those where it is supplied by baryons.

 One can also estimate the radius of a star whose mass is given by (\ref{chandra}).
 One has
\be \epsilon_F\geq mc^2 \sim\frac{\hbar
c}{R}\,N_{max}^{1/3}\,\sim\, \frac{\hbar c}{R}\left(\frac{\hbar
c}{Gm_B^2}\right)^{1/2}\ee and, therefore,
%\beas&& =5\times 10^8 \,cm\ \{m=m_e \}\cr
% R\sim\frac{\hbar
%}{mc}\left(\frac{\hbar c}{Gm_B^2}\right)^{1/2}\Big\{&&\cr
%&&=
%3\times \,10^5 \,cm\{m=m_n\}\eeas
% ......
\bes R\sim\frac{\hbar }{mc}\left(\frac{\hbar c}{Gm_B^2}
\right)^{1/2}\,=\, \Big\{^{\ 5\times 10^8 \,cm\ \{m=m_e \}}_{\
3\times \,10^5 \,cm\ \{m=m_n \}.} \ees
 If a neutron star accretes its mass beyond the  Chandrasekhar
limit nothing can prevent the collapse and it becomes a black
hole\footnote{The exact determination of the mass limit depends on
the model for nuclear forces; for example in \cite{cameron} the
neutron star mass limit is increased beyond $1.4 M_\odot$.}. In
the following table we summarize our discussion; notice that we
report for the various stars also the value of the parameter
$GM/Rc^2$ i.e. the ratio of the Schwarzschild radius to the star's
radius. Its smallness
 measures the validity of the approximation of neglecting the
 general relativity effects; one can see that for the sun and the white
 dwarfs the newtonian treatment of
 gravity represents a fairly good approximation.
\begin{center}
\begin{tabular}{|c|c|c|c|c|}
  \hline &&&& \\ $$ & $M$ & $R$ &$\dd\rho\left(\frac{g}{cm^3}\right)$ &
  $\dd \frac{GM}{Rc^2}$  \\ &&&&\\
  \hline &&&&\\
  Sun & $M_\odot$ & $\dd R_\odot$ & 1 & $10^{-6}$ \\ &&&&\\
  White Dwarf & $\leq M_\odot$ & $\dd 10^{-2}R_\odot$ & $\leq 10^7$&  $10^{-4}$
  \\ &&&&\\
  Neutron Star & $1-3 M_\odot$ & $\dd 10^{-5}R_\odot$ & $\leq 10^{15}$&
 $10^{-1}$  \\ &&&&\\
  Black Hole & arbitrary & $\dd\frac{2GM}{c^2}$ & $\sim \frac{M}{R^3}$&  $1$ \\ &&&&\\ \hline
\end{tabular}\vskip.3cm \par\noindent
Table 5.1 Paramers of different stellar objects.\end{center}
\vskip.5cm

 Neutron stars are the most likely candidate  for the
theoretical description of pulsars. Pulsars are rapidly rotating
stellar objects, discovered in 1967 by  Hewish and collaborators
and identified as rotating neutron stars by Gold \cite{gold}; so
far about 1200 pulsars have been identified.

Pulsars are characterized by the presence of strong magnetic
fields with the magnetic and rotational axis misaligned; therefore
they continuously emit electromagnetic energy (in the form of
radio waves) and constitute indeed a very efficient mean to
convert rotational energy into electromagnetic radiation. The
rotational energy loss is due to dipole radiation and is therefore
given by \be \frac{dE}{dt}= I\omega\frac{d\omega}{dt}=
-\frac{B^2R^6\omega^4\sin^2\theta}{6c^3}\ .\ee Typical values in
this formula are, for the moment of inertia $I\sim R^5\rho\sim
10^{45}$g/cm$^3$, magnetic fields $B\sim 10^{12}$ G, periods
$T=2\pi/\omega$ in the range $1.5$ msec-8.5 sec; these periods
increase slowly \footnote{Rotational period and its derivative can
be used to estimate the pulsar's age by the approximate formula
$\dd \frac{T}{2\frac{dT}{dt}}$, see e.g. \cite{larimer}.}, with
derivatives $\dd\frac{dT}{dt}\sim 10^{-12}$ - $10^{-21}$, and
never decrease except for occasional jumps (called {\it
glitches}).

Glitches were first observed in the Crab and Vela pulsars in 1969;
the variations in the rotational frequency are of the order
$10^{-8}-10^{-6}$.

 This last feature is the most significant
phenomenon pointing to neutron stars as a model of pulsars in
comparison  to other form of hadronic matter, such as strange
quarks. It will be discussed in more detail in par. \ref{glitch},
where we will examine the possible role played by the crystalline
superconducting phase. In the subsequent three paragraphs we will
instead deal with other possible astrophysical implications of
color superconductivity.

\section{Supernovae neutrinos and cooling of neutron stars}
Neutrino diffusion is the single most important mechanism in the
cooling of young neutron stars, i.e. with an age $<10^{5}$ years;
it affects both the early stage and the late time evolution of
these compact stars. To begin with let us consider  the early
evolution of a Type II Supernova.

Type II supernovae are supposed to be born by collapse of massive
($M \sim 8 - 20  M_\odot$) stars \footnote{The other supernovae,
i.e. type I supernovae, result from the complete explosion of a
star with $4M_\odot\leq M \leq 8 M_\odot$ with no remnants.}.
These massive stars have unstable iron cores\footnote{Fusion
processes favor the formation of iron, as the binding energy per
nucleon in nuclei has a maximum for $A\sim 60$.} with masses
 of the order
 of the Chandrasekhar mass. The explosion producing the supernova
originates within the core, while the external mantle of the red
giant star produces remnants that can be analyzed by different
means, optical, radio and X rays. These studies agree with the
hypothesis of a core explosion. The emitted energy ($\sim 10^{51}
erg$) is much less than the total gravitational energy of the
star, which confirms that the remnants are produced by the outer
envelope of the massive star; the bulk of the gravitational
energy, of the order of $10^{53}$ erg, becomes internal energy of
the proto neutron star (PNS). The suggestion that neutron stars
may be formed in supernovae explosions was advanced in 1934 by
Baade and Zwicky \cite{baade} and it has been subsequently
confirmed by the observation of the Crab pulsar in the remnant of
the Crab supernova observed in China in 1054 A.D.

We do not proceed in this description as it is beyond the scope of
this review and we concentrate our attention on the cooling of the
PNS \footnote{See \cite{colgate} for further discussions.}, which
is mostly realized through neutrino diffusion. By this mechanism
one passes from the initial temperature $T\sim 20-30$ MeV to the
cooler temperatures of the neutron star at subsequent stages. This
phase of fast cooling lasts 10-20 secs and the neutrinos emitted
during it have mean energy $\sim 20$ MeV. These properties, that
can be predicted theoretically, are also confirmed by data from SN
1987A.

The role of quark color superconductivity at this stage of the
evolution of the neutron stars has been discussed in \cite{reddy}.
In this paper the neutrino mean free path is computed in a color
super-conducting medium made up by quarks in two flavor (2SC
model). The results obtained indicate that the cooling process by
neutrino emission slows down when the quark matter undergoes the
phase transition to the superconducting phase at the critical
temperature $T_c$, but then accelerates when $T$ decreases below
$T_c$. There should be therefore changes in the neutrino emission
by the PNS and they might be observed in some future supernova
event; this would produce an interesting test for the existence of
a  color superconducting phase in compact stars.

 Let us now
consider the subsequent evolution of the neutron star, which also
depends on neutrino diffusion. The simplest processes of neutrino
production are the so called direct Urca processes \bea
&&f_1+\ell\to f_2+\nu_\ell\ ,\cr &&\cr &&f_2\to f_1+\ell\to
f_2+\bar\nu_\ell\ ;\eea by these reactions, in absence of quark
superconductivity, the interior temperature $T$ of the star drops
below $10^9$ K ($\sim 100$ KeV) in a few minutes and in $10^2$
years to temperatures $\sim 10^7$ K. Generally speaking the effect
of the formation of gaps is to slow down the cooling, as it
reduces
 both the emissivity and the specific heat. However not only quarks, if present in the neutron star, but also
other fermions, such as neutrons, protons or hyperons have gaps,
 as the formation of fermion pairs is unavoidable if there is an
attractive attraction in any channel (see the discussion in par.
\ref{wilson}). Therefore, besides quark color superconductivity,
one has also the phenomenon of baryon superconductivity and
neutron superfluidity, which is the form assumed by this
phenomenon for neutral particles. The analysis is therefore rather
complicated; the thermal evolution of a late time neutron star has
been discussed in \cite{latetime}, but no clear signature for the
presence of color superconductivity seems to emerge from the
theoretical simulations and, therefore, one may tentatively
conclude that the late time evolution of the neutron stars does
not offer a good laboratory to test the existence of color
superconductivity in compact stars.

\section{$R$-mode instabilities in neutron stars and strange stars\label{rmode}}
Rotating relativistic stars are in general unstable against the
rotational mode ($r$-mode instability) \cite{andersson}. The
instability is due to  the emission of angular momentum by
gravitational waves from the mode. Unless it is damped by
viscosity effects, this instability would spin down the star in
relatively short times. More recently it has been realized
\cite{bildstein} that in neutron stars there is an important
viscous interaction damping the $r$-mode i.e. that between the
external metallic crust and the neutron superfluid. The
consequence of this damping is that $r$-modes are significant only
for young neutron stars, with periods $T<2$ msec. For larger
rotating periods the damping of the $r-$mode implies that the
stars slow down only due to magnetic dipole braking.

All this discussion is relevant for the nature of pulsars: Are
they neutron stars or strange stars?

The existence of strange stars, i.e. compact stars made of quarks
$u,\,d,\,s$ in equal ratios would be a consequence of the
existence of stable strangelets. This hypothetical form of nuclear
matter, made of a large number of $u,\,d,\,s$ quarks has been
suggested by Bodmer \cite{bodmer} and Witten \cite{witten} as
energetically favored in comparison to other hadronic phases when
a large baryonic number is involved. The reason is that in this
way the fermions, being of different flavors, could circumvent the
Pauli principle and have a lower energy, in spite of the larger
strange quark mass. If strangelets do exist, basically all the
pulsars should be strange stars because the annihilation of
strange stars, for example from a binary system, would fill the
space around with strangelets that, in turn, would convert
ordinary  nuclear stars into strange stars.

An argument in favor of the identification of pulsars with strange
stars is the scarcity of pulsars with very high frequency ($T<2.5$
msec).

This seems to indicate that indeed the $r-$mode instability is
effective in slowing down the compact stars and
 favors strange stars, where, differently from neutron stars,
 the crust can be absent.
Even in the presence of the external crust, that in a quark star
can be formed by the gas after the supernova explosion or
subsequent accretion, the dampening of the $r-mode$ is less
efficient. As a matter of fact, since electrons are only slightly
bounded, in comparison with quarks that are confined, they tend to
form an atmosphere having a thickness of a few hundred Fermi; this
atmosphere produces a separation  between the nuclear crust and
the inner quark matter and therefore the viscosity is much
smaller.

In quark matter with color superconductivity the presence of gaps
$\Delta\gg T$  exponentially reduces the bulk and shear viscosity,
which renders the $r-$mode unstable. According to Madsen
\cite{madsen,madsen2} this would rule out compact stars entirely
made of quarks in the CFL case (the 2SC model would be marginally
compatible, as there are ungapped quarks in this case). For
example for $\Delta> 1$ MeV any star having $T<10$ msec would be
unstable, which would contradict the observed existence of pulsars
with time period less than 10 msec.

However this conclusion does not rule out the possibility of
neutron stars with a quark core in the color superconducting
state, because as we have stressed already, for them the dampening
of the $r-$mode instability would be provided by the viscous
interaction between the nuclear crust and the neutron superfluid.
\section{Miscellaneous results}
Color superconductivity is a Fermi surface phenomenon and as such,
it does not affect significantly the equation of state of the
compact star. Effects of this phase could be seen in other
astrophysical contexts, such as those considered in the two
previous paragraphs or in relation to the pulsar glitches, which
will be examined in the next paragraph. A few other investigations
have been performed in the quest of possible astrophysical
signatures of color superconductivity; for instance in
\cite{sannino2} it has been suggested that the existence of a 2SC
phase might be partly responsible of the gamma ray bursts, due to
the presence in the two-flavor superconducting phase of a light
glueball that can decay into two photons. Another interesting
possibility is related to the stability of strangelets, because,
as observed in \cite{madsen3}, CFL strangelets, i.e. lumps of
strange quark matter in the CFL phase may be significantly more
stable than strangelets without color superconductivity.

Finally I wish to mention the observation of \cite{alford1}
concerning the evolution of the magnetic field in the interior of
neutron stars. Inside an ordinary neutron star, neutron pairs are
responsible for superfluidity, while proton pairs produce BCS
superconductivity. In this condition magnetic fields experience
the ordinary Meissner effect and are either expelled or restricted
to flux tubes where there is no pairing. In the CFL (and also 2SC)
case, as we know from paragraph \ref{em}, a particular $U(1)$
group generated by \be\tilde Q=
 {\bf 1}\otimes Q+{T}\otimes{\bf 1}\label{u1}\ee
 remains unbroken and plays the role of
electromagnetism. Instead of being totally dragged out or confined
in flux tubes, the magnetic field will partly experience Meissner
effect (the component $\tilde A_\mu$), while the remaining part
will remain free in the star. During the slowing down this
component of the magnetic field should not decay because, even
though the color superconductor is not a BCS conductor for the
group generated by (\ref{u1}), it may be a good conductor due to
the presence of the electrons in the compact star. Therefore it
has been suggested \cite{alford1} that a quark matter core inside
a neutron star may serve as an "anchor" for the magnetic field.
\section{Glitches in neutron stars\label{glitch}}
Glitches are a typical phenomenon of of the pulsars, in the sense
that probably all the pulsar have glitches (for a recent review
see \cite{glitches}). Several models have been proposed to explain
the glitches. Their most popular explanation is based on the idea
that these sudden jumps of the rotational frequency are due to the
the angular momentum stored in the superfluid neutrons in the
inner crust (see Fig. 5.1), more precisely in vortices pinned to
nuclei. When the star slows down, the superfluid neutrons do not
participate in the movement, until the state becomes unstable and
there is a release of angular momentum to the crust, which is seen
as a jump in the rotational frequency.

The presence of glitches is one of the main reasons for the
identification of pulsars with neutron stars; as a matter of fact
neutron stars are supposed to have a dense metallic crust,
differently from quark stars where the crust is absent or, if
present, is much less dense ($\approx 10^{11}$ g cm$^{-3}$).

 A schematic and a somehow exaggerated diagram
of glitches is shown in fig. 5.2.
\begin{figure}[htb]
\epsfxsize=12truecm \centerline{\epsffile{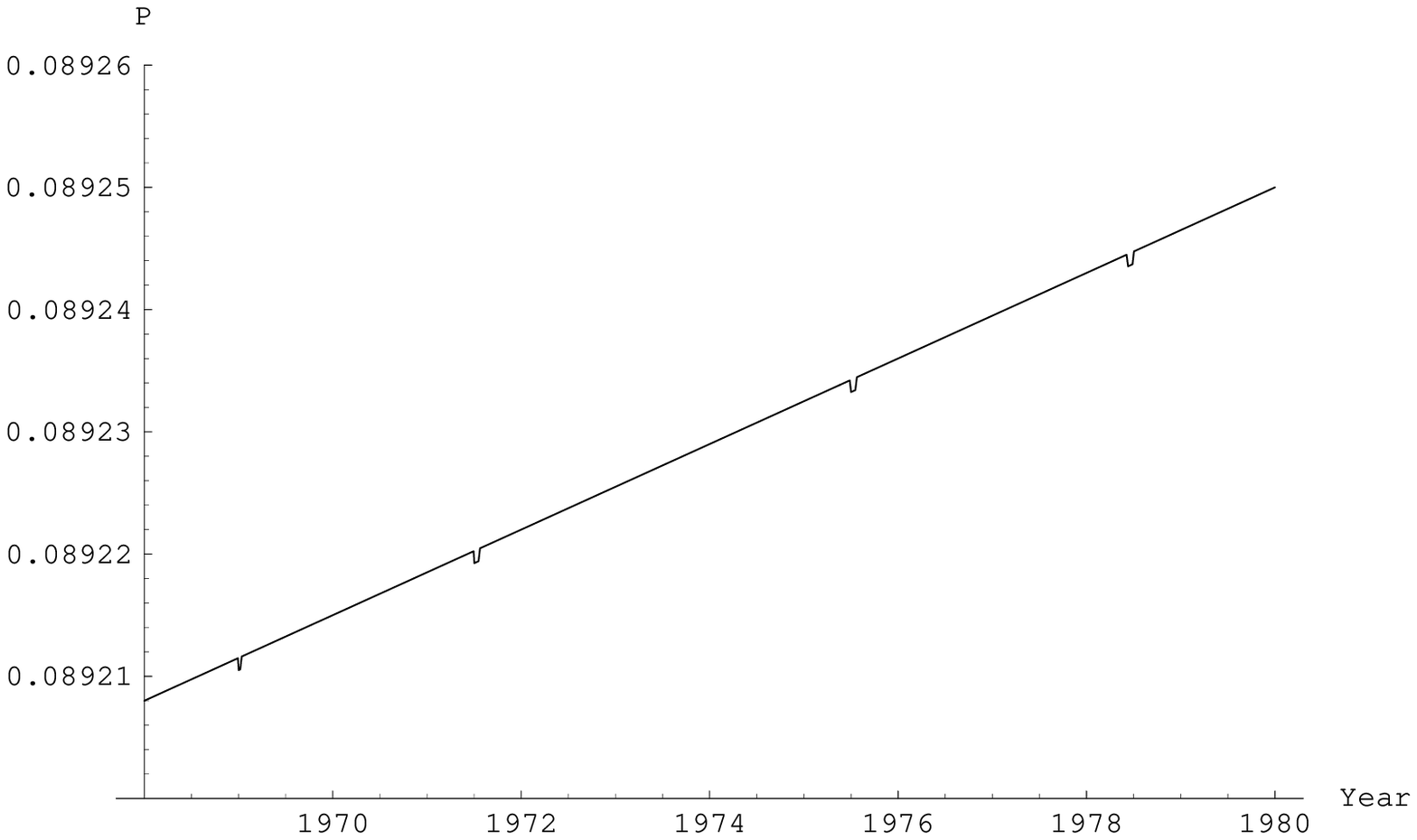}}
\noindent {Fig. 5.2}  {Schematic view of the period $P$ of the PSR
0833-45 (Vela) showing four glitches in the years 1969-1980. Other
glitches have been observed in the subsequent years.}
\end{figure}
The interesting aspect of the LOFF phase discussed in chapter
 \ref{ch5} is that even in quark stars, provided we are in a color
superconductivity phase, one can have a crystal structure able to
produce glitches; it would be given by a lattice characterized by
a geometric array where the gap parameter varies periodically.
This would avoid the objection raised in \cite{Caldwell} that
excludes the existence of strange stars because, if the strange
matter exists, strange stars should be rather common, as we
discussed in par. \ref{rmode}, in contrast with the widespread
appearance of glitches in pulsars. Therefore, if the color
crystalline structure is able to produce glitches, the argument in
favor of the existence of strange stars would be reinforced.

 In a more conservative vein one can
also imagine that the LOFF phase be realized in the inner core of
neutron star; in this case the crystalline color superconductivity
could be partly responsible for the glitches of the pulsar. Some
estimates of the effect are discussed in \cite{LOFF}; a detailed
analysis of this scenario is however premature as one should first
complete the study of the LOFF phase in two directions, first by
including the third quark (all the study in section \ref{ch5} was
relative to the two flavor case) and, second, by sorting out the
exact form of the color lattice, as the plane wave ansatz
discussed in section \ref{ch5} might not correspond to the true
vacuum of the theory.

% ----------------------------------------------------------------
\vskip.6cm\par\noindent
 {\bf Acknowledgements}
\par\noindent
I would like to thank  R. Casalbuoni for reading the manuscript
and for several long-distance, long-lasting calls that helped me
in understanding the subject of this review. I am grateful to R.
Gatto and M. Mannarelli for a very pleasant scientific
collaboration on the themes of this review. It is also a pleasure
to thank K. Rajagopal for many enlightening discussions on color
superconductivity, I. Bombaci for a useful correspondence on the
astrophysical consequences of the LOFF phase and J. R. Pel\'aez
for useful discussions.
  
% ----------------------------------------------------------------


\newpage
\begin{thebibliography}{99}
\bibitem{others}BARROIS B.: {\it Nucl. Phys. B}, {\bf 129},
 (1977) 390; S. FRAUTSCHI, in {\it
 Proceedings of the workshop on hadronic
matter at extreme density, Erice 1978}. See also: BAILIN D. AND
LOVE A.: {\it Phys. Rept.}, {\bf 107}, (1984) 325, and references
therein.
%
\bibitem{alford}ALFORD M., RAJAGOPAL K. AND
WILCZEK F.: {\it Phys. Lett. B}, {\bf 422}, (1998) 247
[arXiv:hep-ph/9711395].
%
\bibitem{wilczekcfl}
ALFORD M., RAJAGOPAL K. AND WILCZEK F: {\it Nucl. Phys. B}, {\bf
537}, (1999) 443 [arXiv:hep-ph/9804403].
\bibitem{wilczekcfl1}
ALFORD M., BERGES J. AND RAJAGOPAL K.: {\it Nucl. Phys. B}, {\bf
558}, (1999) 219 [arXiv:hep-ph/9903502].
%
\bibitem{wilczekcfl2}
SCH\"AFER T. AND WILCZEK F.: {\it Phys. Rev. Lett.},  {\bf 82},
(1999) 3956 [arXiv:hep-ph/9811473].
%
\bibitem{wilczekcfl3}
SCH\"AFER T. AND WILCZEK F.: {\it Phys. Rev. D}, {\bf 60}, (1999)
114033 [arXiv:hep-ph/9906512].
%
\bibitem{wilczekcfl4}
SCH\"AFER T. AND WILCZEK F.: {\it Phys. Rev. D}, {\bf 60}, (1999)
074014 [arXiv:hep-ph/9903503].
%
\bibitem{Schafer:1998na}
SCHAFER T. AND WILCZEK F.: {\it Phys.\ Lett.\ B}, {\bf 450},
(1999)
 325 [arXiv:hep-ph/9810509].
%%CITATION = HEP-PH 9810509;%%
%
\bibitem{generali}CARTER G. W. AND DIAKONOV D:
{\it Phys. Rev. D}, {\bf 60}, (1999) 16004 [arXiv:hep-ph/9812445].
%
\bibitem{shuryak}RAPP R., SCH\"AFER T.,
SHURYAK  E. V. AND M. VELKOVSKY M.: {\it Phys. Rev. Lett.}, {\bf
81}, (1998) 53 [arXiv:nucl-th/9711396].
%
\bibitem{alford1}ALFORD M., BERGES J. AND RAJAGOPAL K.:
{\it Nucl.\ Phys.\ B}, {\bf 571}, (2000) 269
[arXiv:hep-ph/9910254].
%
\bibitem{pisarski}PISARSKI R. D. AND
RISCHKE D.H.: {\it Phys. Rev. Lett.}, {\bf 83},  (1999) 37
[arXiv:nucl-th/9811104]; {\it Phys. Rev. D}, {\bf 61}, (2000)
 051501 [arXiv:nucl-th/9907041].
%
\bibitem{altri}AGASIAN N. O., KERBIKOV  B.O. AND
SCHEVCHENKO  V. I.: {\it Phys. Rep.}, {\bf 320}, (1999)  131
[arXiv:hep-ph/9902335]; SHUSTER E. AND  SON D. T.: {\it Nucl.
Phys.}, {\bf 573}, (2000)  434 [arXiv:hep-ph/9905448]; HONG D. K.,
MIRANSKY V. A.,  SHOVKOVY I. A. AND  WIJEWARDHANA L. C. R.: {\it
Phys. Rev. D}, {\bf 61}, (2000)  056001, erratum, {\it ibidem},
{\bf 62}, (2000) 59903  [arXiv:hep-ph/9906478].
%
\bibitem{rassegne}RAJAGOPAL K. AND F.
WILCZEK F.: in {\it Handbook of QCD}, M.Shifman, ed. (World
Scientific, 2001) [arXiv:hep-ph/0011333].
\bibitem{rassegne2} HONG D. K.: [arXiv:hep-ph/0101025];
ALFORD M.: [arXiv:hep-ph/0102047].
%
\bibitem{hsu} HSU S. D. H.: [arXiv:hep-ph/0003140].
%
\bibitem{bardeen}BARDEEN J.,  COOPER L.N. AND SCHRIEFFER J. R.:
{\it Phys. Rev.}, {\bf 106}, (1957) 162; {\bf 108}, (1957) 1175;
see also SCHRIEFFER J. R.: {\it Theory of Superconductivity}
(Benjamin/Cummings, Menlo Park) 1964.
%
%\cite{Ebert:2001ba}
\bibitem{klimenko}
EBERT D., KHUDYAKOV V.~V., ZHUKOVSKY  V.~C.~AND KLIMENKO K.~G.:
%``The influence of an external chromomagnetic field on color  superconductivity,''
Phys.\ Rev.\ D {\bf 65}, 054024 (2002) [arXiv:hep-ph/0106110];
EBERT D., KHUDYAKOV V.~V.~, KLIMENKO K.~G., TOKI H.~ AND ZHUKOVSKY
V.~C.:
%``Chromomagnetic catalysis of chiral symmetry breaking and color  superconductivity,''
[arXiv:hep-ph/0108185].

%
\bibitem{casalbuonisannino}CASALBUONI R., DUAN Z. AND
SANNINO F., {\it Phys. Rev. D}, {\bf 62},  (2000) 094004
[arXiv:hep-ph/0004207].
%
\bibitem{casalbuonigatto}CASALBUONI R. AND GATTO R.: {\it Phys.
 Lett. B}, {\bf 464},  (1999) 111 [arXiv:hep-ph/9908227].
%
\bibitem{Polchinski:1992ed}
POLCHINSKI J.: in {\it  Proc. of the 1992 TASI}, eds. J. Harvey
and J. Polchinski (Singapore, 1993)
%``Effective field theory and the Fermi surface,''
[arXiv:hep-th/9210046].
%
%\cite{Shankar:1993pf}
\bibitem{Shankar:1993pf}
SHANKAR R.:
%``Renormalization group approach to interacting fermions,''
{\it Rev.\ Mod.\ Phys.\ },  {\bf 66}, (1994) 129.
%
\bibitem{benfatto}BENFATTO G. AND GALLAVOTTI G.:
{\it J. Stat. Phys.}, {\bf 59},  (1990) 541; {\it Phys. Rev. C}
{\bf 42}, (1990)  9967.
%
\bibitem{wilson}WILSON K. G.: {\it Phys. Rev. B}, {\bf 4},
(1981) 3174; {\bf 4}, (1981) 3184.
%
\bibitem{landau}See e.g. ABRIKOSOV A. A., GORKOV L. P. AND
DZYALOSHINSKI I. E.: {\it Methods of Quantum Field Theory in
Statistical Mechanics} (Dover, New York) 1963.
%
\bibitem{Evans:1998ek}
EVANS N., HSU S. D. AND SCHWETZ M.:
%``An effective field theory approach to color superconductivity at high  quark density,''
{\it Nucl.\ Phys.\ B}, {\bf 551}, (1999) 275
[arXiv:hep-ph/9808444].
%
\bibitem{hong}
HONG D. K.: {\it Phys. Lett. B}, {\bf 473}, (2000) 118 [arXiv:
hep-ph/9812510]; HONG D. K.: {\it Nucl. Phys. B}, {\bf 582},
(2000) 451 [arXiv:hep-ph/9905523].
%
\bibitem{beane}
BEANE S. R., BEDAQUE P. F.  AND SAVAGE M. J.:
 {\it Phys. Lett. B},
{\bf 483},  (2000) 131 [arXiv:hep-ph/0002209]
%
\bibitem{cflgatto}
CASALBUONI R., GATTO R. AND NARDULLI G.: {\it Phys. Lett. B}, {\bf
498}, (2001) 179 [arXiv:hep-ph/0010321].
%
\bibitem{HQET}For reviews of the heavy quark effective theory see
for example the textbook: MANOHAR A. V. AND WISE M. B.: {\it Heavy
Quark Physics} (Cambridge Univ. Press, 2000) and CASALBUONI R. ET
AL.: {\it Phys. Rept.}, {\bf 281},  (1997) 145
[arXiv:hep-ph/9605342].
%
\bibitem{2fla}
CASALBUONI R., GATTO R., MANNARELLI M. AND NARDULLI G.: {\it Phys.
Lett. B}, {\bf 524}, (2002) 144 [arXiv:hep-ph/0107024].
%
\bibitem{gattoloff}
CASALBUONI R., GATTO R., MANNARELLI M. AND NARDULLI G.: Phys.
Lett. B {\bf 511}, (2001) 218 [arXiv:hep-ph/0101326].
%
\bibitem{nardulliloff}NARDULLI G.: [arXiv:hep-ph/0111178]
%
\bibitem{gattoloff2}CASALBUONI R., GATTO R., MANNARELLI M. AND NARDULLI G.:
[arXiv:hep-ph/0201059].
%
\bibitem{LOFF}ALFORD M., BOWERS J. A. AND RAJAGOPAL K.;
{\it Phys. Rev. D}, {\bf 63},  (2001) 074016
[arXiv:hep-ph/0008208]; BOWERS J. A.,  KUNDU J., RAJAGOPAL K. AND
SHUSTER E.: {\it Phys. Rev. D}, {\bf 64}, (2001)  014024
[arXiv:hep-ph/0101067].
%
\bibitem{LOFFbis}LEIBOVICH A. K., RAJAGOPAL K., SHUSTER E.:
 {\it Phys. Rev. D}, {\bf 64}, (2001)  094005 [arXiv:hep-ph/0104073].
 %
\bibitem{LOFF3}RAJAGOPAL K.: {\it Acta Phys. Polon. B}, {\bf 31},
(2000) 3021 [arXiv:hep-ph/0009058]; ALFORD M., BOWERS J. A. AND
RAJAGOPAL K.: {\it J. Phys. G}, {\bf 27},  (2001) 541
[arXiv:hep-ph/0009357].
\bibitem{LOFF4}SCH\"AFER T. AND SHURYAK E.: [arXiv:nucl-th/0010049]; K.
RAJAGOPAL K.: in {\it QCD@Work}, P. Colangelo and G. Nardulli eds.
(AIP, 2001) [arXiv:hep-ph/0109135].
%
\bibitem{earlier1} DERYAGIN D. V.,  GRIGORIEV D. YU. AND
RUBAKOV V.A.: {\it Int. J. Mod. Phys. A}, {\bf 7},  (1992) 659.
%
\bibitem{originalloff} LARKIN  A. I. AND
 OVCHINNIKOV YU. N.: {\it Zh. Eksp. Teor. Fiz.}, {\bf 47}, (1964) 1136 (
{\it Sov. Phys. JETP}, {\bf 20}, (1965) 762); FULDE P. AND FERRELL
R. A.: {\it Phys. Rev.}, {\bf 135}, (1964) A550.

\bibitem{Eliashberg}ELIASHBERG G. M.: {\it Zh. Eksp- Teor. Fiz.}, {\bf
34}, 735 (1960) (transl. {\it Sov. Phys. JETP} {\bf 11}, (1960)
696.
%
\bibitem{NJL}NAMBU Y. AND JONA LASINIO G., {\it Phys. Rev.}, {\bf 122},
(1961) 345 ; {\bf 124}, (1961) 246.
%
\bibitem{oneflavor1}SCH\"AFER T.: {\it Phys. Rev. D}, {\bf 62}, (2000)
094007 [arXiv:hep-ph/0006034].
%
\bibitem{bogolubov}BOGOLIUBOV N. N.: {\it Nuovo Cimento} {\bf 7},
(1958) 694; VALATIN J.: {\it Nuovo Cimento}, {\bf 7}, (1958) 843.
%
\bibitem{sonstephanov}SON D. T. AND  STEPHANOV M. A.:
{\it Phys. Rev. D },{\bf 61},  (2000) 074012
[arXiv:hep-ph/9910491]; {\it ibidem} {\bf 62}, (2000) 059902
[arXiv:hep-ph/0004095].
%
\bibitem{zarembo}ZAREMBO K.: {\it Phys. Rev. D}, {\bf 62}, (2000) 054003
[arXiv:hep-ph/0002123].
\bibitem{rho1}RHO M., WIRZBA A. AND ZAHED I.:
{\it Phys. Lett. B}, {\bf 473},  (2000) 126
[arXiv:hep-ph/9910550].
%
\bibitem{rho2}RHO M., SHURYAK E., WIRZBA A. AND ZAHED I.:
{\it Nucl. Phys. A} {\bf 676}, (2000) 273 [arXiv:hep-ph/0001104].
%
\bibitem{honglee}HONG D. K., LEE T. AND MIN D.: {\it Phys. Lett. B}, {\bf
477}, (2000) 137 [arXiv:hep-ph/9912531].
%
\bibitem{tytgat}MANUEL C. AND TYTGAT M. H.: {\it Phys. Lett. B}, {\bf
479}, (2000) 190 [arXiv:hep-ph/0001095].
%
\bibitem{bando}
BALACHANDRAN A.~P., STERN A. AND TRAHERN C.~G.:
%``Nonlinear Models As Gauge Theories,''
{\it Phys.\ Rev.\ D}, {\bf 19},  (1979) 2416; BANDO M., KUGO T.
AND YAMAWAKI K.:
%``Nonlinear Realization And Hidden Local Symmetries,''
{\it Phys.\ Rept.},  {\bf 164}, (1988) 217.
%%CITATION = PRPLC,164,217;%%
%
\bibitem{eguchi}EGUCHI T.: {\it Phys. Rev. D}. {\bf 14},
(1976) 2755.
%
\bibitem{shovkovy}
 GUSYNIN V.P. AND  SHOVKOVY I.A.: [arXiv:hep-ph/0108175].
%
\bibitem{rischke1}
RISCHKE D. H.: {\it Phys. Rev. D},  {\bf 62}, (2000) 034007
[arXiv:nucl-th/0001040].

\bibitem{sonrishke}
 RISCHKE D.H., SON D.T. AND  STEPHANOV M.A.: {\it Phys. Rev. Lett.}, {\bf
87}, (2001) 062001 [arXiv:hep-ph/0011379].
%%
\bibitem{sannino2}OUYED R. AND SANNINO F.: {\it Phys. Lett. B}, {\bf
511}, (2001) 66  [arXiv:hep-ph/0103168].
%
\bibitem{CJT} CORNWALL J. M., JACKIW R. AND  TOMBOULIS E.:
{\it Phys. Rev. D}, {\bf 10}, (1974) 2428.
%
\bibitem{casalbuonisannino2}CASALBUONI R., DUAN Z. AND
SANNINO F.: {\it Phys. Rev. D}, {\bf 63},  (2001) 114026
[arXiv:hep-ph/0011394].
%
\bibitem{continuity}SCH\"AFER T. AND WILCZEK F.: {\it Phys. Rev.
Lett.},
 {\bf 82},  (1999) 3956 [arXiv:hep-ph/9811473].
%
\bibitem{son}SON D. T.:
{\it Phys. Rev. D}, {\bf 59},  (1999) 094019
[arXiv:hep-ph/9812287].
%
\bibitem{rischke}PISARSKI R. D. AND  RISCHKE D. H.:
{\it Phys. Rev. D}, {\bf 61},  (2000) 051501
[arXiv:hep-ph/9812287].
%
\bibitem{Barrois}BARROIS B.: {\it Nonperturbative effects in dense quark
matter}, preprint UMI 79-04847.mc (1979).
%
\bibitem{Kogut:eq}
~KOGUT J.~B., LOMBARDO M.~P. AND SINCLAIR D.~K.:
%``Quenched QCD At Finite Density,''
{\it Phys.\ Rev.\ D}, {\bf 51},  (1995) 1282
[arXiv:hep-lat/9401039].
%%CITATION = HEP-LAT 9401039;%%
%
\bibitem{Kogut:1994wf}
~KOGUT J.~B., LOMBARDO M.~P. AND SINCLAIR D.~K.:
%``Quenched QCD at finite density: g = 1 and g = infinity,''
{\it Nucl.\ Phys.\ Proc.\ Suppl.},  {\bf 42},  (1995) 514
[arXiv:hep-lat/9412057].
%%CITATION = HEP-LAT 9412057;%%
%
\bibitem{isospin}SON D. T. AND STEPHANOV M. A.:
{\it Phys. Rev. Lett.}, {\bf 86},  (2001) 592
[arXiv:hep-ph/0005225].
%
\bibitem{Alford:1998sd}
ALFORD M., KAPUSTIN A. AND WILCZEK F.:
%``Imaginary chemical potential and finite fermion density on the lattice,''
{\it Phys.\ Rev.\ D}, {\bf 59},  (1999) 054502
[arXiv:hep-lat/9807039].
%%CITATION = HEP-LAT 9807039;%%
%
\bibitem{wong}WONG S. K.: {\it Nuovo Cimento}, {\bf 65 A},  (1970) 689.
%
\bibitem{heinz}HEINZ U.: {\it Phys. Lett. B}, {\bf 144}, (1984) 228.
%
\bibitem{lucchesi}KELLY P.F., LIU Q., LUCCHESI C. AND
MANUEL C.: [arXiv:hep-ph/9406285].
%
\bibitem{litim}
LITIM D. F. AND MANUEL C.:
%``Transport theory for a two-flavor color superconductor,''
{\it Phys.\ Rev.\ Lett.},  {\bf 87},  (2001) 052002
[arXiv:hep-ph/0103092].
%%CITATION = HEP-PH 0103092;%%

%\cite{Manuel:1995td}
\bibitem{Manuel:1995td}
MANUEL C.:
%``Hard Dense Loops in a Cold Non-Abelian Plasma,''
{\it Phys.\ Rev.\ D}, {\bf 53},  (1996) 5866
[arXiv:hep-ph/9512365].
%%CITATION = HEP-PH 9512365;%%
\bibitem{Braaten:1989mz}
PISARSKI R.~D.:
%``Scattering Amplitudes In Hot Gauge Theories,''
{\it Phys.\ Rev.\ Lett.},  {\bf 63},  (1989) 1129; BRAATEN E. AND
PISARSKI R.~D.:
%``Soft Amplitudes In Hot Gauge Theories: A General Analysis,''
{\it Nucl.\ Phys.\ B}, {\bf 337},  (1990) 569; BLAIZOT J.~P. AND
~IANCU E.:
%``Kinetic equations for long wavelength excitations of the quark - gluon plasma,''
{\it Phys.\ Rev.\ Lett.},  {\bf 70}, (1993) 3376
[arXiv:hep-ph/9301236] and {\it Nucl.\ Phys.\ B}, {\bf 417},
(1994) 608 [arXiv:hep-ph/9306294].
%%CITATION = HEP-PH 9306294;%%
%%CITATION = NUPHA,B337,569;%%;
%%CITATION = PRLTA,63,1129;%%
%
\bibitem{ll}
LIFSHITZ E. AND PITAEVSKII L.: {\it Physical Kinetics}, (Pergamon
Press, Oxford), 1981.
%
\bibitem{manohar}LOW I. AND  MANOHAR A. V.: [arXiv:hep-ph/0110285].
%
\bibitem{Shapiro}
SHAPIRO S. L. AND  TEUKOLSKI S. A.: {\it Black Holes, White Dwarfs
and  Neutron Stars: The Physics of Compact Objects},
(Wiley-Interscience), 1983.
%
\bibitem{generaliNS}S. TSURUTA S.: {\it Phys. Rep.}, {\bf 292}, (1998) 1;
PAGE D.: in {\it The Many Faces of Neutron Stars}, R. Bucheri, J.
van Paradijs and M. A. Alpar eds. (Kluwer Academic Publishers,
Dordrecht) 1998, p. 539.; HEISELBERG H. AND  PANDHARIPANDE V.:
[arXiv:astro-ph/0003276].
%
\bibitem{cameron}CAMERON A. G. W. AND CANUTO V.: {\it  Proc. 16th Solvay
Conf. on Astrophysics and Gravitation}, Universit\`e de Bruxelles,
 (1974) p. 221.
%
\bibitem{gold}GOLD T.: {\it Nature}, {\bf 221}, (1969) 25.
%
\bibitem{larimer} LORIMER D. R.: [astro-ph/9911519].
%
\bibitem{baade}BAADE W. AND  ZWICKY F.: {\it  Proc. Nat. Acad. Sci.}, {\bf
20}, (1934) 255.
%
\bibitem{colgate}COLGATE S. A. AND  WHITE R. H.:
{\it Astrophys. Journ.}, {\bf 143}  (1966) 626;  IMSSHENNIK V. S.
AND  NADYOZHIN D. K.: {\it Sov. Phys. JETP}, {\bf 36}, (1973) 821;
BURROWS A. AND LATTIMER J.M.: {\it Astrophys. Journ.}, {\bf 307},
(1986) 178; KEIL W. AND  JANKA H.T.: {\it Astronomy and
Astrophysics}, {\bf 296}, (1995) 145; PONS J. ET AL.: {\it
Astrophys. Journ.}, {\bf 513}, (1999) 513.
%
\bibitem{reddy}CARTER G. AND REDDY S.: {\it Phys. Rev. D}, {\bf 62}, (2000) 103002,
[arXiv:hep-ph/0005228].
%
\bibitem{latetime}PAGE D.,  PRAKASH M., LATTIMER J. AND STEINER A.: {\it
Phys. Rev. Lett.}, {\bf 85}, (2000) 2048 [arXiv:hep-ph/0005094];
BLASCHKE D., KL\"AHN T. AND  VOSKRESENSKY D. N.:
[arXiv:astro-ph/9908334].
%
\bibitem{andersson}N. ANDERSSON, {\it Astrophys. J.}, {\bf 502},
(1998) 708.
%
\bibitem{bildstein}BILDSTEIN L. AND
USHOMISRKY G.: [arXiv:astro-ph/9911155]; ANDERSSON N.:
[arXiv:astro-ph/0002114].
%
\bibitem{bodmer} BODMER A. R.: {\it Phys. Rev D}, {\bf 4}, (1971) 1601.
%
\bibitem{witten}WITTEN E.: {\it Phys. Rev D}, {\bf 30}, (1984) 272.
%
\bibitem{madsen}
MADSEN J.: {\it  Phys. Rev. Lett.}, {\bf 85},  (2000) 10
[arXiv:astro-ph/9912418].
%
\bibitem{madsen2}MADSEN J.: [arXiv:astro-ph/0111417].
%
\bibitem{madsen3}MADSEN J.:  [arXiv:hep-ph/0108036].
%
\bibitem{glitches}LINK B., EPSTEIN R. AND  LATTIMER J. M.:
[arXiv:astro-ph/0001245].
%
\bibitem{Caldwell}CALDWELL R. R. AND  FRIEDMAN
J. L.: {\it Phys. Lett. B}, {\bf 264},  (1991) 143.
%
\end{thebibliography}
\end{document}